\documentclass[usenatbib]{mn2e}
\usepackage{times}
\usepackage{epsfig}
\usepackage{appendix}
\usepackage{fixltx2e}

\newcommand{\bmodel}{$\beta$-$\rm model$}
\newcommand{\rc}{$r_{\rm core}$}
\newcommand{\ffh}{$0.105\times \rm R_{500}$}
\newcommand{\rfh}{$ \rm R_{500}$}
\newcommand{\fc}{$f_{c}$}
\def\gsim{\mathrel{\rlap{\lower4pt\hbox{\hskip1pt$\sim$}}
    \raise1pt\hbox{$>$}}}                

\title[Evolution of the X-ray Profiles of Poor Clusters from the XMM-LSS Survey]
{Evolution of the X-ray Profiles of Poor Clusters from the XMM-LSS Survey}
\author[A. Alshino et al.]
  {Abdulmonem ~Alshino,$^1$\thanks{E-mail: alshino@star.sr.bham.ac.uk}
  Trevor ~Ponman,$^1$ 
  Florian Pacaud$^2$ and  Marguerite ~Pierre$^3$ \\
  $^1$School of Physics and Astronomy, The University of
  Birmingham, Birmingham B15 2TT, UK. \\
  $^2$Argelander-Institut f\"ur Astronomie, University of Bonn, Auf dem H\"ugel 71, 53121 Bonn, Germany. \\
  $^3$DAPNIA/SAp CEA Saclay, 91191 Gif sur Yvette, France.\\
  \newauthor
  }
\begin{document}

\date{Accepted XXXX Xxxxx XX. Received XXXX Xxxxx XX; in original from XXXX Xxxxx XX}

\pagerange{\pageref{firstpage}--\pageref{lastpage}} \pubyear{xxxx}

\maketitle

\label{firstpage}

\begin{abstract}
A sample consisting of 27 X-ray selected galaxy clusters from the
\textit{XMM-LSS} survey is used to study the evolution in the 
X-ray surface brightness profiles of the hot intracluster plasma.
These systems are mostly groups and poor clusters, with 
temperatures 0.6-4.8 keV, spanning the redshift range
0.05 to 1.05. Comparing the profiles with a standard \bmodel\ motivated
by studies of low redshift groups, we find 54\% of our systems to
possess a central excess, which we identify with a cuspy cool core.
Fitting \bmodel\ profiles, allowing for blurring by the XMM point spread
function, we investigate trends with both temperature
and redshift in the outer slope ($\beta$) of the X-ray surface 
brightness, and 
in the incidence of cuspy cores. Fits to individual cluster profiles
and to profiles stacked in bands of redshift and temperature
indicate that the incidence of cuspy cores does
not decline at high redshifts, as has been reported in rich clusters.
Rather such cores become more prominent with increasing redshift. 
$\beta$ shows a positive correlation with both
redshift and temperature. Given the $\beta$-$T$ trend seen in local
systems, we assume that temperature is the primary driver for this
trend. Our results then demonstrate that this correlation is still
present at $z\gsim0.3$, where most of our clusters reside.
\end{abstract}


\section{Introduction}

Clusters of galaxies, as the largest virialised gravitationally-bound
products of the process of hierarchical structure formation,
are powerful probes for both testing cosmological
models and tracing structural evolution (e.g., \citealt{Voit05}). One
of the most important properties of galaxy clusters is their mass.
Since cluster mass cannot be directly
observed, it is studied indirectly through observables such as
X-ray radiation emitted by the intracluster medium (ICM) which
represents 80\% of the total baryonic component of galaxy
clusters at z=0 (\citealt{Ettori04}) and accounts for about 10\% of the total
(including dark) mass content of clusters
(\citealt{Sarazin86}). The study of the ICM can provide
important insights into the evolution and dynamics of cluster and
their member galaxies.

Observationally, there are two distinct classes of clusters: cool core (CC)
clusters with dense gaseous core regions in which gas temperature drops
inwards, and non-cool core (NCC)
clusters with shallower core profiles which often exhibit more internal
structure (e.g., \citealt{Jones84}; \citealt{Ota04}; \citealt{Peres98}
and \citealt{Schuecker01}). Cool core clusters have sharply peaked
X-ray emission at their centres due to the rise in central gas density
which accompanies the central cooling. However, the gas is not observed to cool
to very low temperatures at the rates naively expected from the observed core
X-ray luminosities, and a consensus has now emerged that
this is due to the effects of feedback from a central active galactic nucleus,
which limits the effects of cooling through processes which are still not
very well understood. For reviews of cool cores in clusters, see for example, 
\citealt{Fabian94}, \citealt{Donahue04} and \citealt{Peterson06}.

In the local Universe, some studies have found that nearly two thirds of
clusters have cool cores (e.g. \citealt{Peres98}, \citealt{White97} and
\citealt{Vikhlinin07}). However, other studies gave different values:
\cite{Edge92} found a CC fraction as high as 90\%, while the
results of \cite{Chen07} indicated that 49\% of local clusters host cool cores. 
These differences relate to both the selection of the cluster sample, and the
way in which cool cores are identified within them.

The evolution of cooling within clusters provides an important probe of
the history of cosmic feedback (\citealt{Voit05}).
At intermediate redshifts ($z\approx0.15-0.4$), \cite{Bauer05},
using a sample of 38 X-ray-luminous clusters, found that cool cores
appeared still to be common, with an incidence
nearly identical to that in luminous low-redshift clusters. Consequently, 
they suggested that heating and cooling processes
must have stabilised in massive clusters since $z\sim0.4$.

At higher redshifts, \cite{Vikhlinin07} reported that the 
fraction of clusters with cuspy X-ray cores dropped from $\sim$70\% at $z\sim0$ to
$\sim$15\% at $z>0.5$.
\cite{Santos08} compared the fraction of clusters with non-cool cores, moderate cool
cores and strong cool cores in nearby ($0.143\leq z
\leq0.3$) and high redshift ($0.7\leq z < 1.4$) clusters. These authors
detected a significant fraction of clusters harbouring moderate cool
cores out to z=1.4, similar to the fraction in their
low-redshift sample. However, they noticed an absence of clusters with
strong-cool cores at redshift $z>0.7$.


Regarding the spatial distribution of the ICM, \cite{Cavaliere76}
introduced the \bmodel\ profile, motivated by the distribution expected
for an isothermal plasma in hydrostatic equilibrium with a
virialised mass distribution. Although it is now known that the gas
is rarely isothermal, the \bmodel\ is generally found to give a reasonable 
representation of X-ray surface brightness profiles (e.g. \citealt{Neumann99}).
However, an additional central component is usually required to fit
the inner regions of CC clusters (e.g. \citealt{Pratt02}), and 
detailed studies of surface brightness profiles of clusters extending to large
radii have shown that the logarithmic slope continues to increase slowly
towards larger radii (e.g. \citealt{Vikhlinin99}, \citealt{Croston08} and 
\citealt{Maughan08}). It is less clear whether this progressive
steepening is also present in galaxy groups. For example, \cite{Rasmussen04}
traced the surface brightness out to \rfh\ in two rich groups and found
them to be well fitted by simple \bmodel s the whole way.

For \bmodel\ fits,
the $\beta$ parameter, which characterises the outer slope of the 
surface brightness profile, has a value of
$\approx 2/3$ for rich clusters (\citealt{Jones84}) and lower values
for poor clusters and galaxy groups (\citealt{Finoguenov01},
\citealt{Helsdon00} and \citealt{Horner99}). Several studies
have shown that $\beta$ has a mild
positive trend with the average temperature of the ICM in nearby
clusters (\citealt{Vikhlinin99}, \citealt{Croston08} and 
\citealt{Maughan08}) and that the value in poorer galaxy
groups is lower (i.e. flatter surface brightness slope) than that
in clusters (\citealt{Osmond04b}). In terms of evolution, 
a study of Chandra data for
115 clusters spanning the range $0.1<z<1.3$ by \cite{Maughan08} shows
some indication that the $\beta$-$T$ correlation is weaker for clusters at
$z>0.5$.

Low-mass galaxy clusters or groups, with ICM temperatures less than
2-3 keV, play an important role in the evolution of galactic systems
because they lie at a transition between the field environment and
rich cluster environments, and also because non-gravitational processes
have a larger impact in groups than in rich clusters (e.g.,
\citealt{Zabludoff98}, \citealt{Ponman03} and \citealt{Sun09}). However,
these poor systems, and in particular the evolution of their properties,
have received rather little attention. This is mainly due to the difficulty in
detecting and studying them, especially at large redshifts, 
due to their faint X-ray emission and small complement of galaxies.

For these reasons, research on the evolution of galactic systems in the regime
of groups and poor clusters has only started recently, as a result of 
improvements in observing capabilities in both the X-ray and optical.
By comparing optically-selected systems at
$0.3 \leq z \leq 0.55$ with nearby groups, \cite{Wilman05} showed that the
fraction of group members undergoing significant star formation increases
strongly with redshift out to $z\sim0.45$. However, the study of 
X-ray selected groups
by \cite{Jeltema07} showed a contrary result: they did not
observe significant evolution in the morphology or star formation of the
galaxy populations in their $0.2<z<0.6$ groups compared to
low-redshift X-ray luminous groups.
They argued that this discrepancy could be due
to different selection methods, since optically-selected systems are
typically lower in mass and contain more spiral galaxies and therefore a
stronger evolution in the galaxies is expected. They also found that their
moderate redshift groups had galaxy populations similar to clusters at the
same redshift; in particular, a large fraction of early-type galaxies and a
low fraction of galaxies with significant star formation.
However, in contrast to the situation in low redshift X-ray bright
groups, a significant fraction of these intermediate redshift groups
were found (\citealt{Mulchaey06}, \citealt{Jeltema07}) to have no 
bright early-type galaxy at the centre of the X-ray emission, or to
have a central galaxy with multiple nuclei. This was taken as evidence
for the dynamical youth of many of these groups.

The small number of studies which have addressed the evolution in the X-ray 
properties of galaxy groups have found little convincing evidence for any.
\cite{Jeltema06}, in a multiwavelength study of six galaxy groups and
poor clusters at intermediate redshift (0.2-0.6), found that they
appear to follow the scaling relations between luminosity, temperature, and
velocity dispersion defined by low-redshift groups and clusters. This is also
true (\citealt{Jeltema09}) for three higher redshift poor clusters
from the AEGIS survey. A study of evolution in the $L$-$T$ relation 
based on the present
XMM-LSS cluster sample by \cite{Pacaud07}, taking into account the selection
function of the survey, found that the range of models consistent with the
data included self-similar evolution, and also (marginally) a no-evolution
model. \cite{Finoguenov07} extracted a larger sample of 72 
clusters (mostly poor ones) from the XMM-Newton observations of the COSMOS
field, and found no evidence for evolution in the luminosity function
of these systems out to $z\sim 1$, though the quality of their data did
not permit them to study the morphology of the X-ray emission.

Motivated by the the paucity of information available for the evolution
of the ICM in the important environment of low-mass galaxy clusters, 
we aim in this study to shed light on the 
spatial distribution of the ICM in X-ray selected clusters covering 
a wide redshift range ($z\sim 0-1$), paying special attention to 
trends in the slope and central cuspiness of the X-ray emission.

The paper is constructed as follows: in section 3.2, we describe the
data and briefly introduce the properties of the cluster sample; then
we describe the data reduction used to produce X-ray surface
brightness profiles. In section 3.3, we present our results, starting
with the individual cluster profiles, and then profiles of
redshift-stacked and temperature-stacked clusters. In section
3.4, we discuss the implication of our results and compare them with
other studies. Finally, in section 3.5, we summarise our conclusions.

Throughout this paper, we adopt  the cosmological
parameters from the five-year WMAP data presented by \cite{Hinshaw09}, namely:
$H_0=70.5$~km~s$^{-1}$~Mpc$^{-1}$, $\Omega_m=0.27$, $\Omega_\Lambda=0.73$.

\section{Data}

\subsection{The sample}

Our sample is based on the 29 Class 1 (C1) clusters from the X-ray
Multi-Mirror Large-Scale Structure (XMM-LSS) survey. The XMM-LSS C1
cluster sample is a well-controlled X-ray selected and
spectroscopically confirmed cluster sample. The criteria used to
select the members of this sample guarantee negligible contamination
of point-like sources. The observations of the clusters were performed
in a homogeneous way (10-20 ks exposures). For full details of the C1
sample, see \cite{Pacaud07}. Detailed information on the selection
function of the C1 sample can be found in \cite{Pacaud06}.

The C1 sample is dominated by groups and poor clusters
with temperatures of $0.63\leq kT \leq 4.80$ keV,
spanning a redshift range $0.05\leq z \leq 1.05$.
Typically, we have a few hundreds X-ray counts for each cluster,
with only a few having over a thousand detected photons.
Two of the  29 clusters with less than 80 counts had to be excluded
from our analysis because their data were inadequate
for useful profiles to be extracted.  The excluded
clusters are XLSSC clusters 39 and 48. Hence our sample
consists of 27 clusters. Cluster 47, with 81 counts,
was a marginal case. We were unable to constrain a fit to its
individual profile, but 
its data were included in the analysis of the stacked profiles.
Key properties of the sample are presented in Table \ref{C1prop}.

\begin{table*}
\begin{center}
\begin{tabular}{cccccccccccccc}
\hline

XLSSC	&	R.A.	&	Dec.	&	Redshift	&	$kT$	&	r$_{500}$	&	$\beta$	&	\rc$/R_{500}$	&	$\beta$	&	Central Excess	&	Counts	\\
number	&	(J2000)	&	(J2000)	&		&	($keV$)	&	(Mpc)	&	Fitted \rc	&		&	Fixed \rc	&	Factor ($f_c$)	&		\\
\hline																					
11	&	36.5413	&	-4.9682	&	0.05	&	0.64	&	0.290	&	$0.45_{-0.02}^{+0.03}$	&	$0.08_{-0.03}^{+0.03}$	&	$0.47_{-0.02}^{+0.01}$	&	$1.14  \pm 0.14$	&	795	\\
52	&	36.5681	&	-2.6660	&	0.06	&	0.63	&	0.285	&	$0.69_{-0.05}^{+0.07}$	&	$0.14_{-0.02}^{+0.03}$	&	$0.62_{-0.01}^{+0.02}$	&	$0.95  \pm 0.09$	&	561	\\
21	&	36.2345	&	-5.1339	&	0.08	&	0.68	&	0.297	&	$0.65_{-0.09}^{+0.27}$	&	$0.04_{-0.02}^{+0.03}$	&	$1.34_{-0.22}^{+0.29}$	&	$1.04  \pm 0.12$	&	87	\\
41	&	36.3777	&	-4.2391	&	0.14	&	1.34	&	0.440	&	$0.47_{-0.01}^{+0.02}$	&	$0.05_{-0.01}^{+0.02}$	&	$0.51_{-0.01}^{+0.01}$	&	$1.70  \pm 0.22$	&	656	\\
50	&	36.4233	&	-3.1895	&	0.14	&	3.50	&	0.804	&	$0.78_{-0.05}^{+0.06}$	&	$1.12_{-0.09}^{+0.10}$	&	$0.37_{-0.01}^{+0.01}$	&	$0.59  \pm 0.07$	&	4387	\\
35	&	35.9507	&	-2.8588	&	0.17	&	1.20	&	0.394	&	$0.44_{-0.04}^{+0.05}$	&	$0.13_{-0.06}^{+0.07}$	&	$0.42_{-0.02}^{+0.02}$	&	$0.67  \pm 0.37$	&	422	\\
25	&	36.3531	&	-4.6776	&	0.26	&	2.00	&	0.533	&	$0.58_{-0.04}^{+0.05}$	&	$0.07_{-0.02}^{+0.02}$	&	$0.65_{-0.02}^{+0.03}$	&	$1.42  \pm 0.18$	&	683	\\
44	&	36.1411	&	-4.2347	&	0.26	&	1.30	&	0.399	&	$1.37_{-0.61}^{+4.48}$	&	$0.63_{-0.31}^{+0.25}$	&	$0.50_{-0.02}^{+0.03}$	&	$1.02  \pm 0.41$	&	276	\\
51	&	36.4982	&	-2.8265	&	0.28	&	1.20	&	0.518	&	$1.81_{-0.84}^{+***}$	&	$0.77_{-0.31}^{+0.12}$	&	$0.47_{-0.03}^{+0.05}$	&	$2.18  \pm 1.09$	&	160	\\
22	&	36.9165	&	-4.8576	&	0.29	&	1.70	&	0.471	&	$0.57_{-0.02}^{+0.02}$	&	$0.04_{-0.01}^{+0.01}$	&	$0.71_{-0.02}^{+0.02}$	&	$1.46  \pm 0.16$	&	1234	\\
27	&	37.0143	&	-4.8510	&	0.29	&	2.80	&	0.653	&	$0.70_{-0.10}^{+0.16}$	&	$0.40_{-0.10}^{+0.13}$	&	$0.45_{-0.01}^{+0.02}$	&	$0.86  \pm 0.23$	&	653	\\
8	&	36.3370	&	-3.8015	&	0.30	&	1.30	&	0.396	&	$0.53_{-0.07}^{+0.11}$	&	$0.04_{-0.03}^{+0.04}$	&	$0.67_{-0.06}^{+0.08}$	&	$1.54  \pm 0.47$	&	160	\\
28	&	35.9878	&	-3.0991	&	0.30	&	1.30	&	0.399	&	$0.43_{-0.05}^{+0.08}$	&	$0.08_{-0.06}^{+0.09}$	&	$0.45_{-0.04}^{+0.04}$	&	$3.65  \pm 1.42$	&	245	\\
13	&	36.8586	&	-4.5380	&	0.31	&	1.00	&	0.340	&	$0.58_{-0.14}^{+1.25}$	&	$0.06_{-0.05}^{+0.20}$	&	$0.73_{-0.12}^{+0.20}$	&	$1.54  \pm 1.01$	&	120	\\
18	&	36.0087	&	-5.0904	&	0.32	&	2.00	&	0.521	&	$0.94_{-0.22}^{+0.76}$	&	$0.19_{-0.06}^{+0.14}$	&	$0.68_{-0.04}^{+0.04}$	&	$1.02  \pm 0.27$	&	89	\\
40	&	35.5272	&	-4.5431	&	0.32	&	1.60	&	0.442	&	$0.44_{-0.02}^{+0.04}$	&	$0.01_{-0.01}^{+0.02}$	&	$0.55_{-0.04}^{+0.05}$	&	$2.70  \pm 0.74$	&	223	\\
10	&	36.8435	&	-3.3623	&	0.33	&	2.40	&	0.574	&	$0.57_{-0.05}^{+0.08}$	&	$0.08_{-0.03}^{+0.05}$	&	$0.61_{-0.02}^{+0.03}$	&	$1.94  \pm 0.35$	&	327	\\
23	&	35.1894	&	-3.4328	&	0.33	&	1.70	&	0.457	&	$0.54_{-0.03}^{+0.03}$	&	$0.05_{-0.01}^{+0.01}$	&	$0.65_{-0.03}^{+0.03}$	&	$1.19  \pm 0.27$	&	394	\\
6	&	35.4385	&	-3.7715	&	0.43	&	4.80	&	0.838	&	$0.68_{-0.03}^{+0.03}$	&	$0.19_{-0.02}^{+0.02}$	&	$0.570_{-0.011}^{+0.003}$	&	$0.84  \pm 0.09$	&	1394	\\
36	&	35.5280	&	-3.0539	&	0.49	&	3.60	&	0.676	&	$0.60_{-0.04}^{+0.06}$	&	$0.08_{-0.02}^{+0.03}$	&	$0.67_{-0.02}^{+0.03}$	&	$1.37  \pm 0.22$	&	481	\\
49	&	35.9892	&	-4.5883	&	0.49	&	2.20	&	0.493	&	$2.20_{-0.92}^{+***}$	&	$0.38_{-0.13}^{+0.04}$	&	$0.67_{-0.05}^{+0.06}$	&	$1.59  \pm 0.55$	&	157	\\
1	&	36.2367	&	-3.8131	&	0.61	&	3.20	&	0.584	&	$0.62_{-0.04}^{+0.05}$	&	$0.11_{-0.02}^{+0.02}$	&	$0.62_{-0.02}^{+0.02}$	&	$2.25  \pm 0.30$	&	595	\\
2	&	36.3844	&	-3.9200	&	0.77	&	2.80	&	0.493	&	$0.83_{-0.18}^{+0.57}$	&	$0.09_{-0.04}^{+0.08}$	&	$0.92_{-0.11}^{+0.14}$	&	$3.03  \pm 0.68$	&	136	\\
47	&	35.5461	&	-2.6783	&	0.79	&	3.90	&	0.592	&		&		&		&		&	81	\\
3	&	36.9098	&	-3.2996	&	0.84	&	3.30	&	0.518	&	$0.48_{-0.04}^{+0.05}$	&	$0.03_{-0.02}^{+0.02}$	&	$0.64_{-0.05}^{+0.07}$	&	$1.15  \pm 0.54$	&	193	\\
5	&	36.7854	&	-4.3003	&	1.05	&	3.70	&	0.489	&	$2.42_{-1.33}^{+***}$	&	$0.33_{-0.17}^{+0.05}$	&	$0.80_{-0.09}^{+0.13}$	&	$2.46  \pm 0.94$	&	130	\\
29	&	36.0172	&	-4.2251	&	1.05	&	4.10	&	0.524	&	$0.76_{-0.16}^{+0.37}$	&	$0.11_{-0.05}^{+0.08}$	&	$0.76_{-0.06}^{+0.07}$	&	$1.86  \pm 0.59$	&	233	\\

\hline

\end{tabular}
\caption{List of the properties of the 29 C1 galaxy clusters sample sorted 
according to their redshifts (\protect \citealt{Pacaud07}) with the fitted 
parameters of the 26 successfully constrained clusters. The $\beta$ and 
\rc\ values in the seventh and eighth columns are for the the \bmodel\ 
with both \rc\ and $\beta$ freely fitted. The $\beta$ values in the 
ninth column are for the \bmodel\ with fixed \rc =\ffh. All errors 
are 1$\sigma$ errors. 
Clusters 47 failed to converge but its data were not excluded from the stacked 
analysis.  The central excess factor, $f_c$ (tenth column), is calculated 
by dividing the observed surface brightness by the predicted surface 
brightness (from the fitted model) for the innermost radial bin, that 
is 0.05 $\times$ \rfh. If $f_c$ is greater than 1 then this is considered 
as an indication that the cluster has a cool core and vice versa. 
The stars (***) denote unconstrained errors.} 
\label{C1prop}
\end{center}
\end{table*}


\subsection{Data analysis}

To construct the X-ray surface brightness profiles for each cluster,
we used three X-ray FITS images, three exposure maps and one
segmentation map, all produced using the production pipe line
described in \cite{Pacaud06}. The images were taken by the MOS1, MOS2
and PN imagers on board the XMM-Newton satellite in the energy band
[0.5-2.0] keV with exposure times ranging from 10 to 20 ks. The
exposure maps are FITS images containing the vignetting-corrected
exposure of the clusters as a function of the sky position. A single
segmentation map, generated by \textsc{SExtractor} was used for each
cluster to remove contaminating sources.

The right ascension (RA) and declination (Dec) values of the centres
were determined as outlined in \cite{Pacaud06}. But when we examined
the X-ray profiles of the clusters, some showed dips at the
centre. For these clusters, we mosaicked the three images, smoothed
the resulting image and took the coordinates of the pixel
with the maximum photon counts and modified the cluster centre
accordingly. The modified centres at most are only $\sim$14 arcseconds
from the original values but remove the central dips in the
profiles. The clusters with modified centres have XLSSC numbers:
50,28,40,1,47 and 5. The RAs and Decs in Table \ref{C1prop} are the
modified centres and the original coordinates can be found in Table 1
in \cite{Pacaud07}. 

Since the angular size of our clusters is small, background removal 
using a local estimate works well. The background was taken from 
an annulus extending from 2$\times$\rfh\ to 3$\times$\rfh\ about each 
cluster, where \rfh\ is the radius within which the mean cluster mass
density is 500 times the critical density of the Universe at the
cluster redshift. As in \cite{Pacaud07}, the \rfh\ values were
calculated using \begin{equation} \rm R_{500}=0.388\times T^{0.63}\times
h_{70.5}(z)^{-1} Mpc, \end{equation} where $T$ is the ICM
temperature in keV and $h_{70.5}$ is the Hubble constant in units of
70.5 $\rm{kms}^{-1} \rm{Mpc}^{-1}$. This formula was originally
derived from M-T relation of \cite{Finoguenov01}.

Two background components were evaluated: the photon background
component and the particle background component. These were separated 
using the fact that photons are vignetted, whilst particles are
not. Hence the relationship between count rate and effective area for 
pixels in the background annulus gives an
estimate of the photon background component from the slope, and 
of the particle background component from the intercept. 


Surface brightness profiles were extracted from a standard set of 
annuli (as a fraction of \rfh\ to facilitate later stacking), extending 
to 3$\times$\rfh.
For each annulus we removed the particle background and
computed the the vignetting-corrected count rate (in ct/s/pix) and
its error. The MOS1, MOS2 and PN profiles,
generated in this way were then combined, and their errors
added in quadrature.
Profiles were extracted up to 3$\times$\rfh\ where they
flattened and reached the photon background values. The final column in
Table \ref{C1prop} is the total X-ray counts within 3$\times$\rfh\
after subtracting from it both the photon and the particle background.

The X-ray cameras on the XMM-Newton satellite have a point spread function
(PSF) of $\sim$ 6 arcsec FWHM, see \cite{Struder01} and
\cite{Turner01}. Correction for PSF blurring is important to avoid
biased estimation of the parameters of the cluster's radial profile,
especially, the core radius, \rc. We applied the PSF correction method
used by \cite{Arnaud02} (and described in detail therein) which
analytically computes a photon redistribution matrix (RDM) based on
the properties of the three XMM-Newton cameras and depends on the
energy band used and off-axis angle between the centre of the camera
and the cluster position. The PSF matrices for the three cameras
were weighted by the source counts in each camera and combined
to produce a matrix appropriate to the coadded profiles.

We fitted the surface brightness profile with a \bmodel\
(\citealt{Cavaliere76}) \begin{equation} S(r)=S_{0}(1+(r/r_{\rm core})^{2})^{-3\beta
+0.5},\end{equation} where $S_{0}$ is central brightness (cts/s/pix) and \rc\ is
the core radius (in units of $ \rm R_{500}$). The model was blurred
with the PSF redistribution matrix, and fitted to the 
surface brightness data. The best values of \rc\ and
$\beta$ were estimated by computing the minimum $\chi^{2}$ value on an
adaptively refined \rc--$\beta$ grid. $1\sigma$ errors were computed
for \rc\ and $\beta$, and
$1\sigma$, $2\sigma$ and $3\sigma$ error regions computed
in the \rc--$\beta$ plane for each fit.


As will be seen below, a number of systems show a central excess above the
fitted \bmodel. Such a central cusp suggests the possible presence of a
cool core. Since a central excess may distort the \bmodel\ fit, we
attempted to fit a model with the central bin excluded, but given the
limited statistical quality of our data, loss of the central bin resulted
in poorly behaved fits in many cases. We therefore adopted the
approach of fitting a \bmodel\ with core radius fixed at a value (as a
fraction of \rfh) motivated by the observed profiles of local groups in
which detailed modelling of the surface brightness has been possible. A
central excess above this model then indicates the presence of a cuspy
core.

The fixed value of the core radius we adopt as a canonical value for
poor clusters is taken from \cite{Helsdon00}, who studied 24
X-ray-bright galaxy groups. For half of their
sample they found that two-component \bmodel s were required to
give acceptable fits to the surface brightness distribution.
The outer component represented the intragroup gas, 
whilst core emission could be
distinguished by a clear shoulder in the profile in many cases, and
was fitted by the inner \bmodel\ component. The
median value of \rc\ for the outer component in the 12 
clusters was found to be 60 kpc.
Correcting to our value of $H_{0}$, this median \rc\ would be 42.6 kpc. 
The 12
systems in \cite{Helsdon00} had an average temperature of 1.07
keV. To calculate this \rc\ as a fraction of $R_{500}$, we used
the $R_{500}(T)$ equation above,
which gives \rc=\ffh. This value of \rc\ was therefore used in 
our fixed-core fits. 

To quantitatively determine whether a profile of a cluster (or a
stacked set of clusters) has a central brightness excess, and
therefore a CC, we define the central excess factor ($f_c$) as
the ratio between the observed surface brightness and predicted
surface brightness (from the fitted fixed-core model) for the innermost radial
bin, at $r=0.05 \times$ \rfh. If $f_c$ is greater than 1 then
this was considered as an indication that the cluster has a CC
and vice versa. The error on $f_c$ is derived simply from the error on 
the innermost surface brightness value. We checked the effects
of excluding the central bin for the fixed-core fits. 
This has little effect on the fitted $\beta$ value,
but in cases where there is a central excess, it results in a somewhat
lower normalisation for the fitted model, and hence a slightly (up
to 10\%) higher value of $f_c$.

\subsection{Stacked profiles}

The statistical quality of our individual profiles is limited,
so we stacked
the observed profiles of clusters with similar redshifts and
temperatures, producing higher quality profiles which might
highlight any trends with temperature or redshift.

Since each cluster has a different \rfh, we extracted the profiles in fixed
radial bins in units of \rfh\ up to 3$\times$\rfh. The range from
(0-1)$\times$\rfh\ was divided into 20 equally spaced bins, from
(1-2)$\times$\rfh\ into 15 equally spaced bins and the
range from (2-3)$\times$\rfh\ into 10 equally spaced bins. These
different bin widths allow for the decline in flux with radius
whilst keeping the inner bins
sufficiently fine to resolve the core. This distribution of bins was chosen
after some experimentation to obtain the best fit constraints.

Before the stacking, the profiles were multiplied by the scale factor:
\begin{equation} \frac{1}{A \times B \times C}, \end{equation}
where \begin{equation} A=\rm R_{500} \end{equation} to account for the cluster line of
sight depth, \begin{equation} B=\left( \frac{\rho_c(z)}{\rho_c(z=0)}\right) ^2 \end{equation} to
correct for the change in critical density of the Universe and
\begin{equation} C=(1+z)^{-4} \end{equation} to eliminate the effect of cosmological dimming.
The aim of the scaling is to allow for the effects of variable
cluster depth and for cosmological factors, so that
all profiles would be similar (to within the rather weak temperature
dependence of the X-ray emissivity) in the case where clusters are simple
self-similar systems, evolving with the critical density of the Universe.

The profiles for each component cluster were then added bin by bin and 
their errors quadratically summed to generate a stacked profile.
The photon background values for each cluster were scaled by the 
same factors as the source profiles,  before being combined.
This coadded photon background was then 
included in the fitted model as a fixed background level.
The PSF redistribution matrices were weighted by the scaled
count rate for each cluster
before being combined to produce a composite matrix.

\section{Results}

\subsection{X-ray surface brightness profiles of individual C1 clusters}

Most individual C1 clusters profiles fit successfully with a free--\rc\
\bmodel, and nearly all have well-constrained fixed--\rc\ \bmodel\
fits. The profile of cluster 47, with only 81 counts, is the only
one for which we could not achieve a useful \bmodel\ fit.
For clusters 51 (160 counts), 49 (157 counts) and
5 (130 counts), although best fit models were obtained,
the upper bound of
the free--\rc\ $\beta$ values were not constrained, and  
no \rc--$\beta$ contour plots could be produced.

\begin{figure*}
\center
\epsfig{file=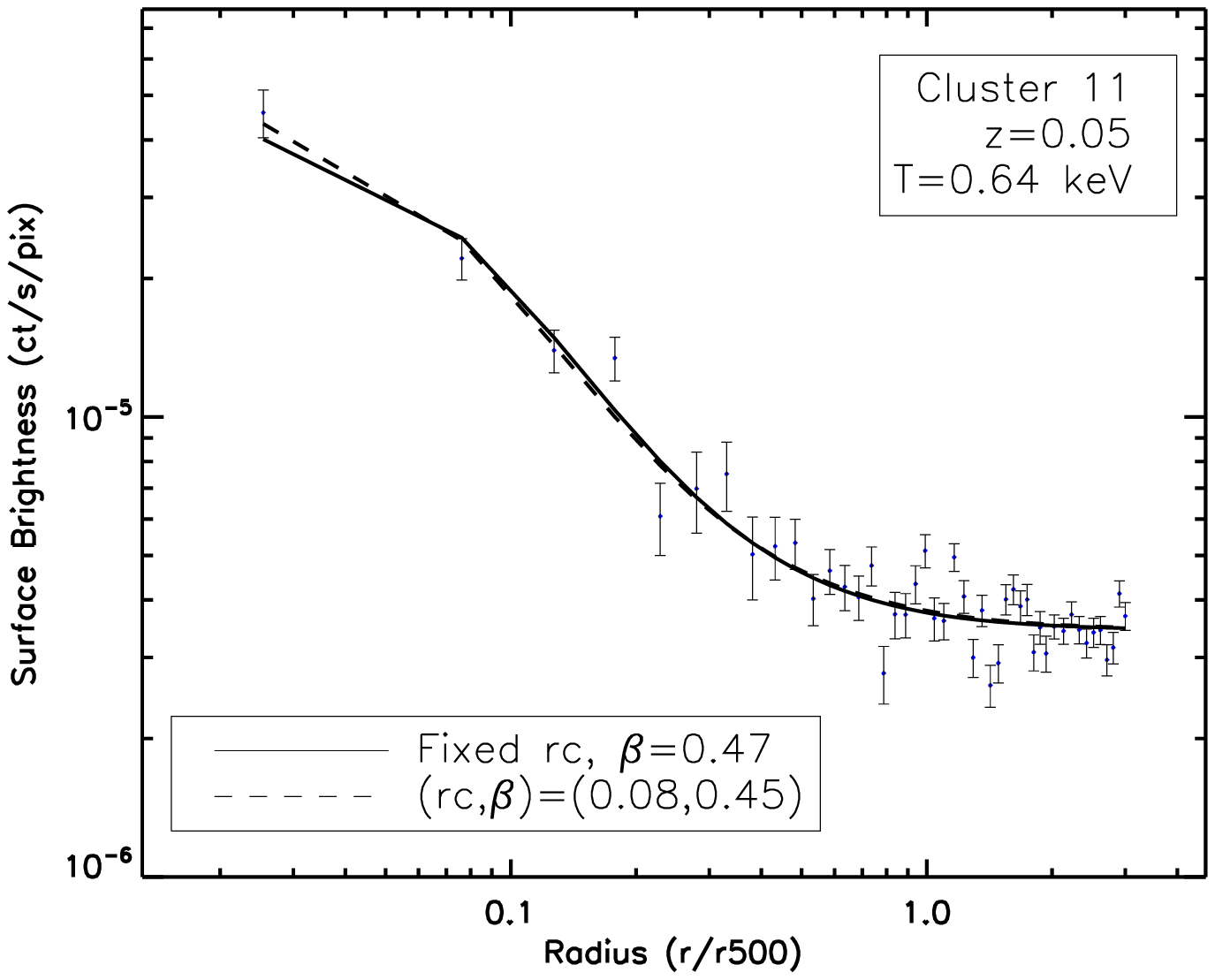,width=5.5cm,height=5cm}
\epsfig{file=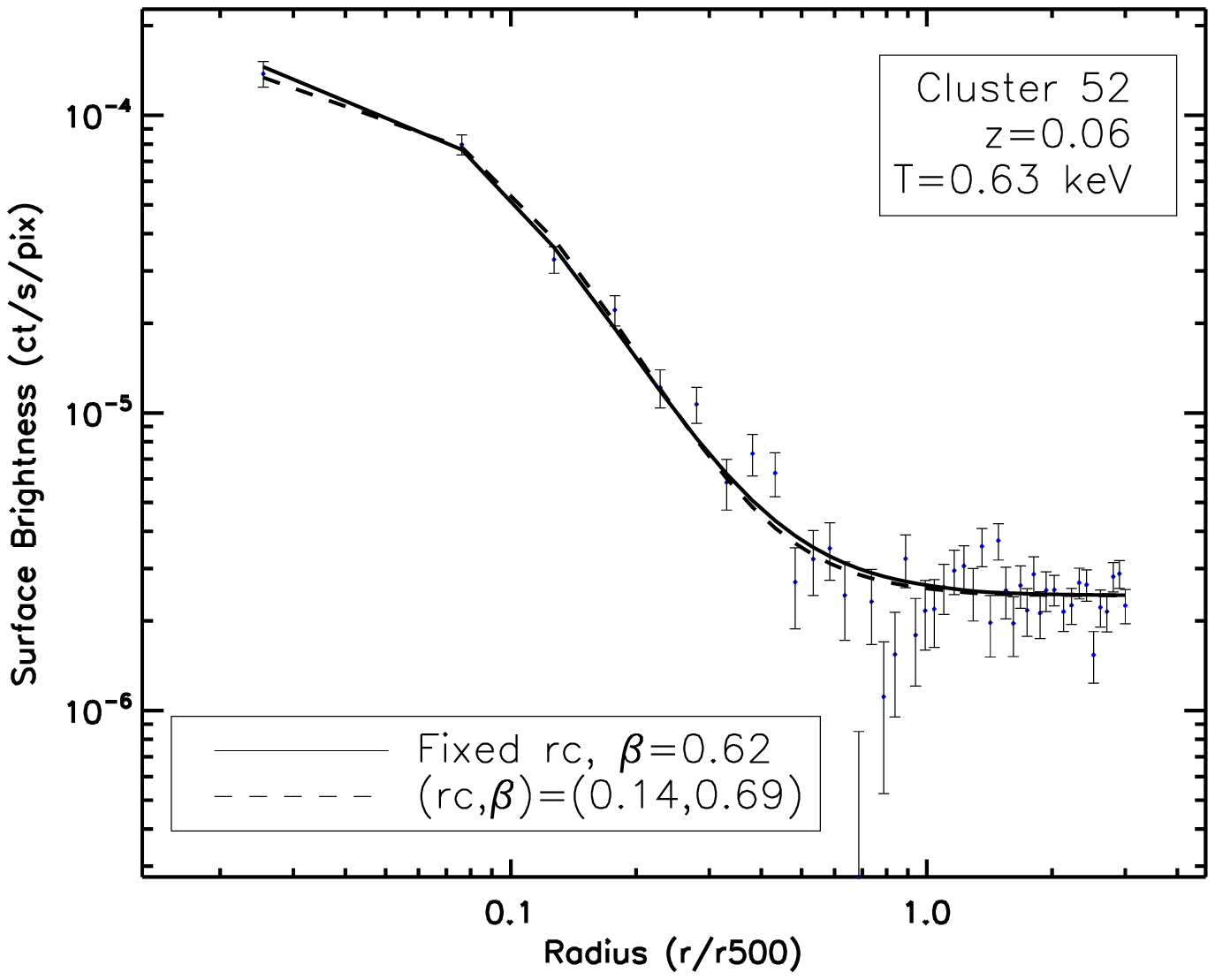,width=5.5cm,height=5cm}
\epsfig{file=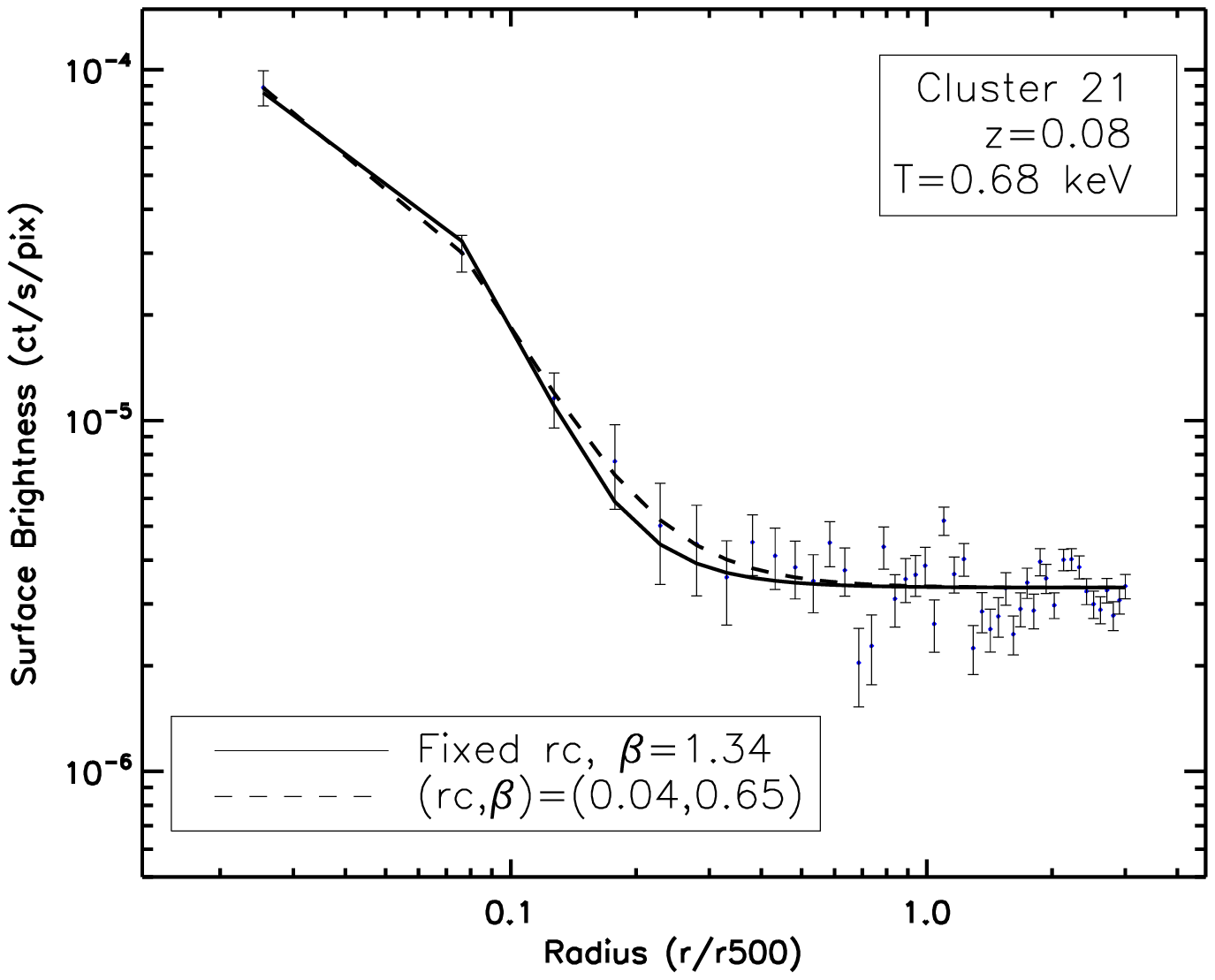,width=5.5cm,height=5cm}
\epsfig{file=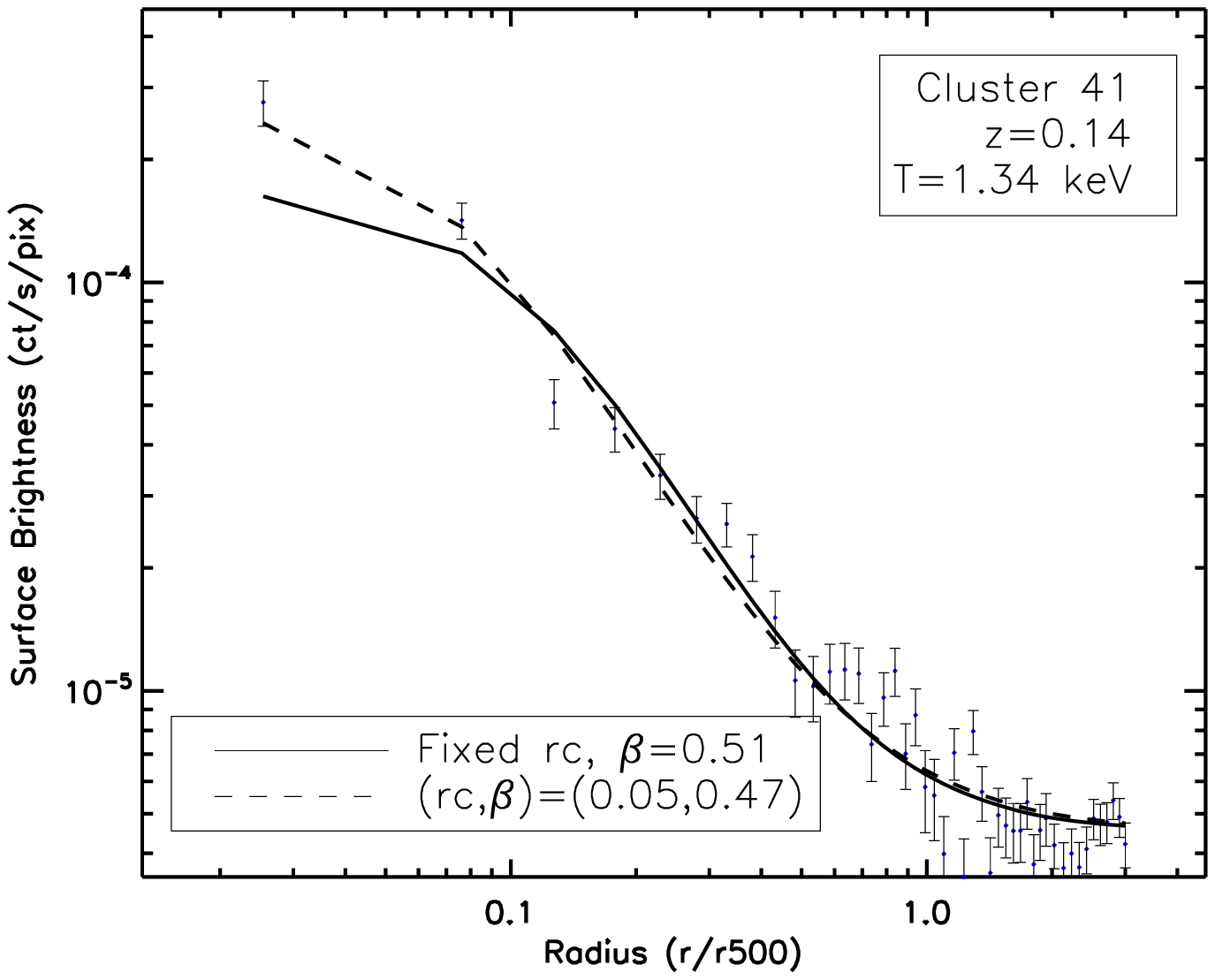,width=5.5cm,height=5cm}
\epsfig{file=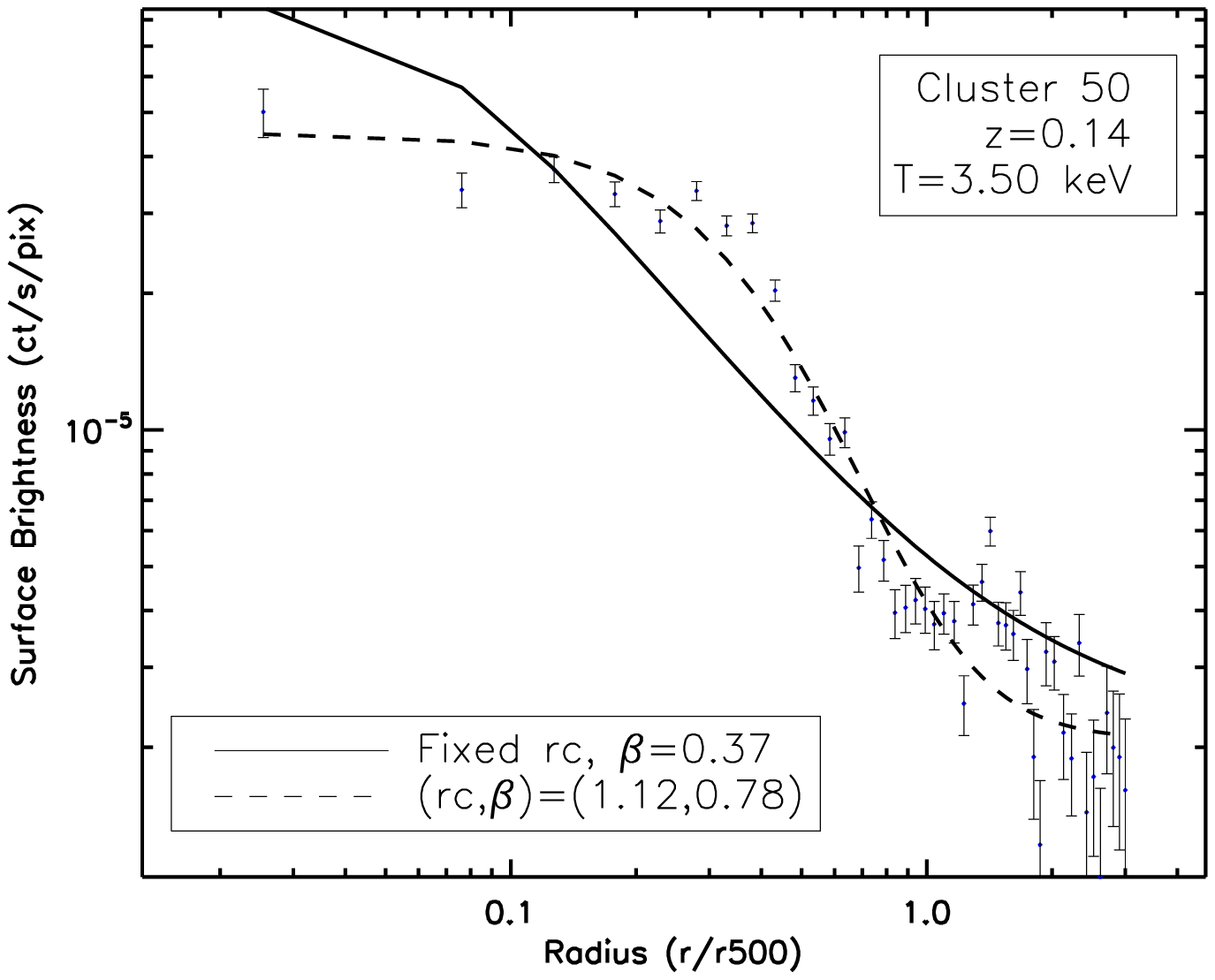,width=5.5cm,height=5cm}
\epsfig{file=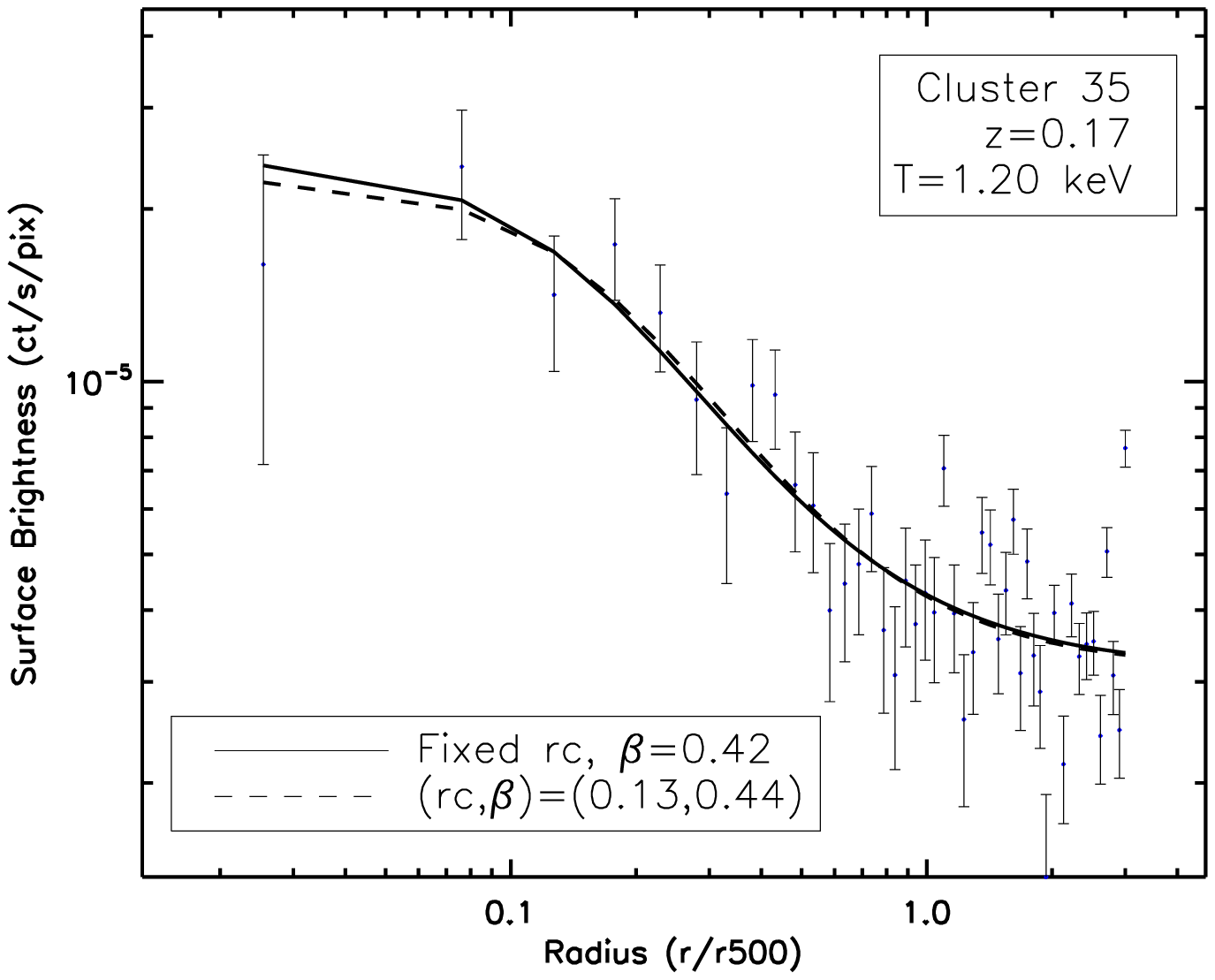,width=5.5cm,height=5cm}
\epsfig{file=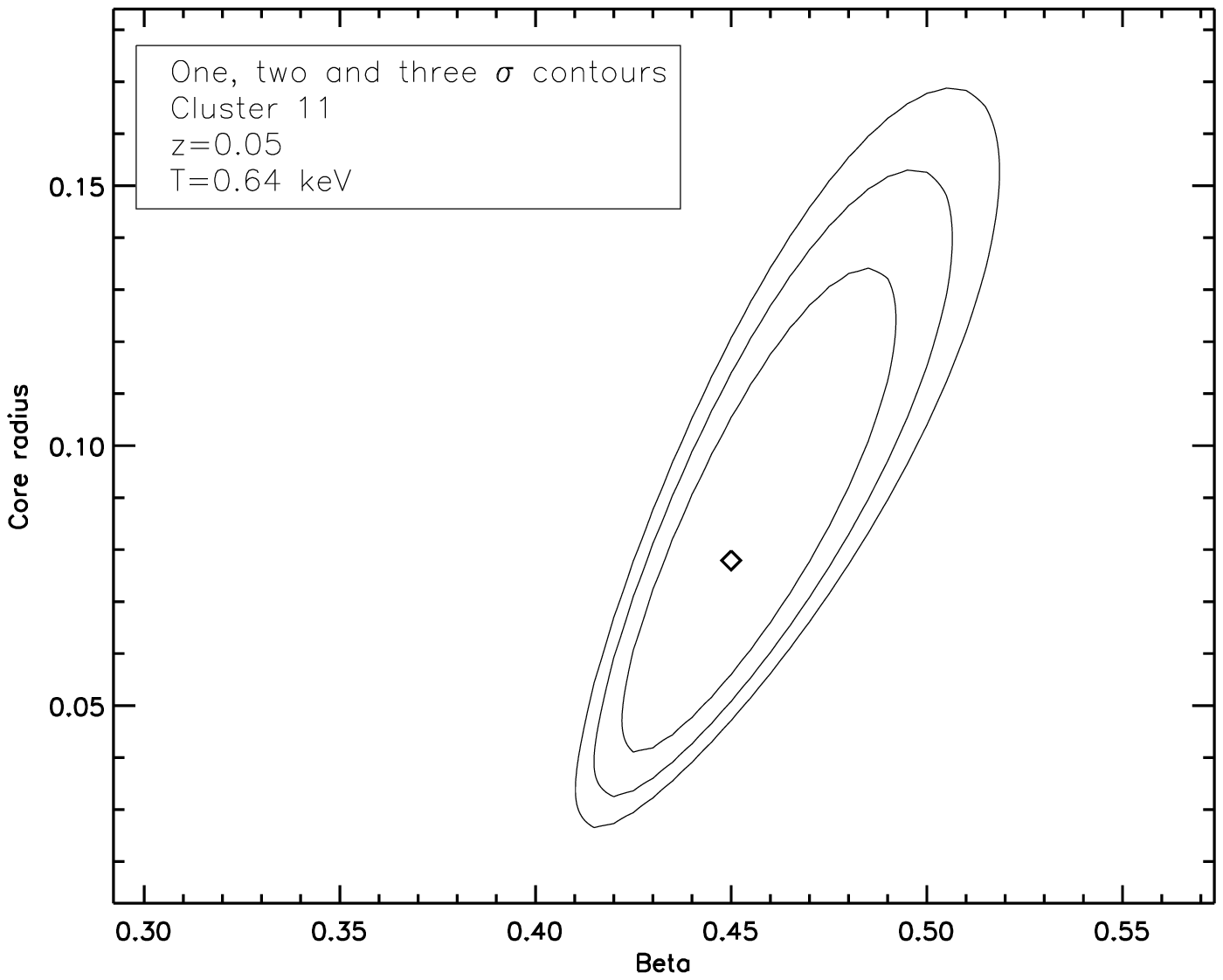,width=5.5cm,height=5cm}
\epsfig{file=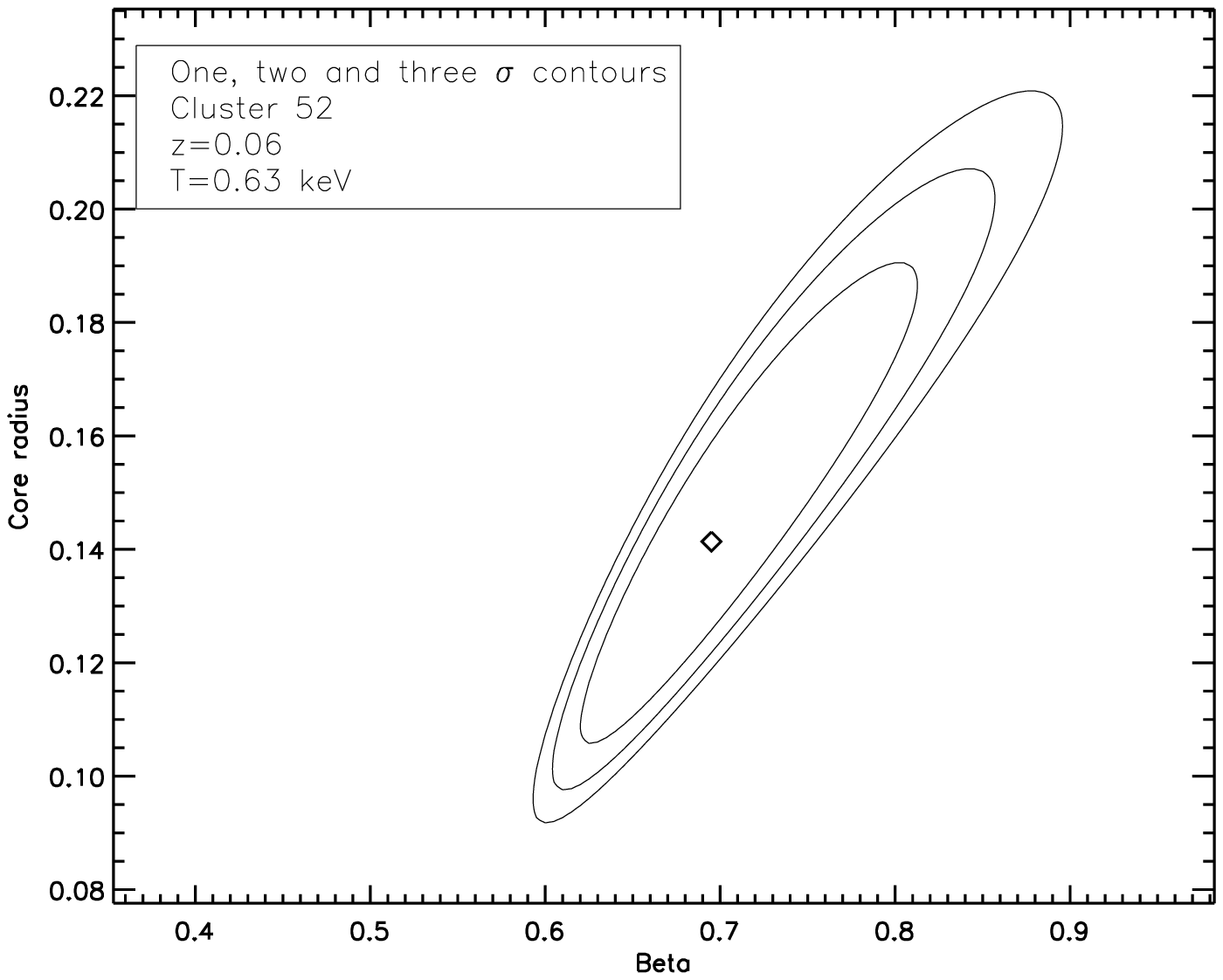,width=5.5cm,height=5cm}
\epsfig{file=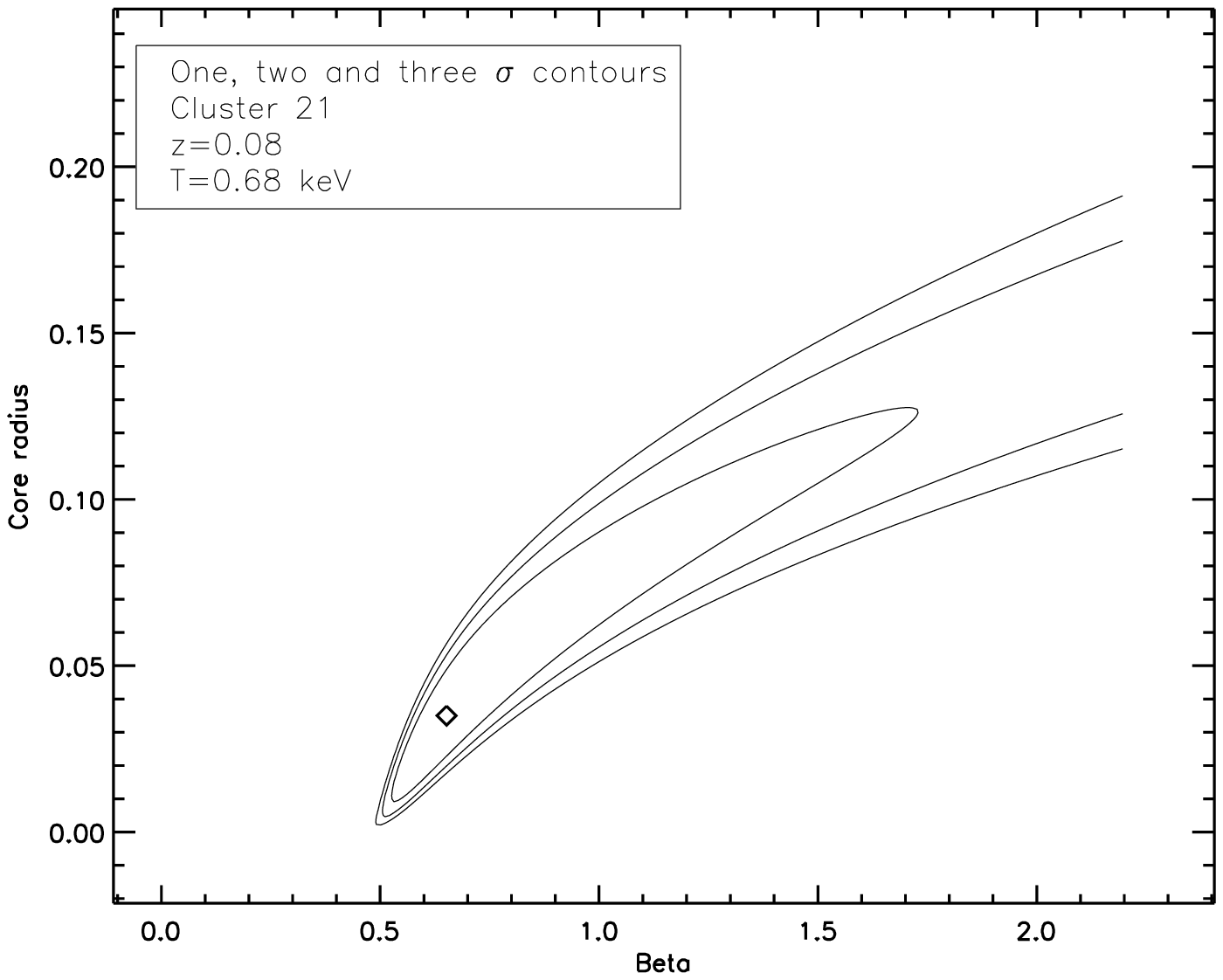,width=5.5cm,height=5cm}
\epsfig{file=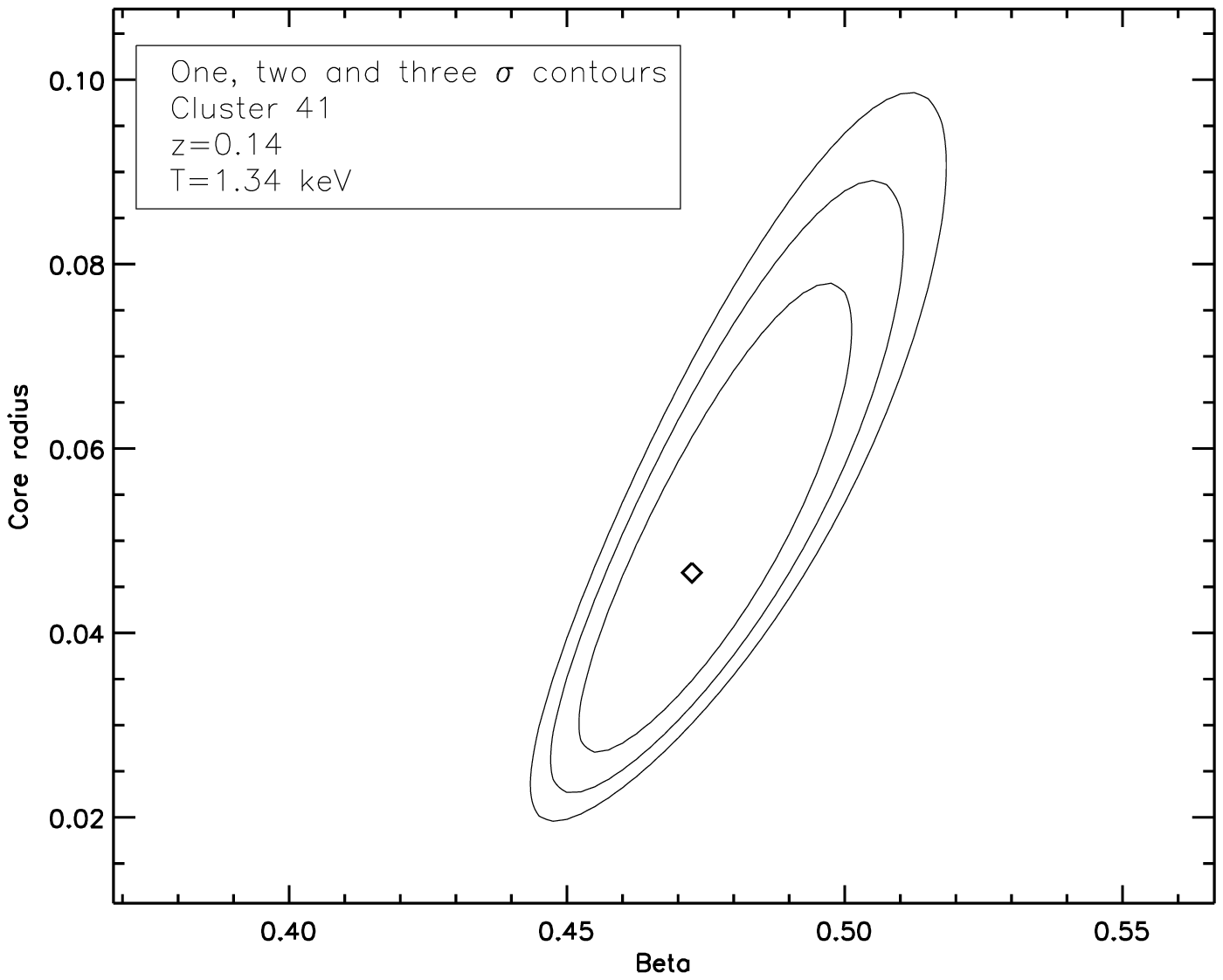,width=5.5cm,height=5cm}
\epsfig{file=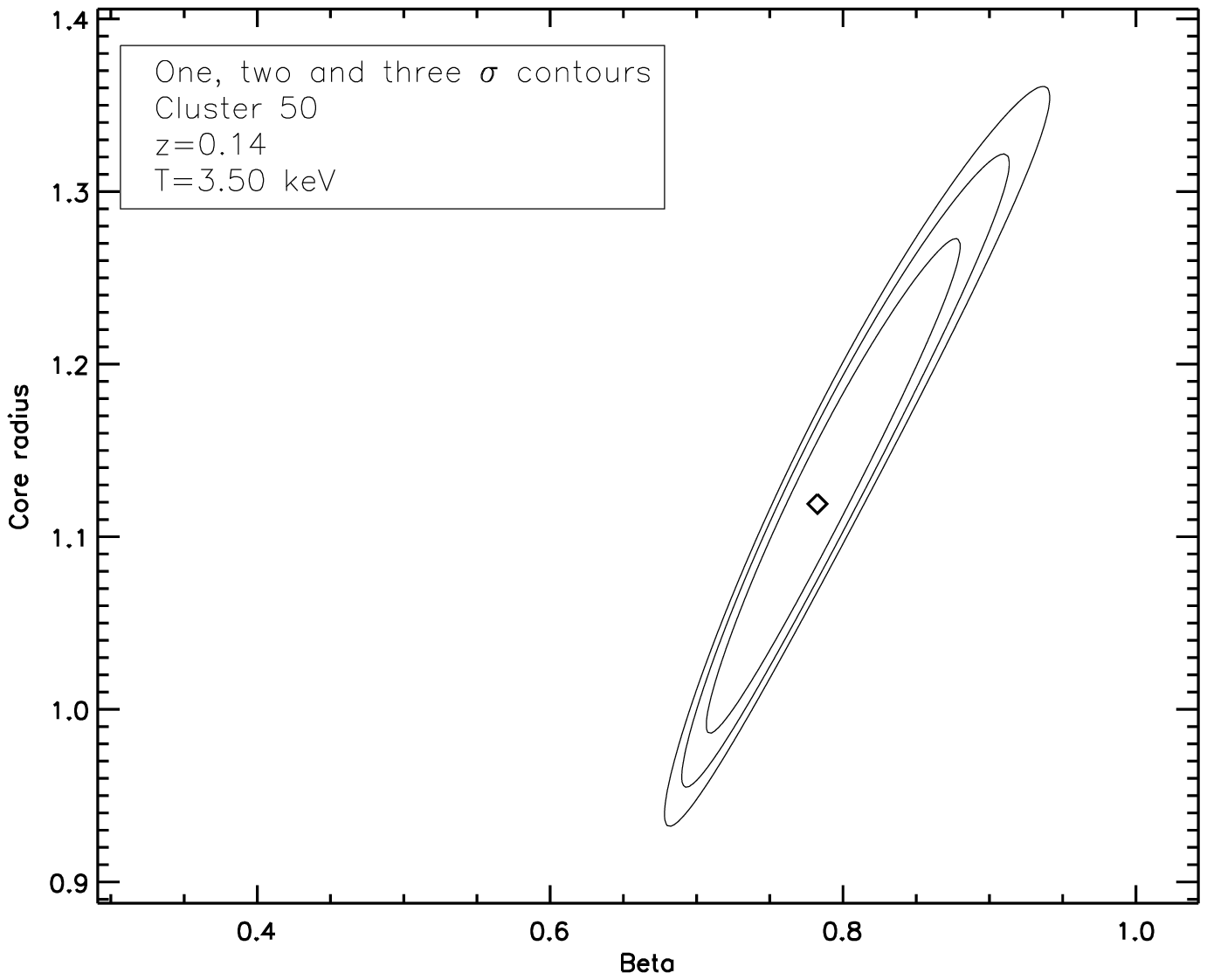,width=5.5cm,height=5cm}
\epsfig{file=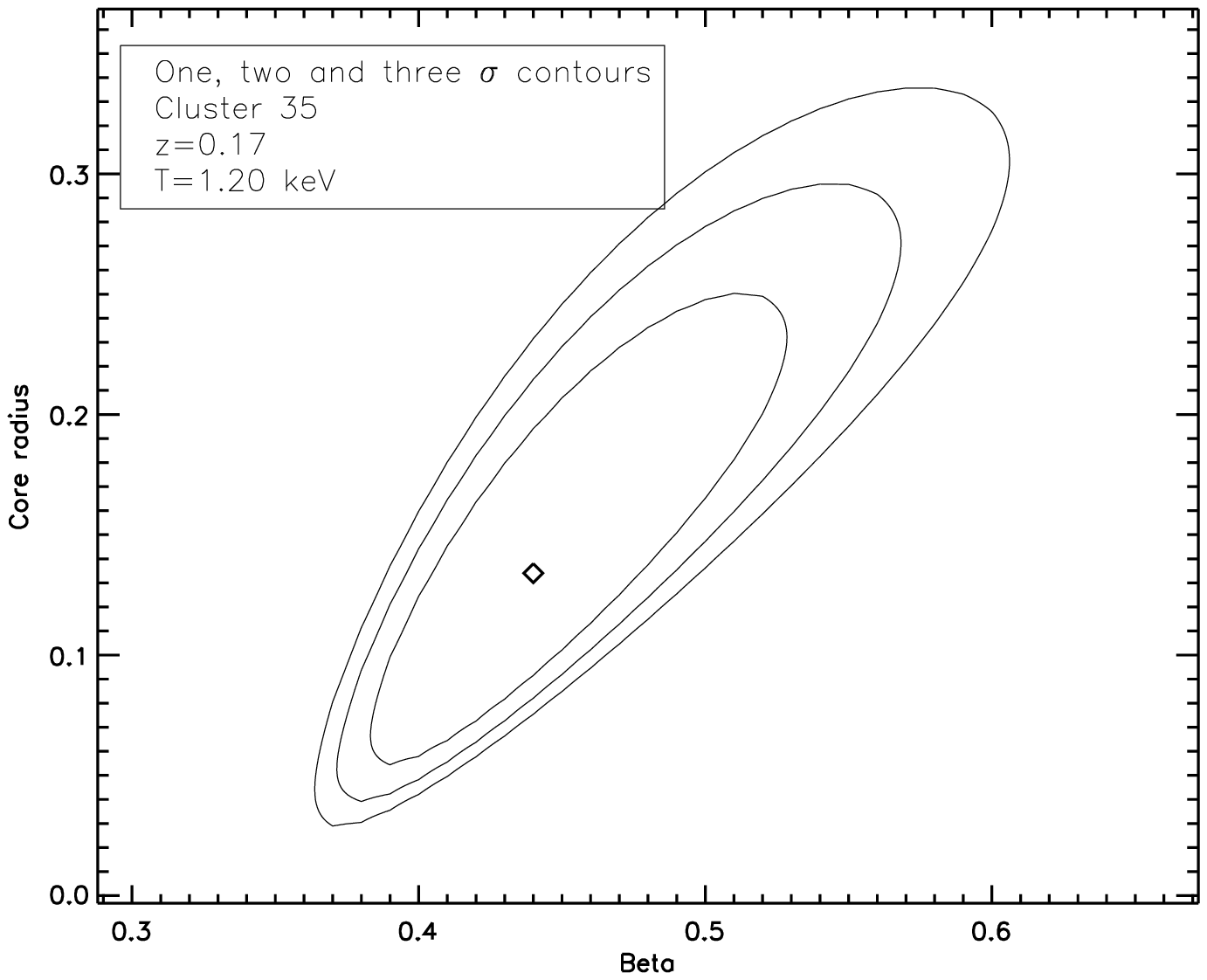,width=5.5cm,height=5cm}

\caption{X-ray surface brightness profiles of the individual C1 clusters with redshift $0.05 \leq z \leq 0.17$, ordered according to redshift and the associated $1 \sigma, 2 \sigma$ and $3 \sigma$ contours. The dashed lines are the fitted \bmodel\ profiles with both \rc\ and $\beta$ freely fitted, while the solid lines are for the fitted profiles with free $\beta$ and \rc\ fixed to \ffh. }
\label{ind_prof1}
\end{figure*}

\begin{figure*}
\center

\epsfig{file=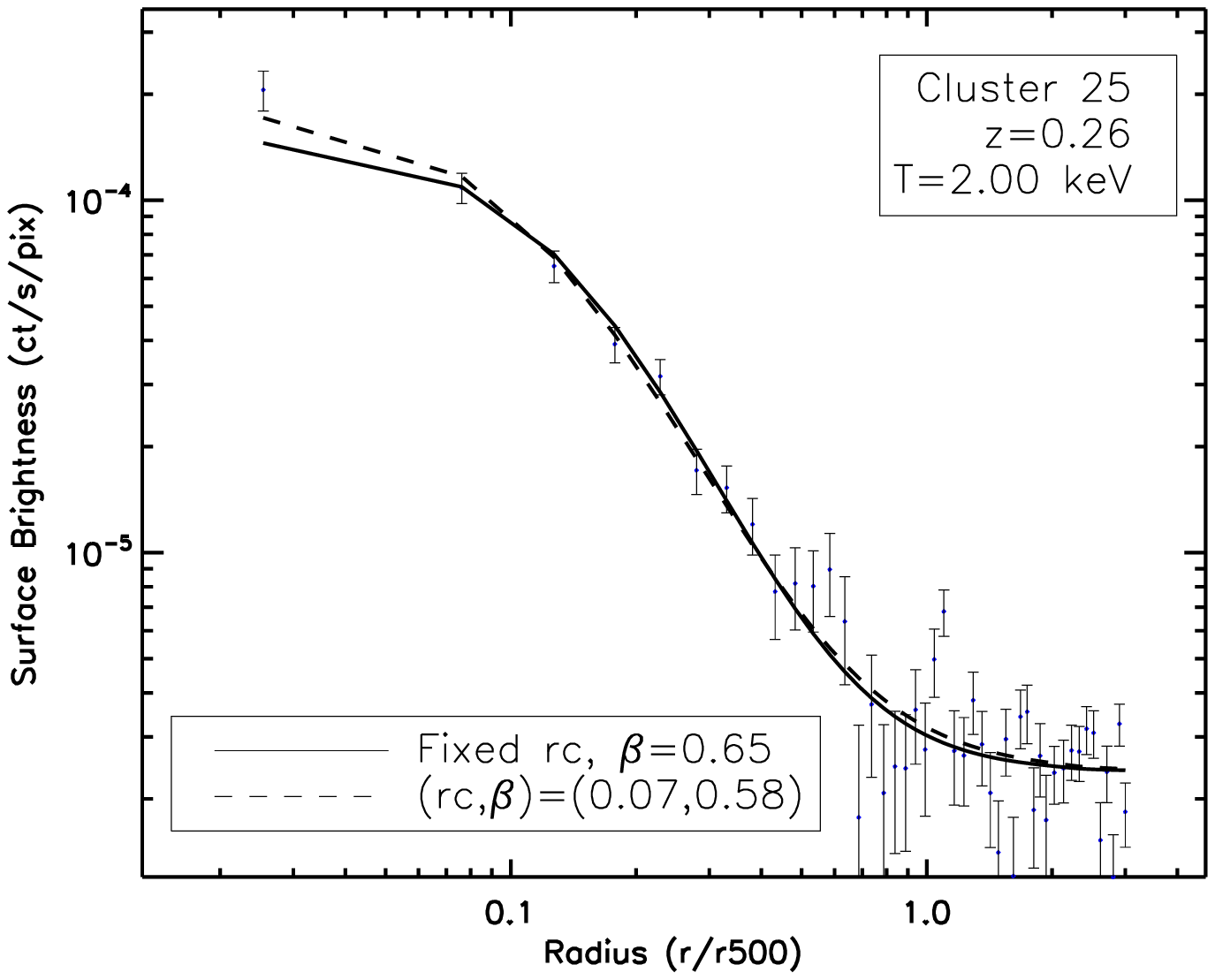,width=3.8cm,height=3.6cm}
\epsfig{file=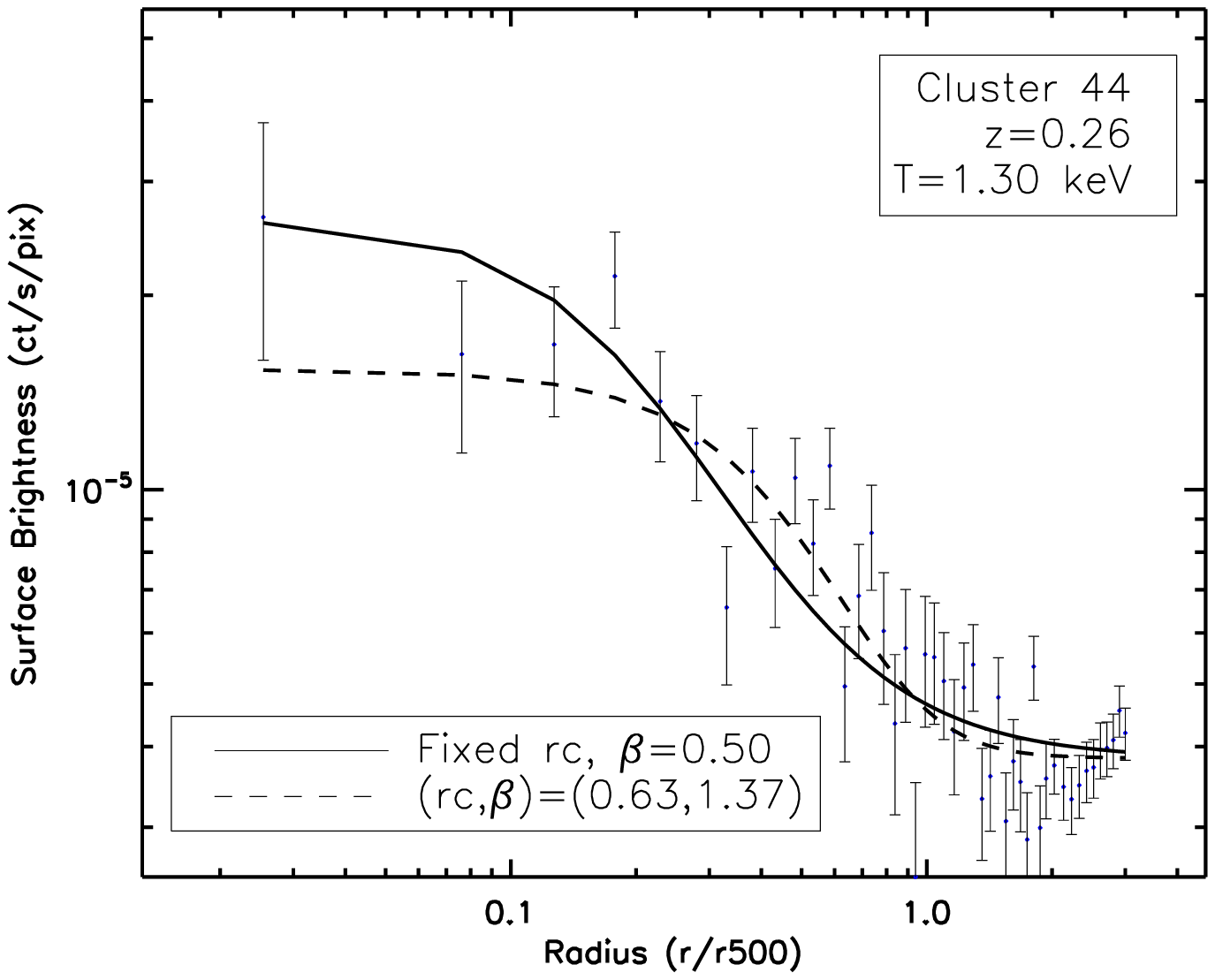,width=3.8cm,height=3.6cm}
\epsfig{file=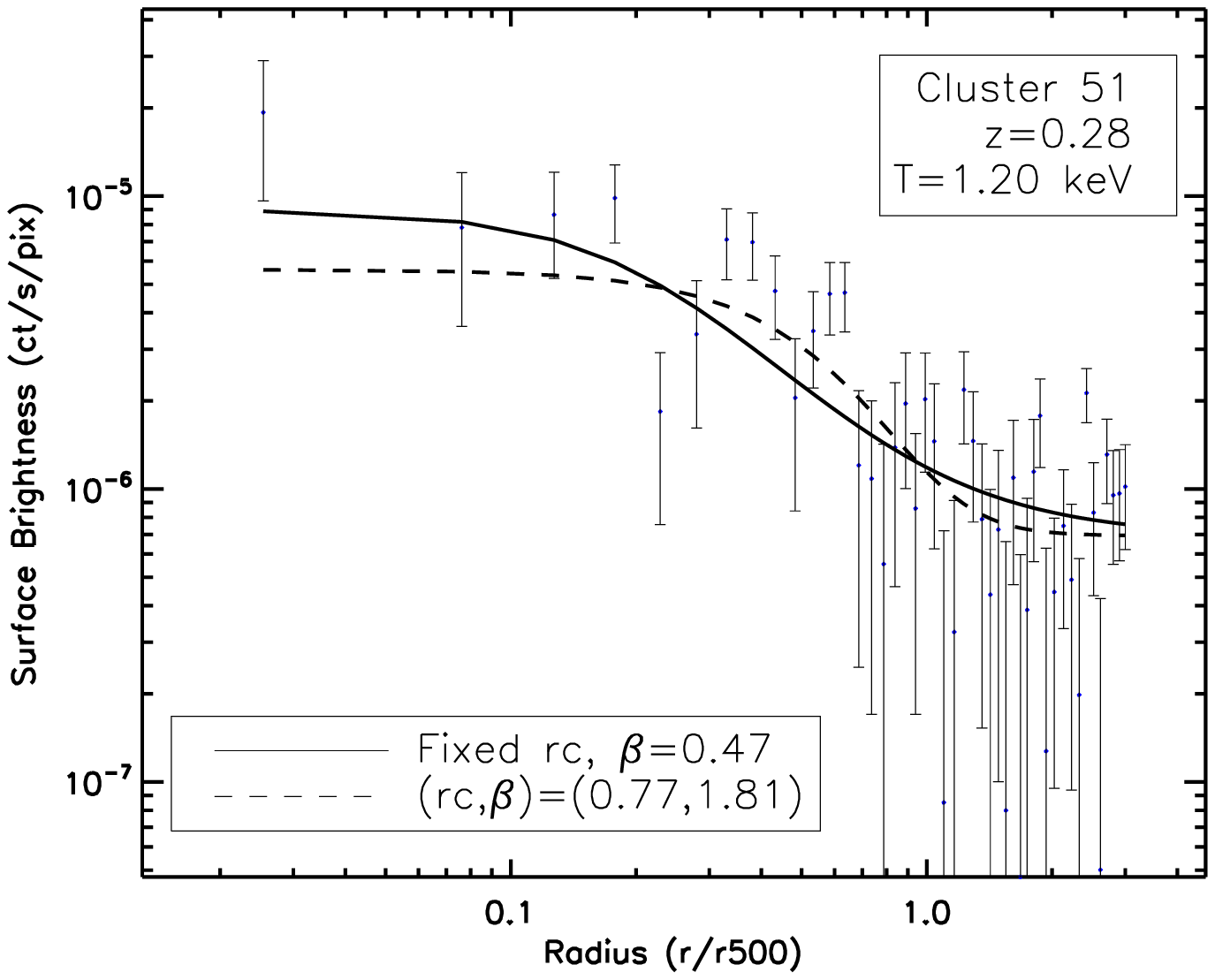,width=3.8cm,height=3.6cm}
\epsfig{file=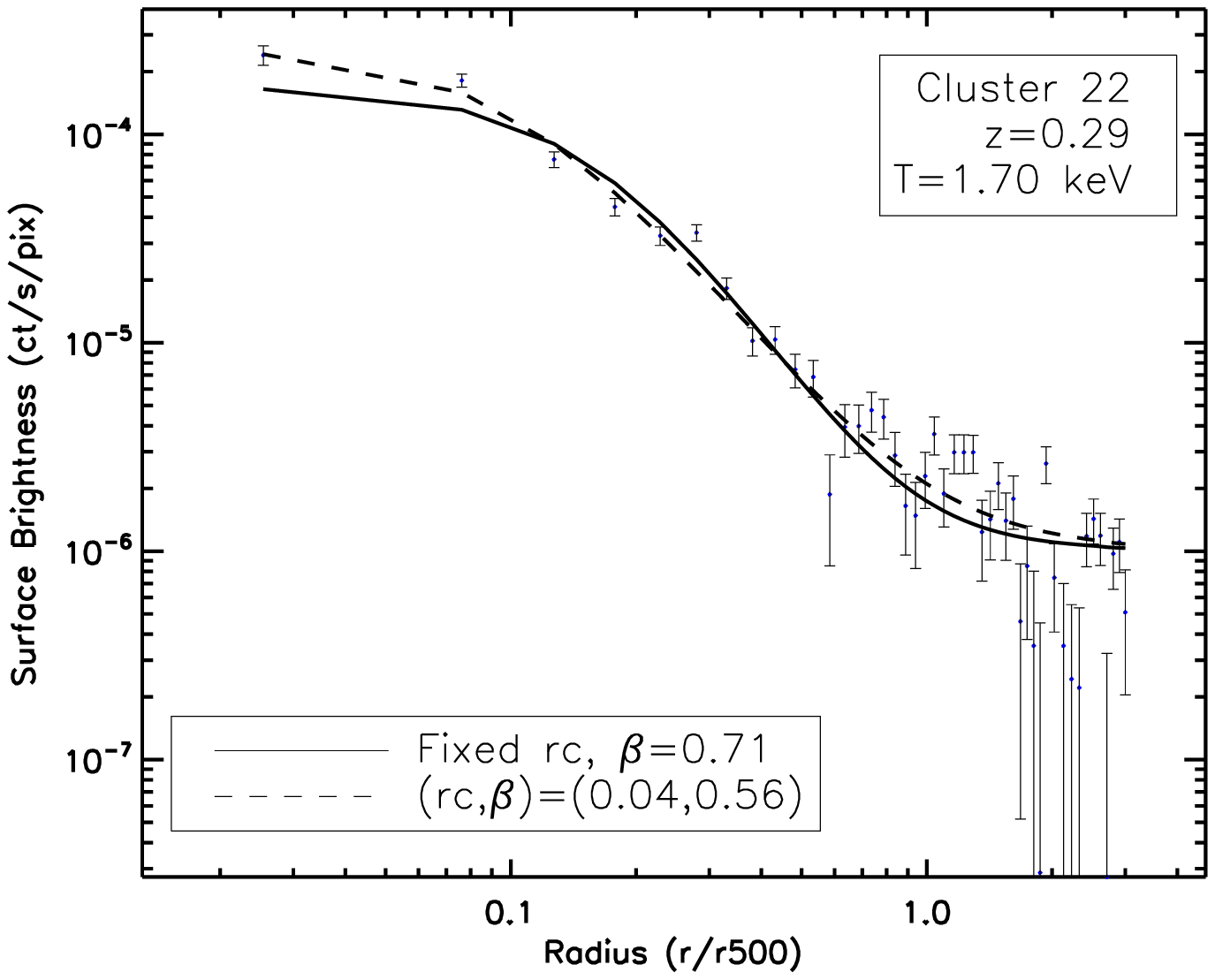,width=3.8cm,height=3.6cm}
\epsfig{file=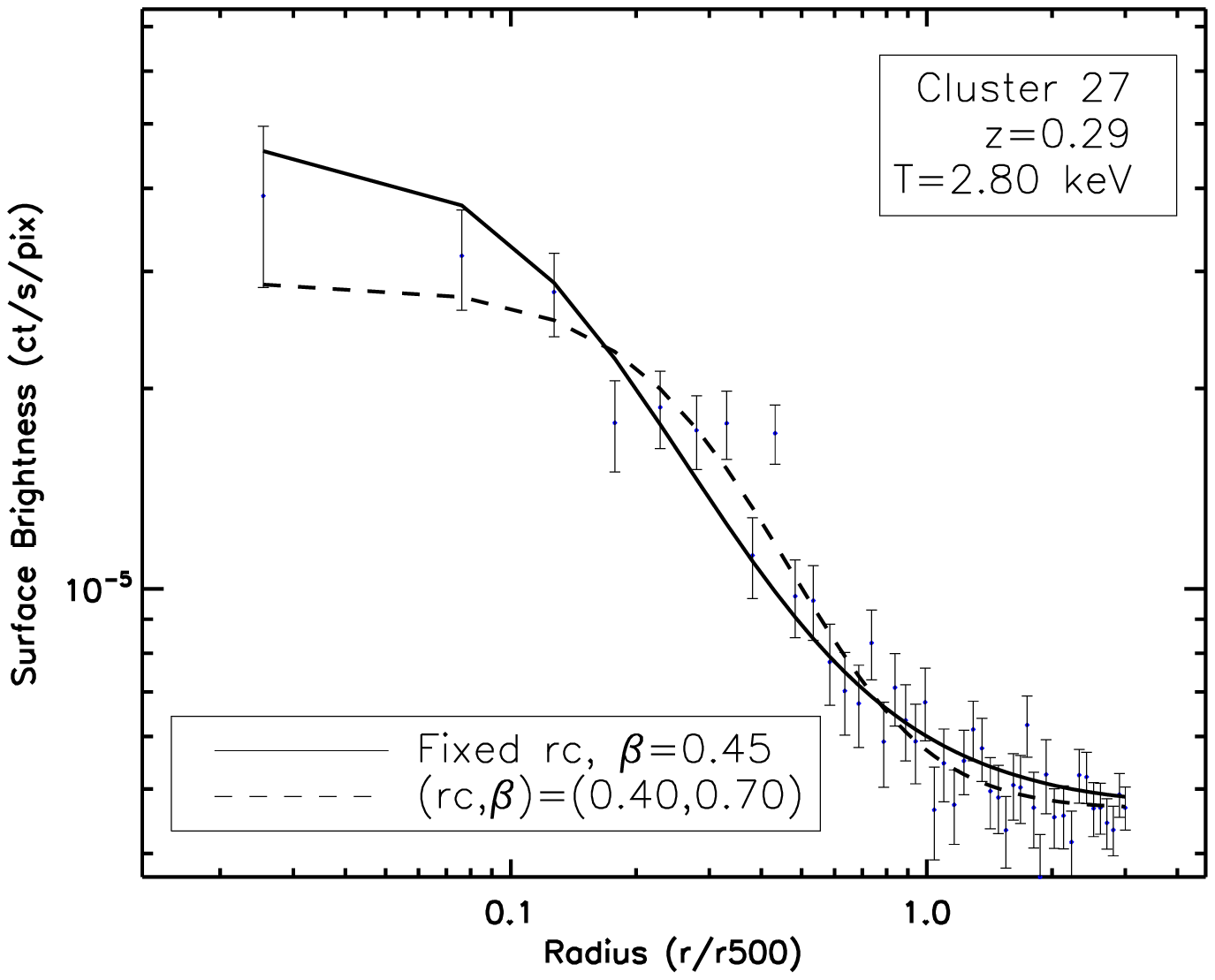,width=3.8cm,height=3.6cm}
\epsfig{file=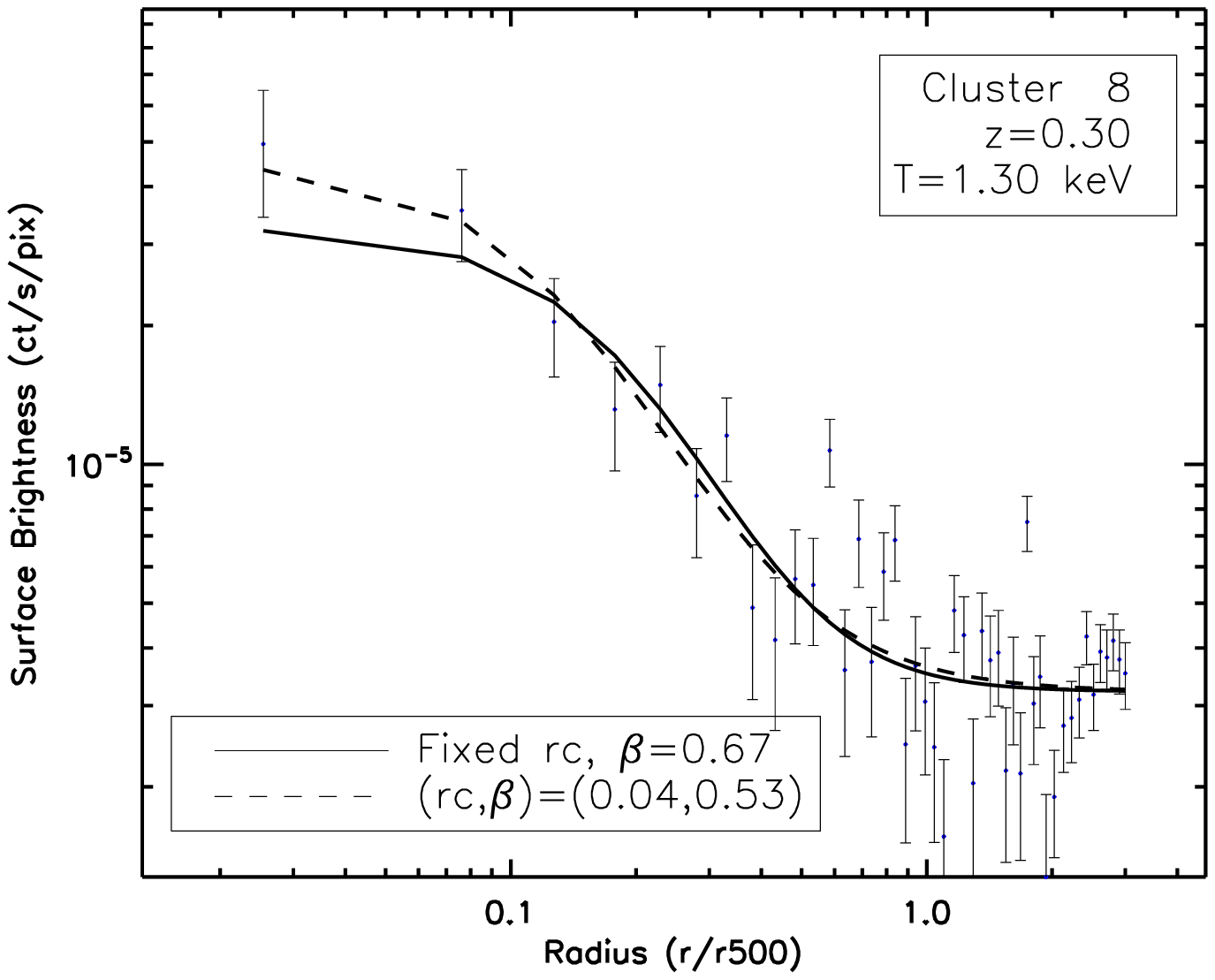,width=3.8cm,height=3.6cm}
\epsfig{file=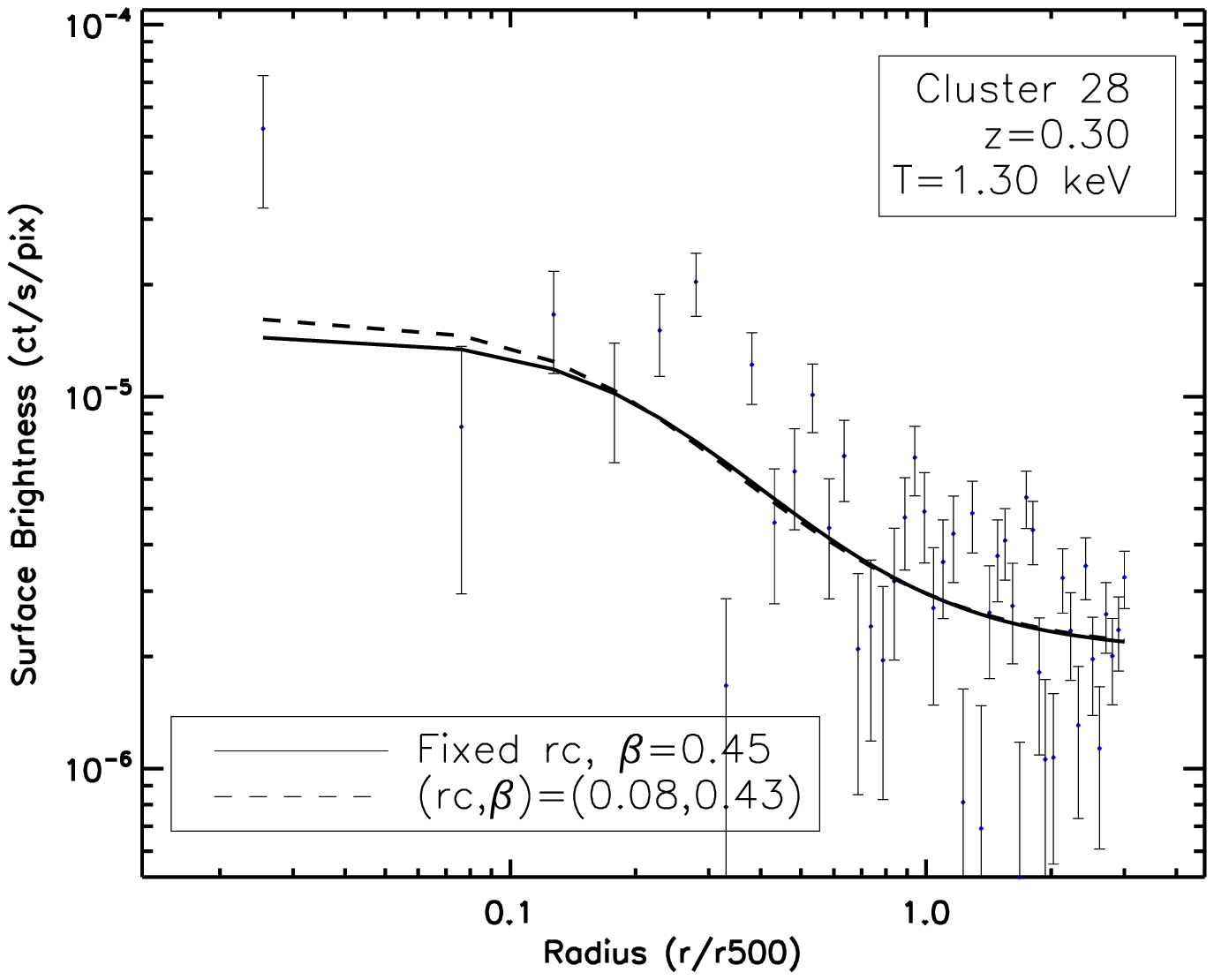,width=3.8cm,height=3.6cm}
\epsfig{file=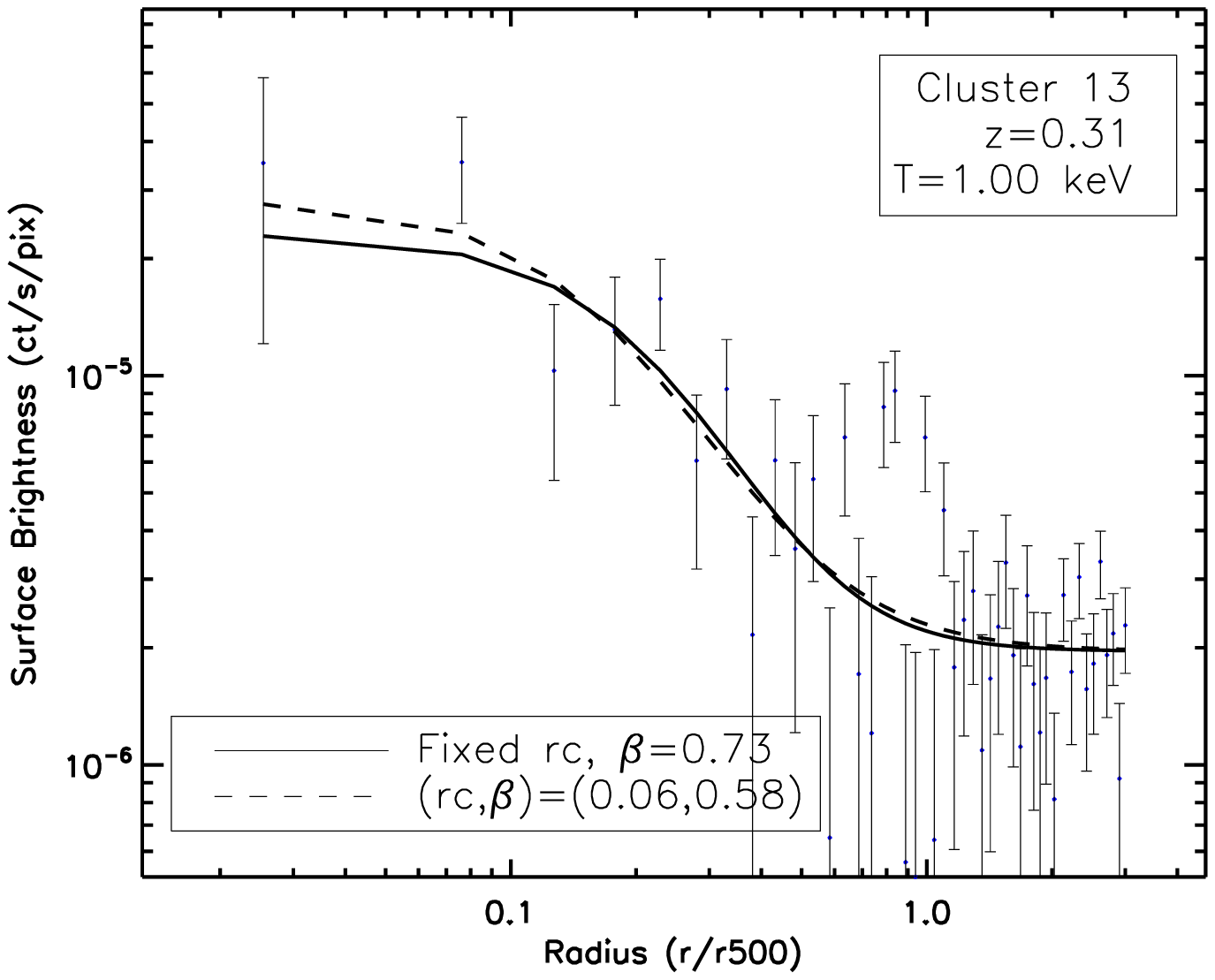,width=3.8cm,height=3.6cm}
\epsfig{file=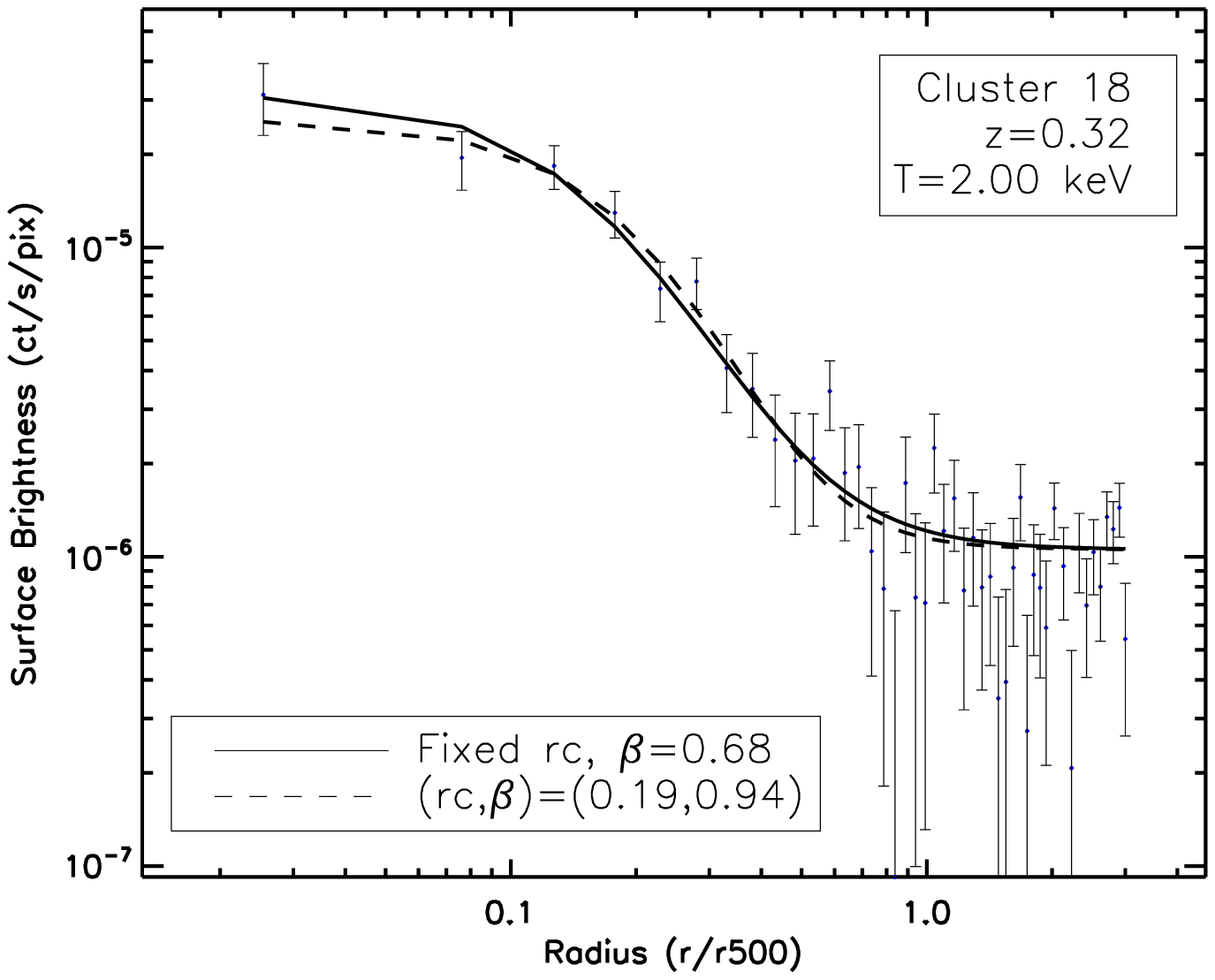,width=3.8cm,height=3.6cm}
\epsfig{file=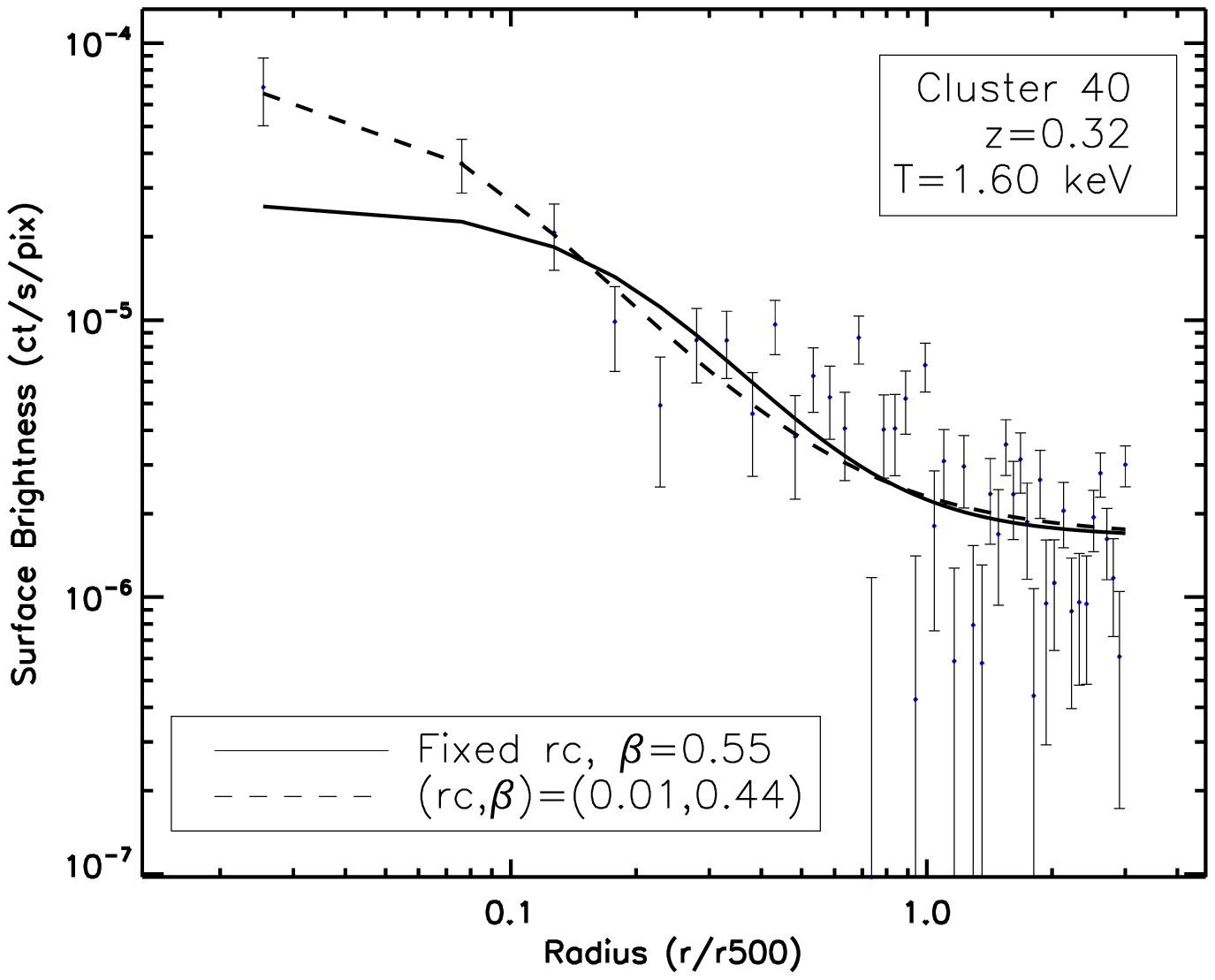,width=3.8cm,height=3.6cm}
\epsfig{file=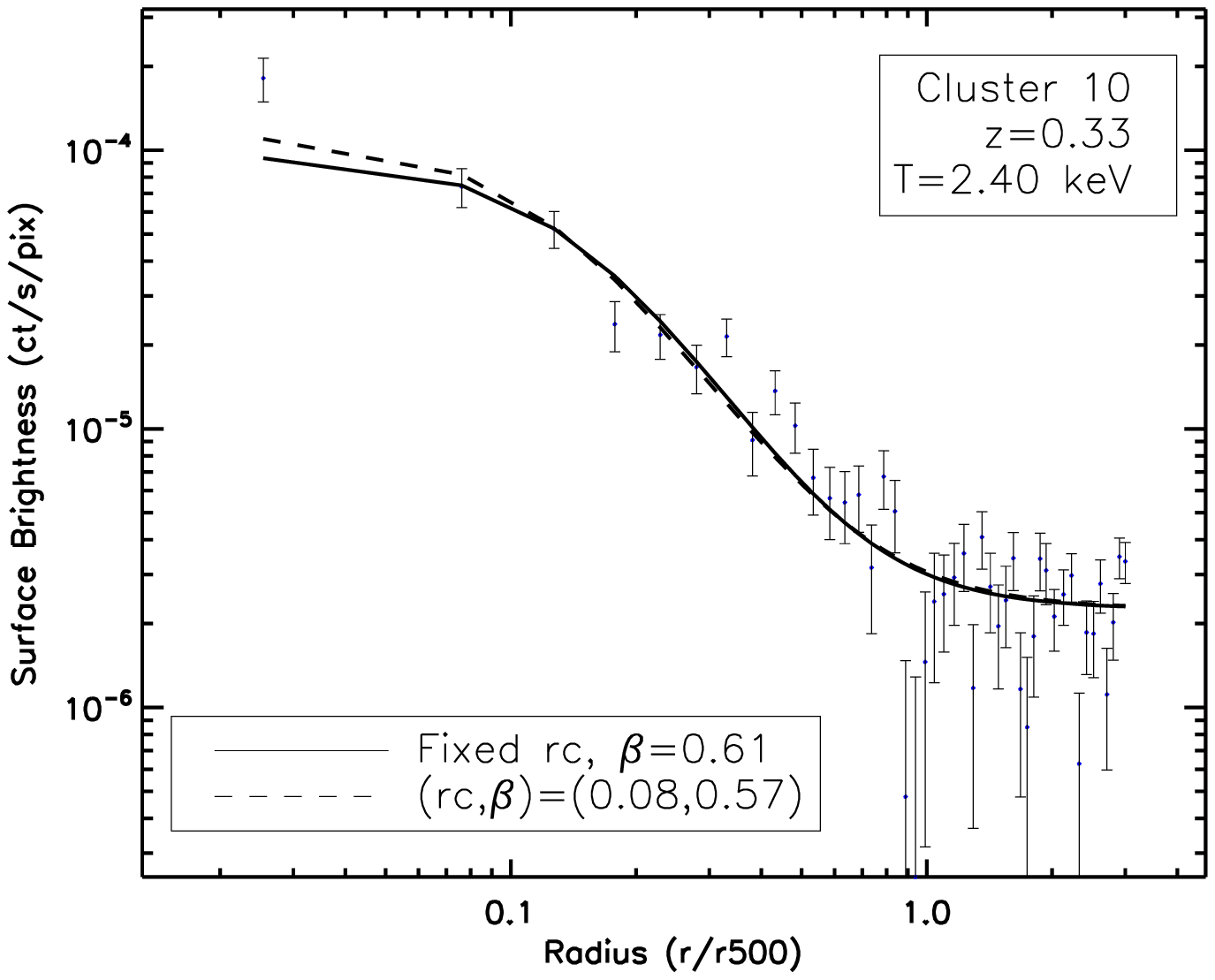,width=3.8cm,height=3.6cm}
\epsfig{file=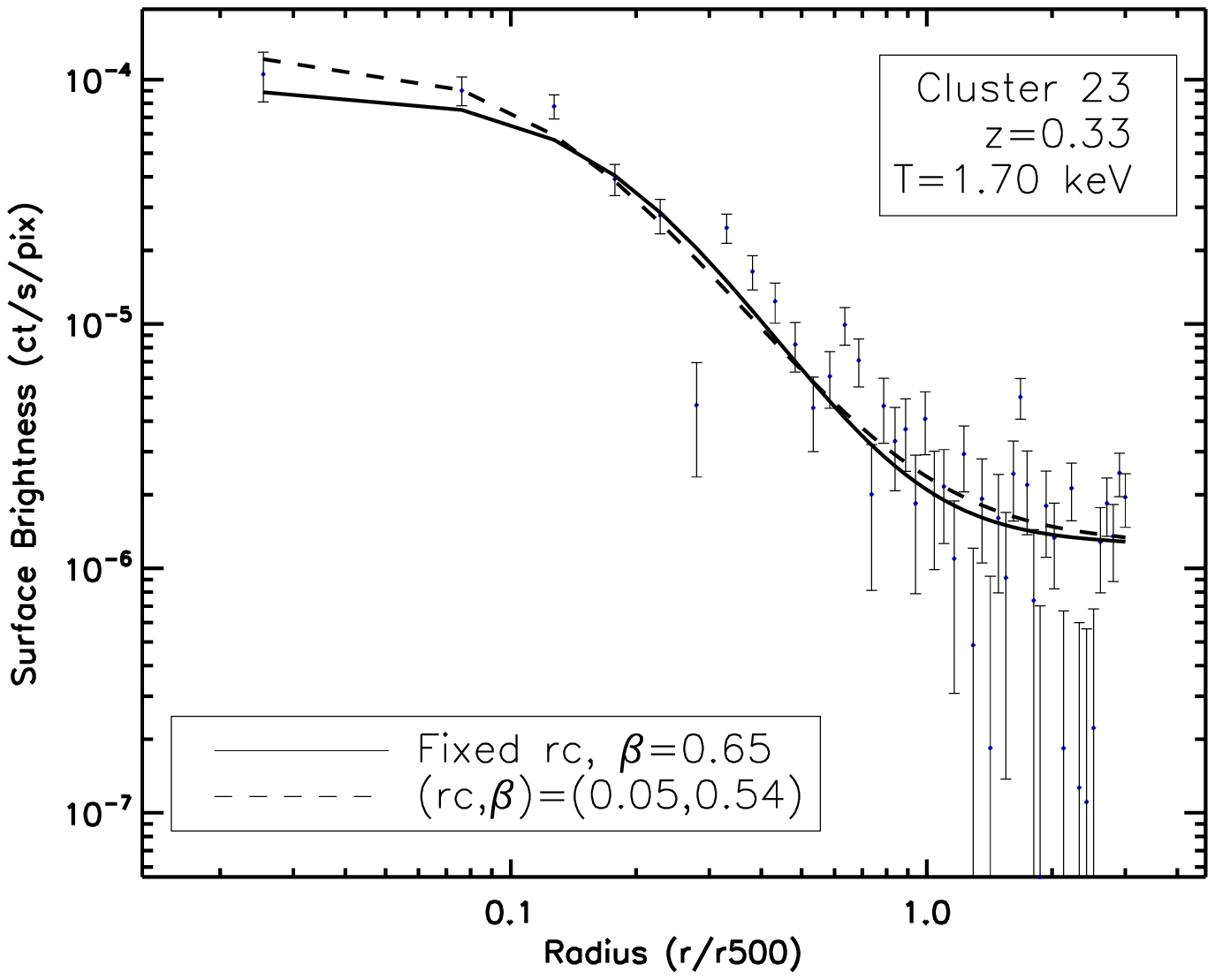,width=3.8cm,height=3.6cm}
\epsfig{file=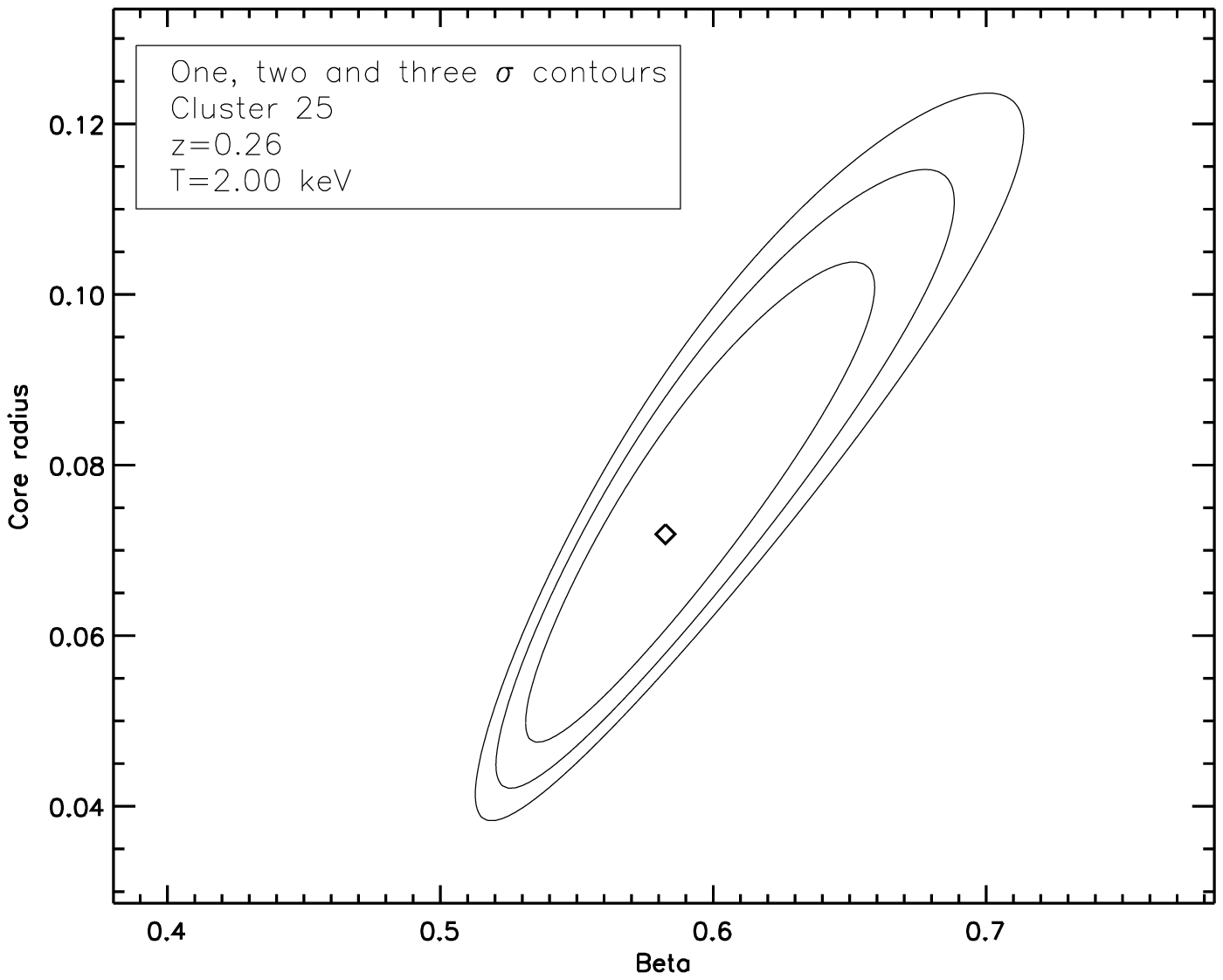,width=3.8cm,height=3.6cm}
\epsfig{file=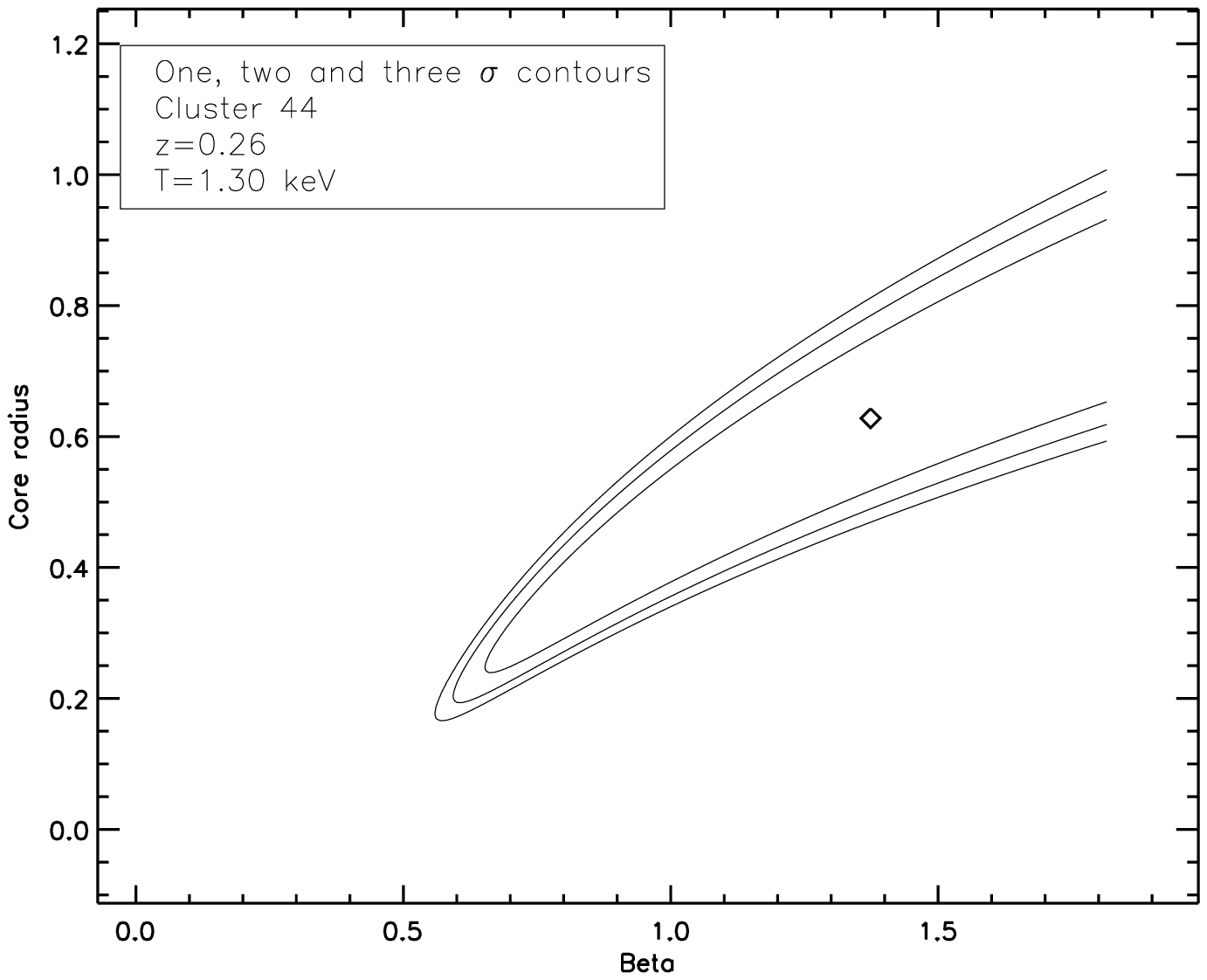,width=3.8cm,height=3.6cm}
\epsfig{file=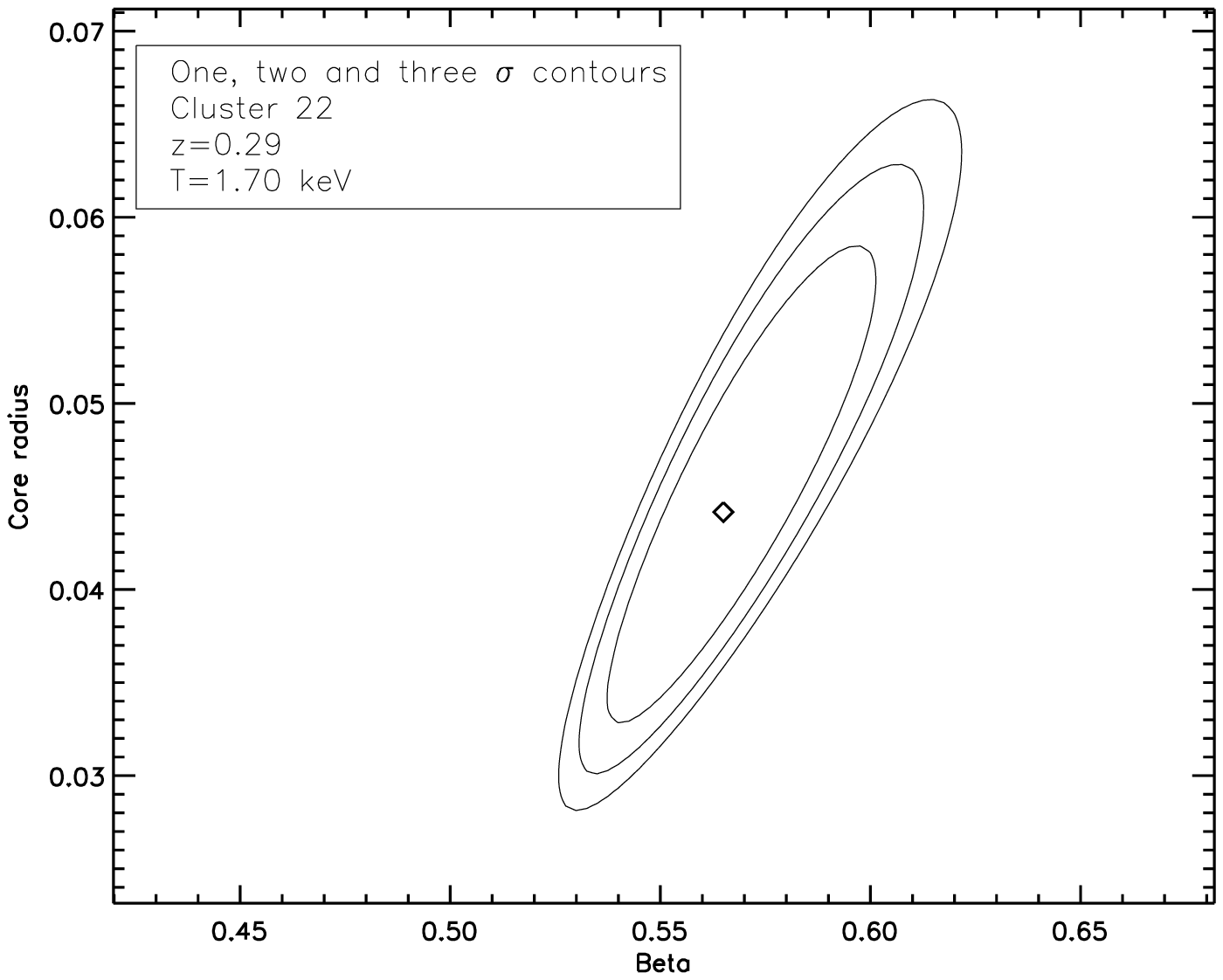,width=3.8cm,height=3.6cm}
\epsfig{file=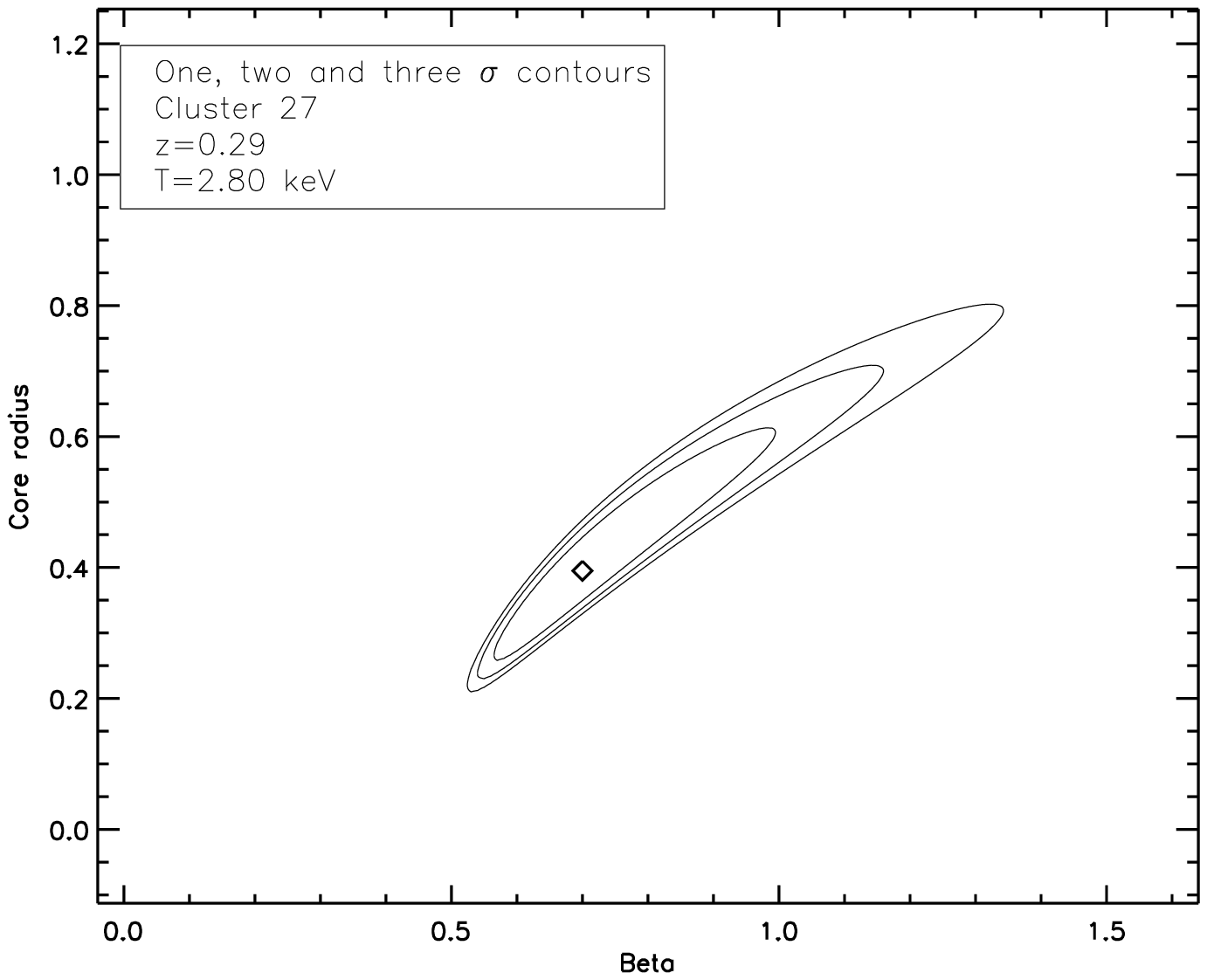,width=3.8cm,height=3.6cm}
\epsfig{file=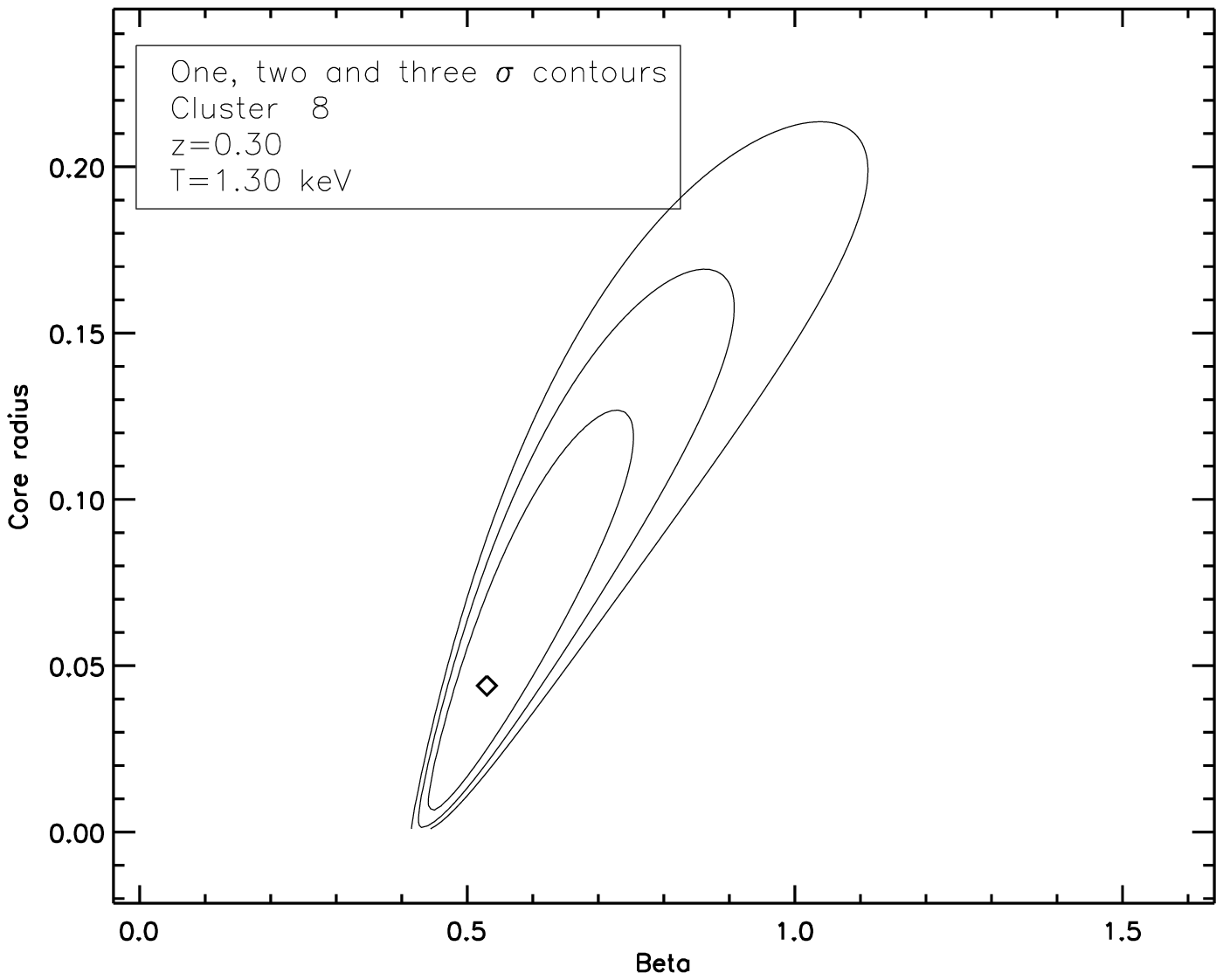,width=3.8cm,height=3.6cm}
\epsfig{file=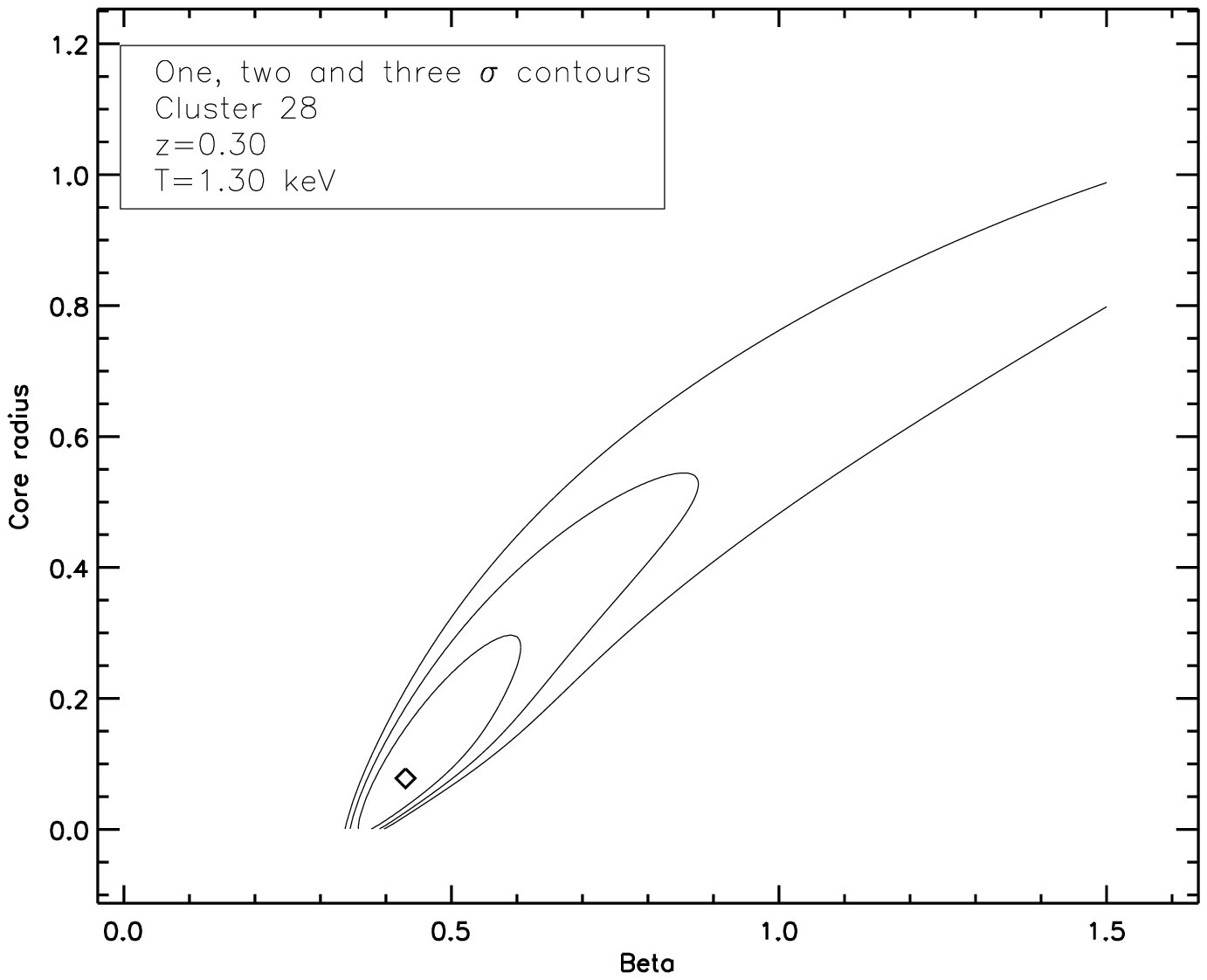,width=3.8cm,height=3.6cm}
\epsfig{file=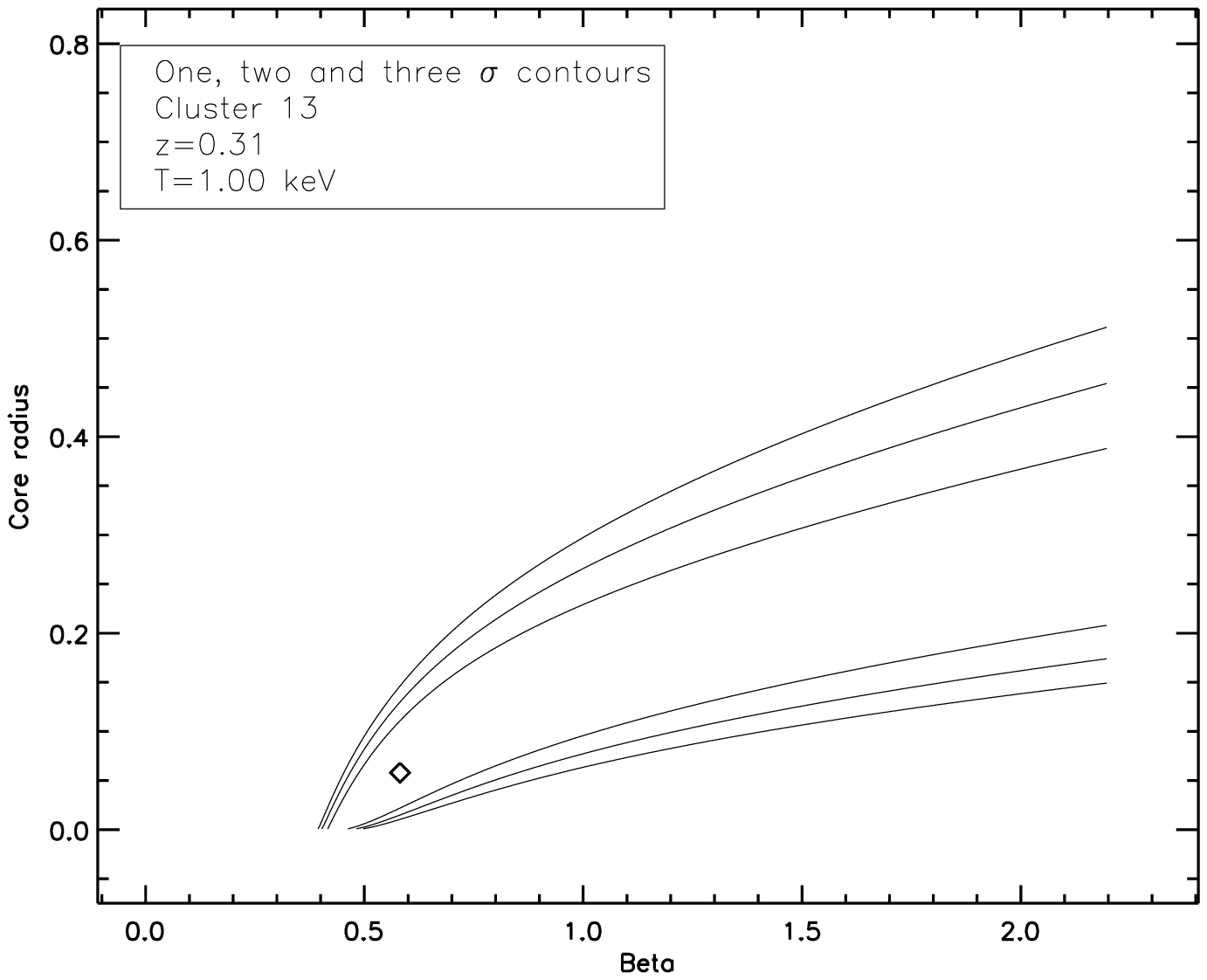,width=3.8cm,height=3.6cm}
\epsfig{file=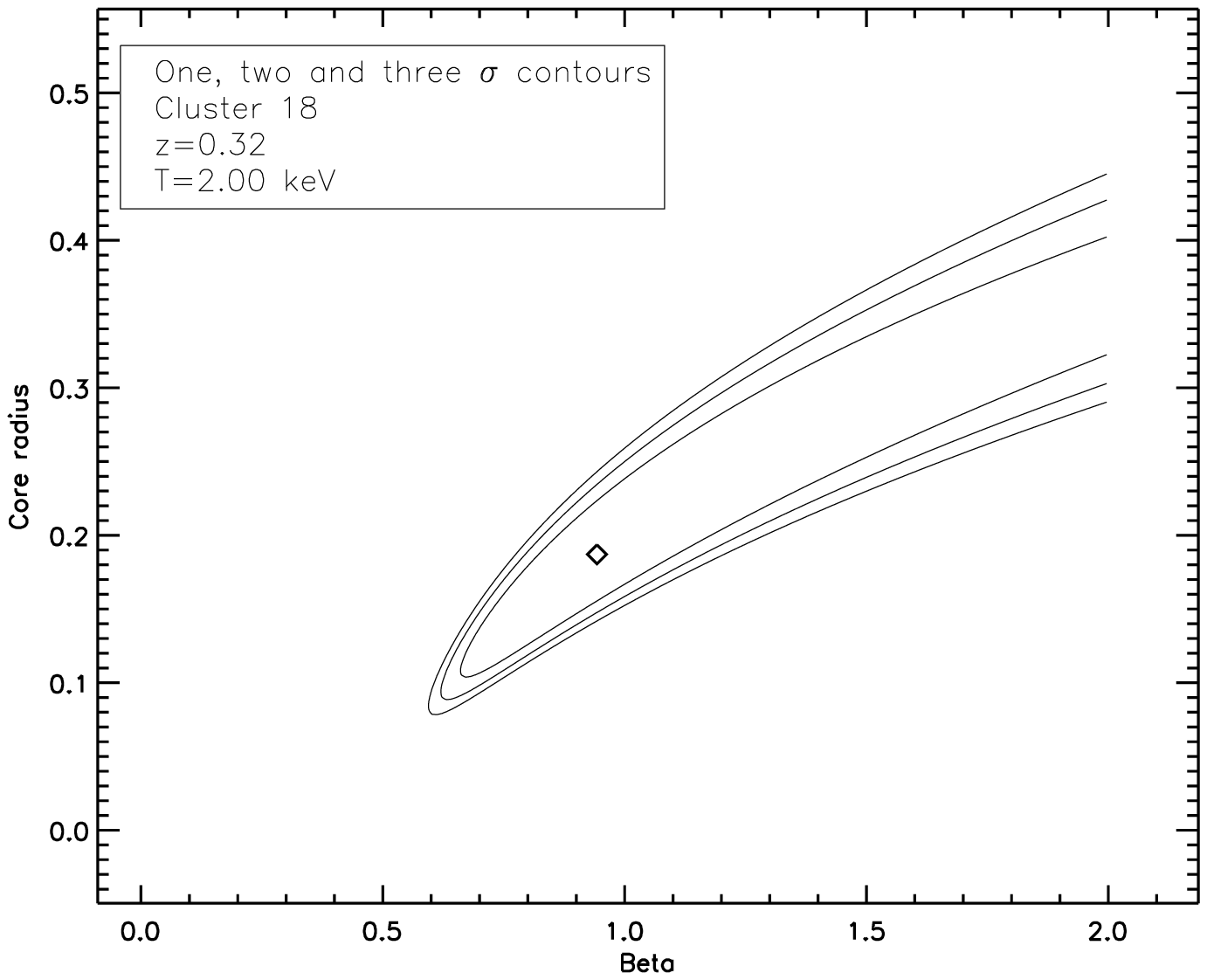,width=3.8cm,height=3.6cm}
\epsfig{file=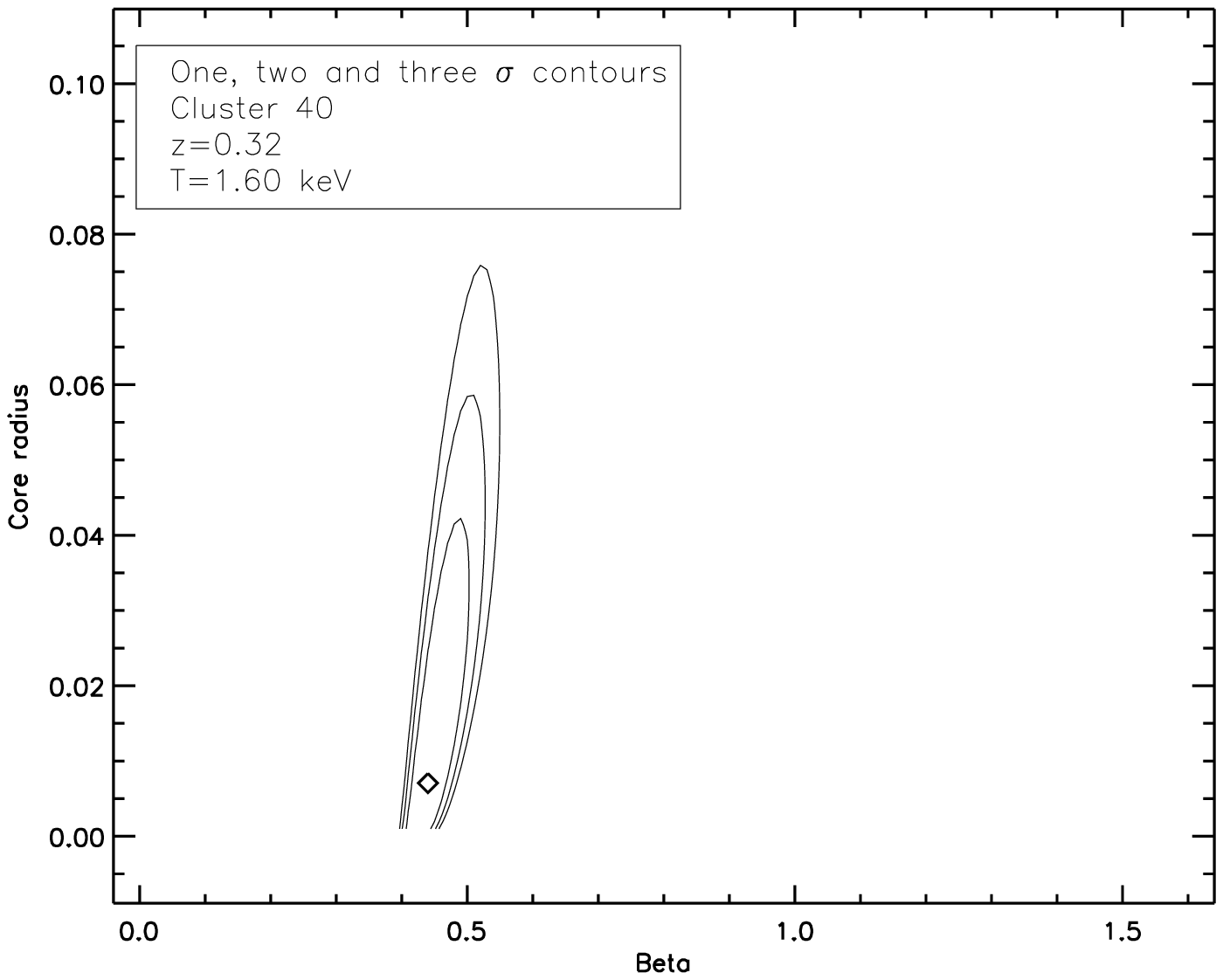,width=3.8cm,height=3.6cm}
\epsfig{file=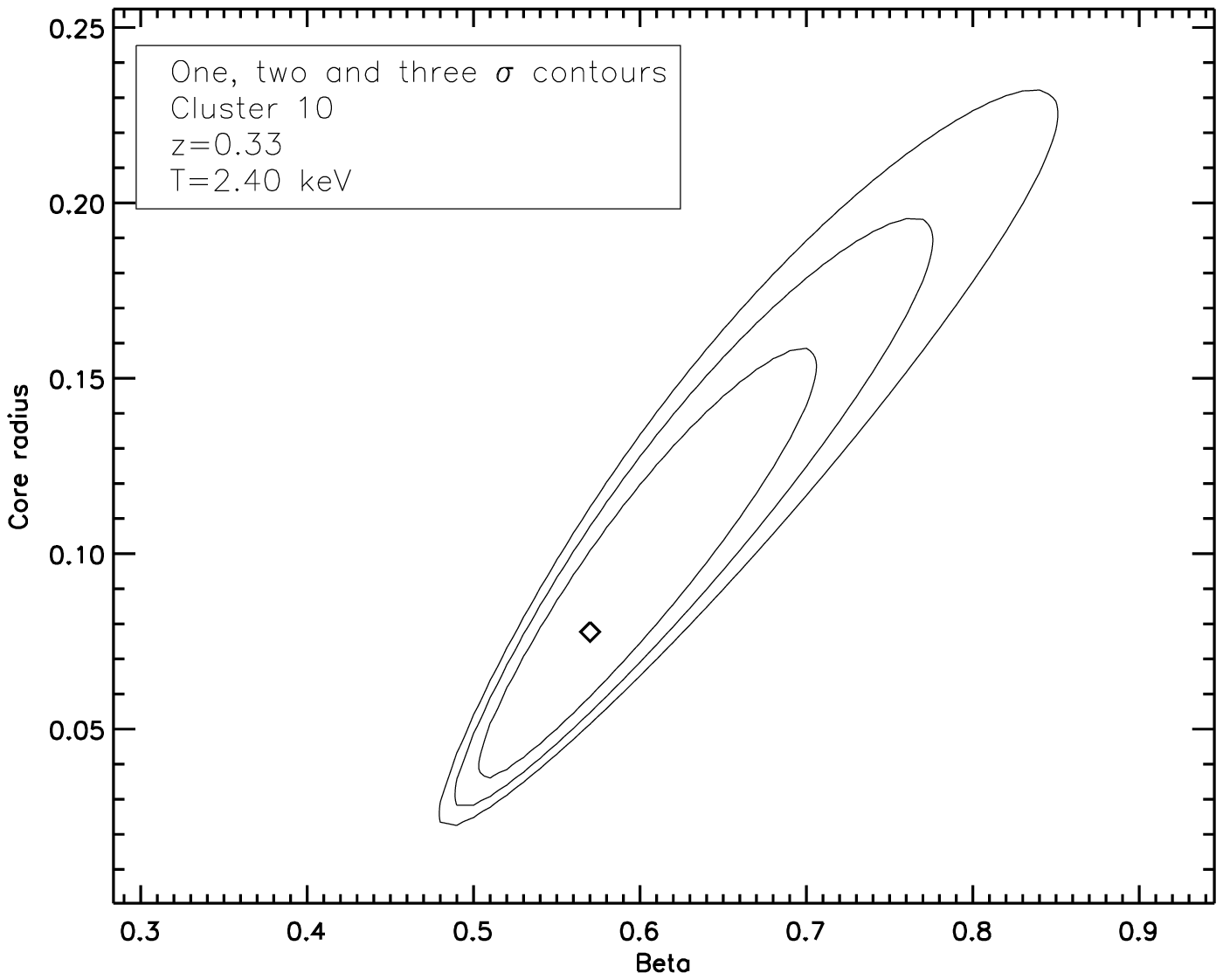,width=3.8cm,height=3.6cm}
\epsfig{file=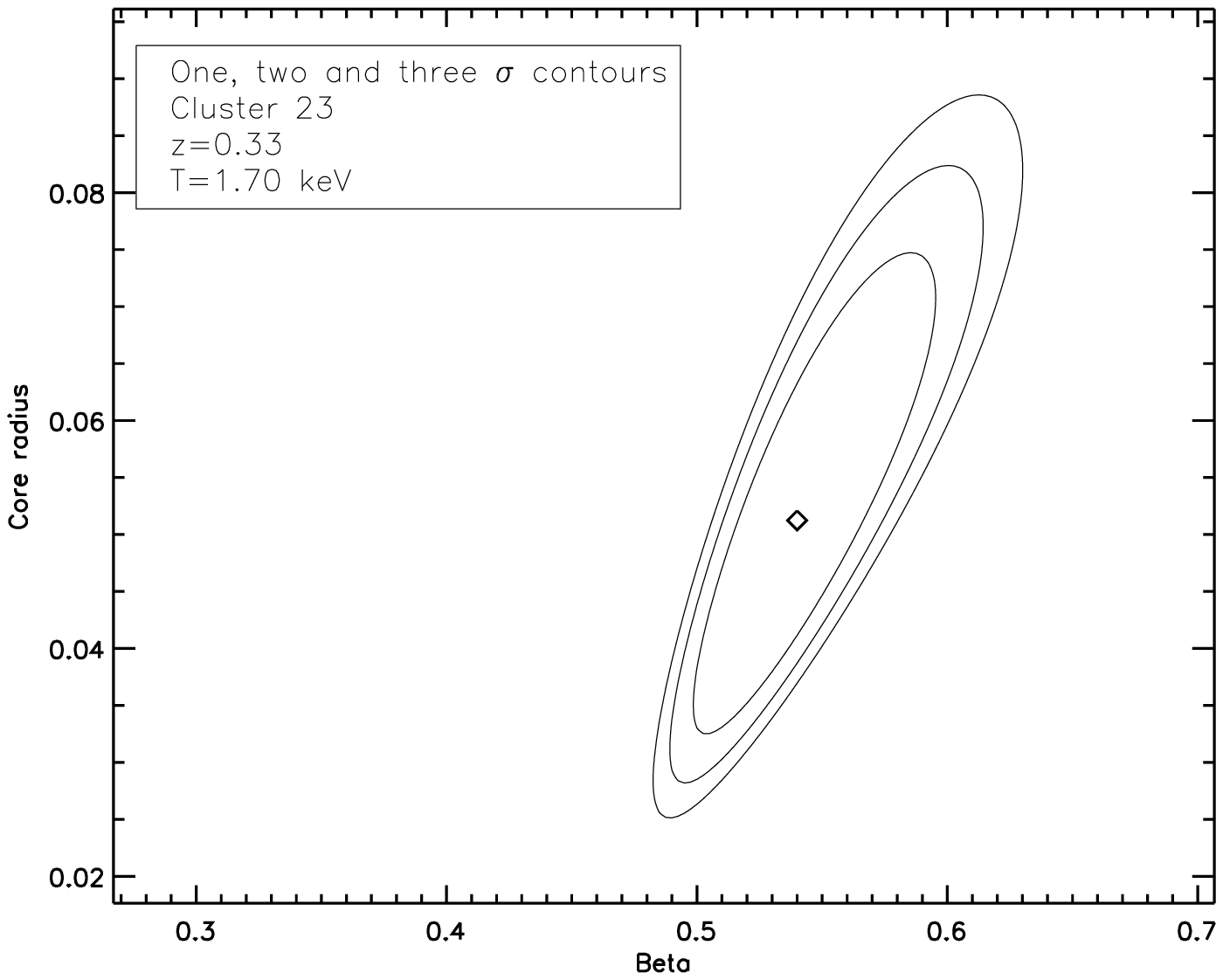,width=3.8cm,height=3.6cm}

\caption{X-ray surface brightness profiles of the individual C1 clusters with redshift $0.26 \leq z \leq 0.33$, ordered according to redshift and the associated constrained $1 \sigma, 2 \sigma$ and $3 \sigma$ contours.  The dashed lines are the fitted \bmodel\ profiles with both \rc\ and $\beta$ freely fitted, while the solid lines are for the fitted profiles with free $\beta$ and \rc\ fixed to \ffh.}
\label{ind_prof2}
\end{figure*}

\begin{figure*}
\center
\epsfig{file=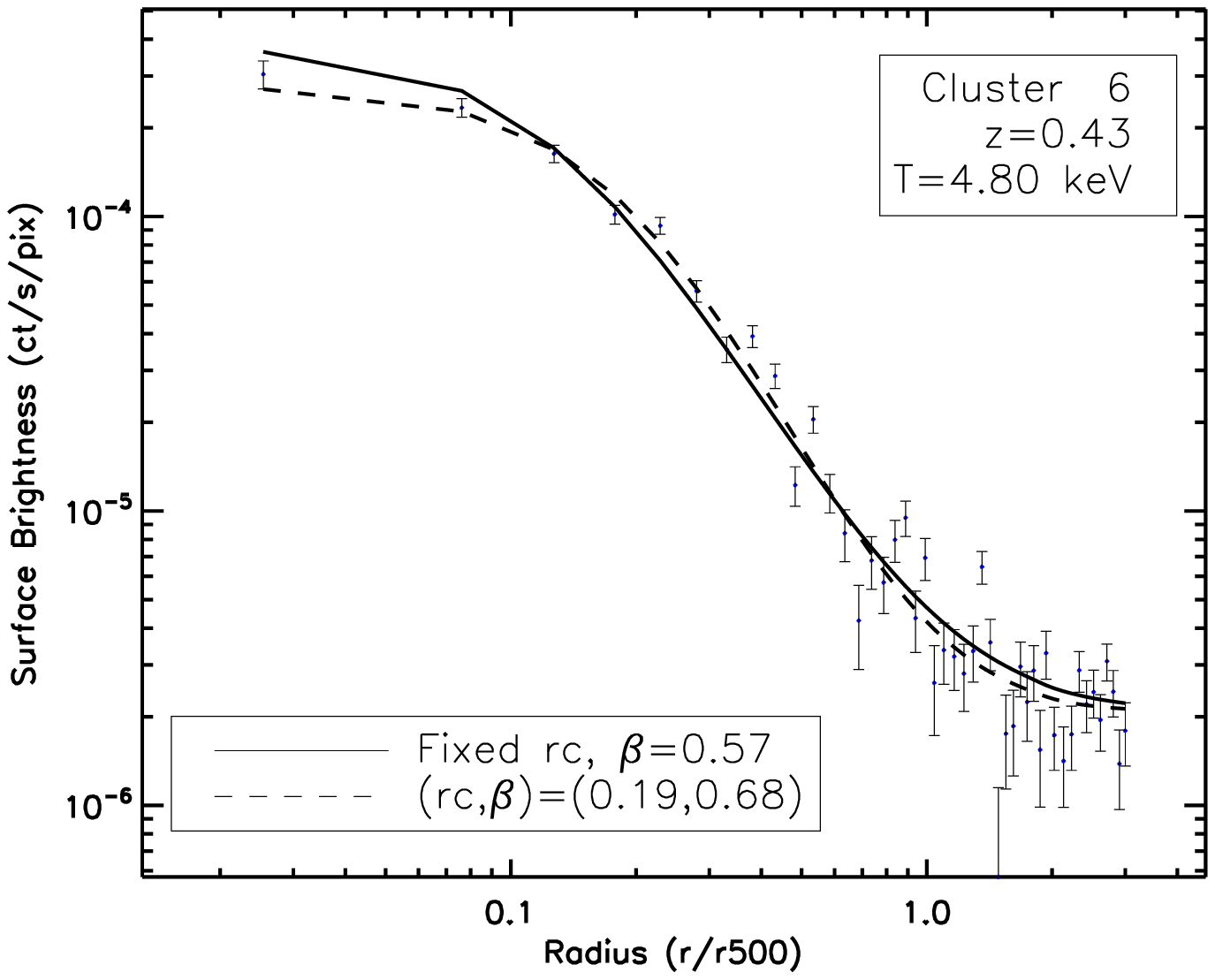,width=5.5cm,height=4.3cm}
\epsfig{file=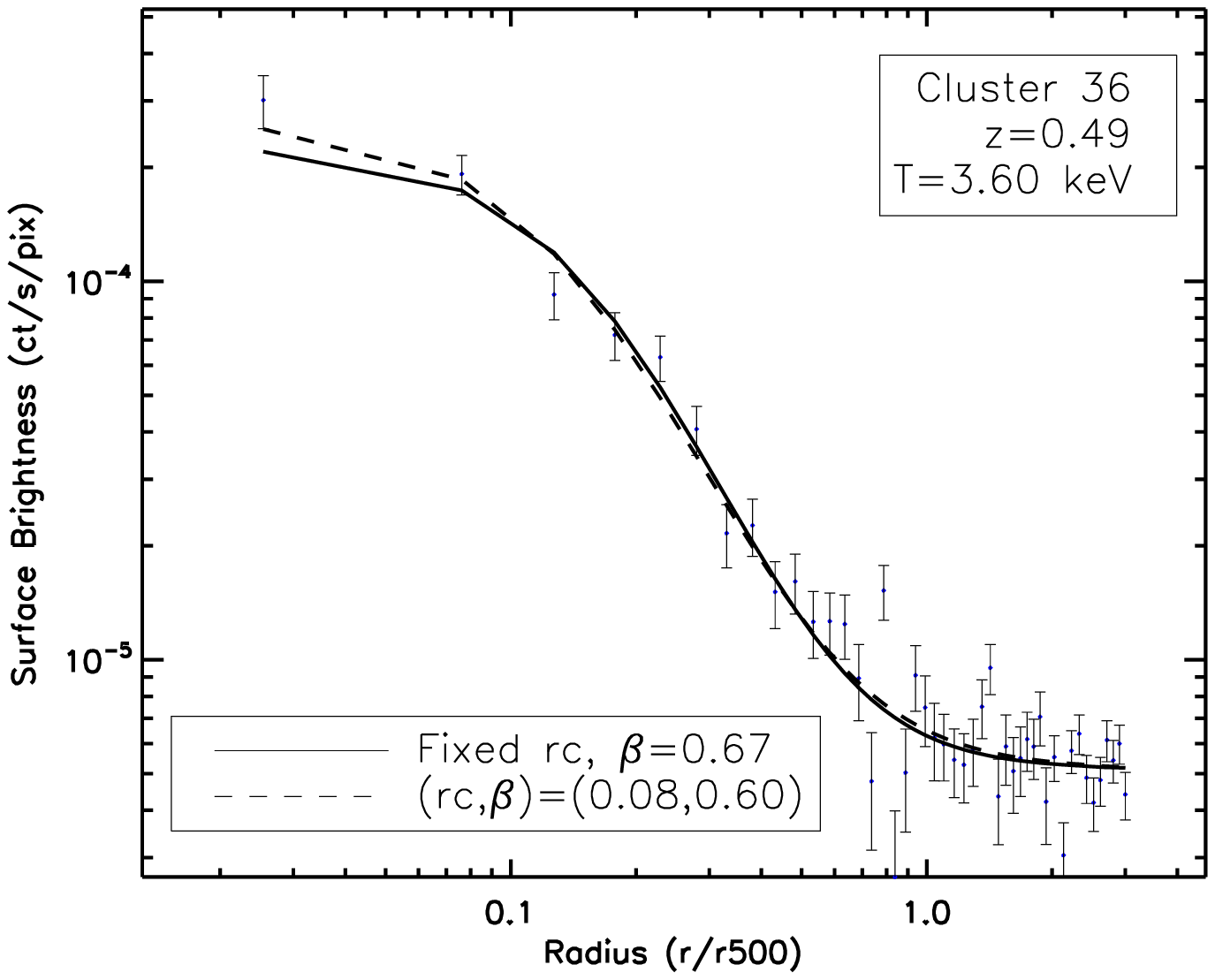,width=5.5cm,height=4.3cm}
\epsfig{file=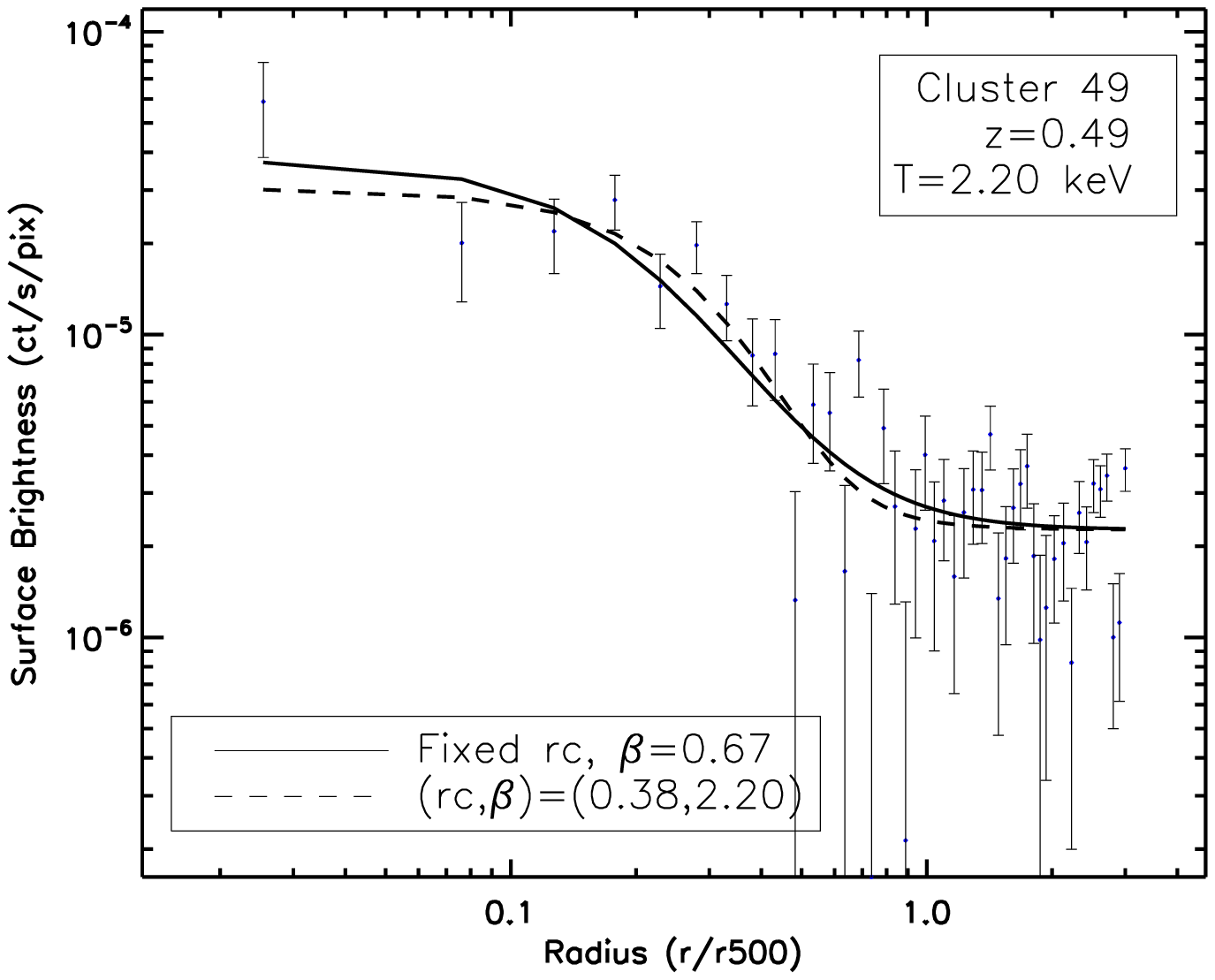,width=5.5cm,height=4.3cm}
\epsfig{file=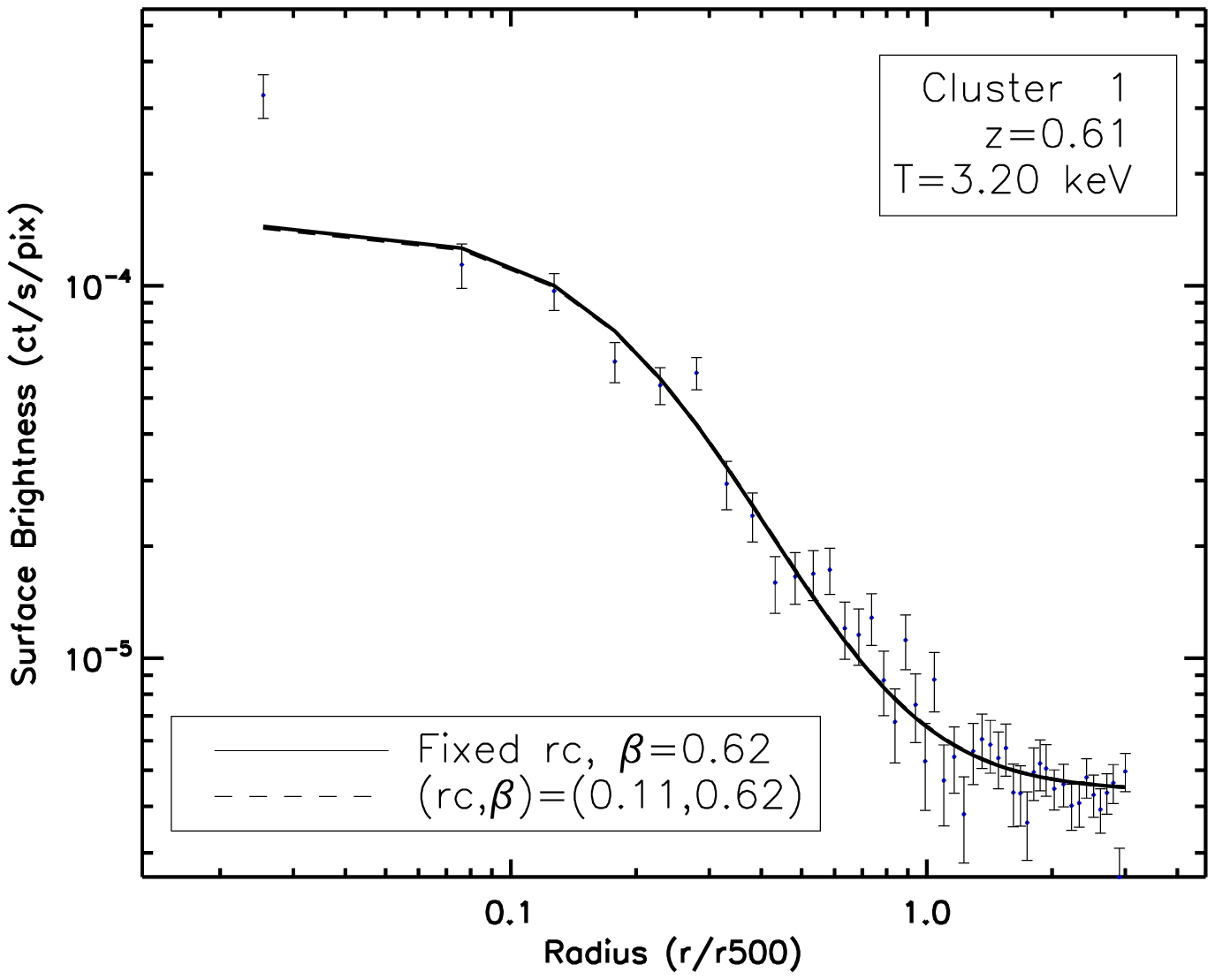,width=5.5cm,height=4.3cm}
\epsfig{file=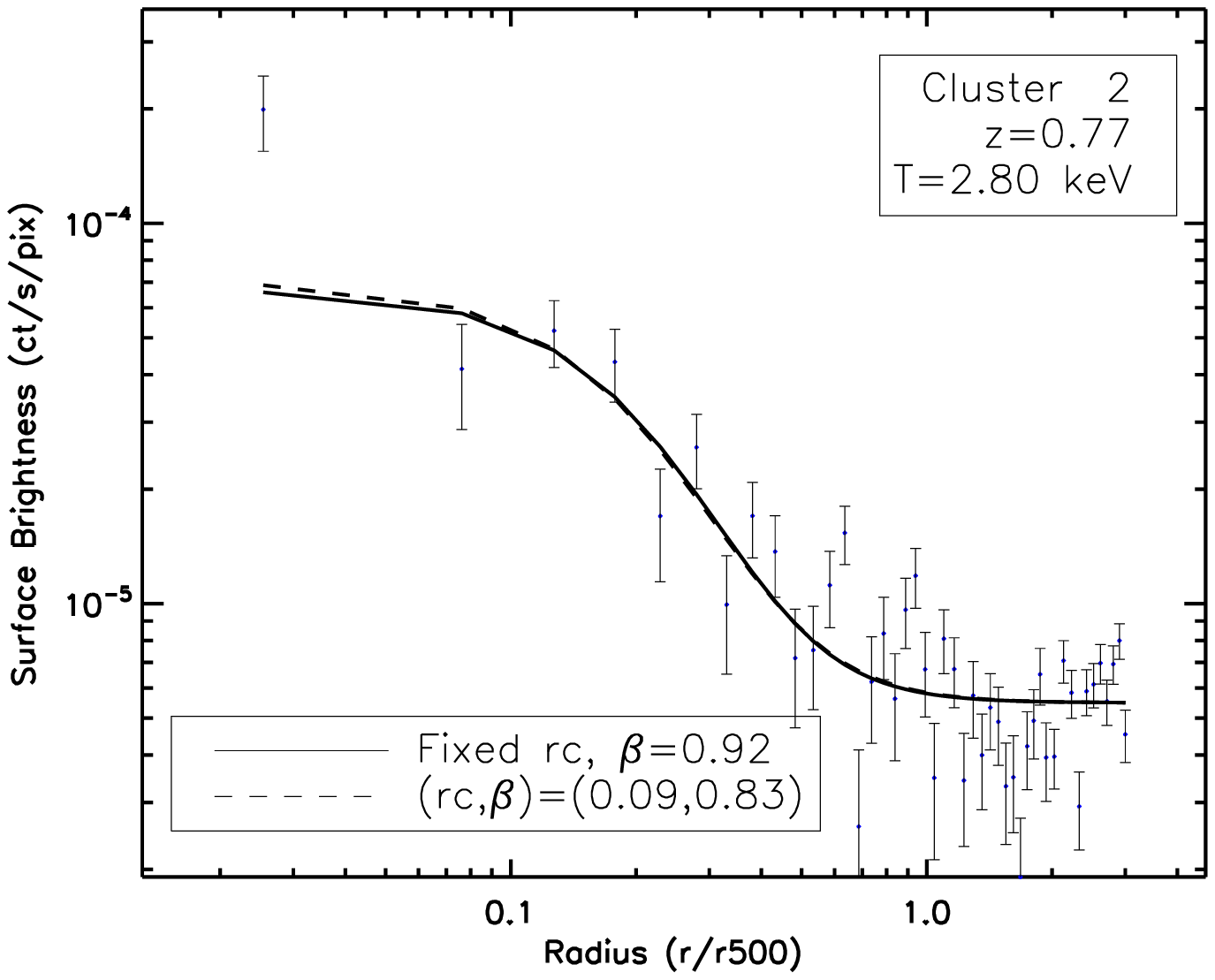,width=5.5cm,height=4.3cm}
\epsfig{file=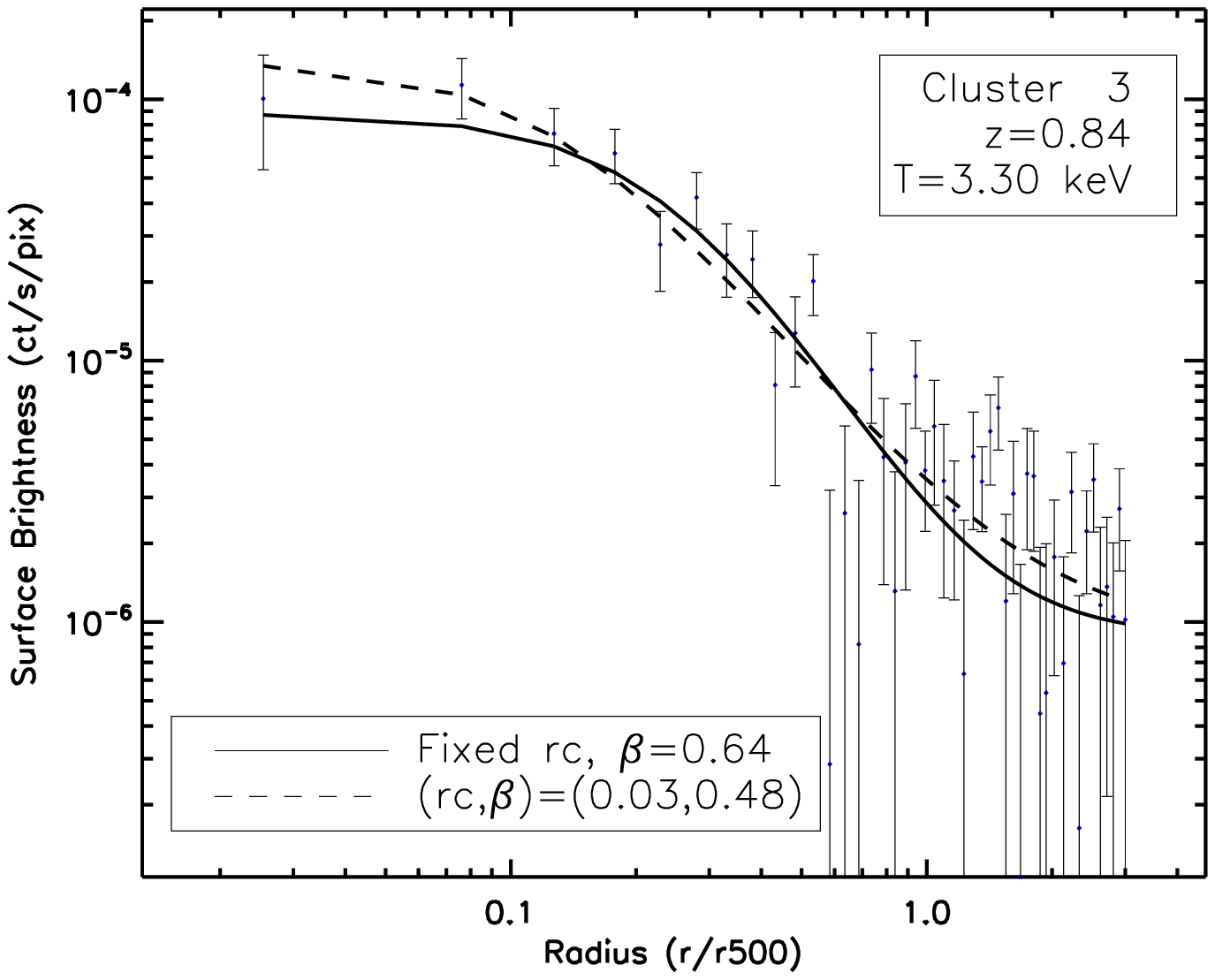,width=5.5cm,height=4.3cm}
\epsfig{file=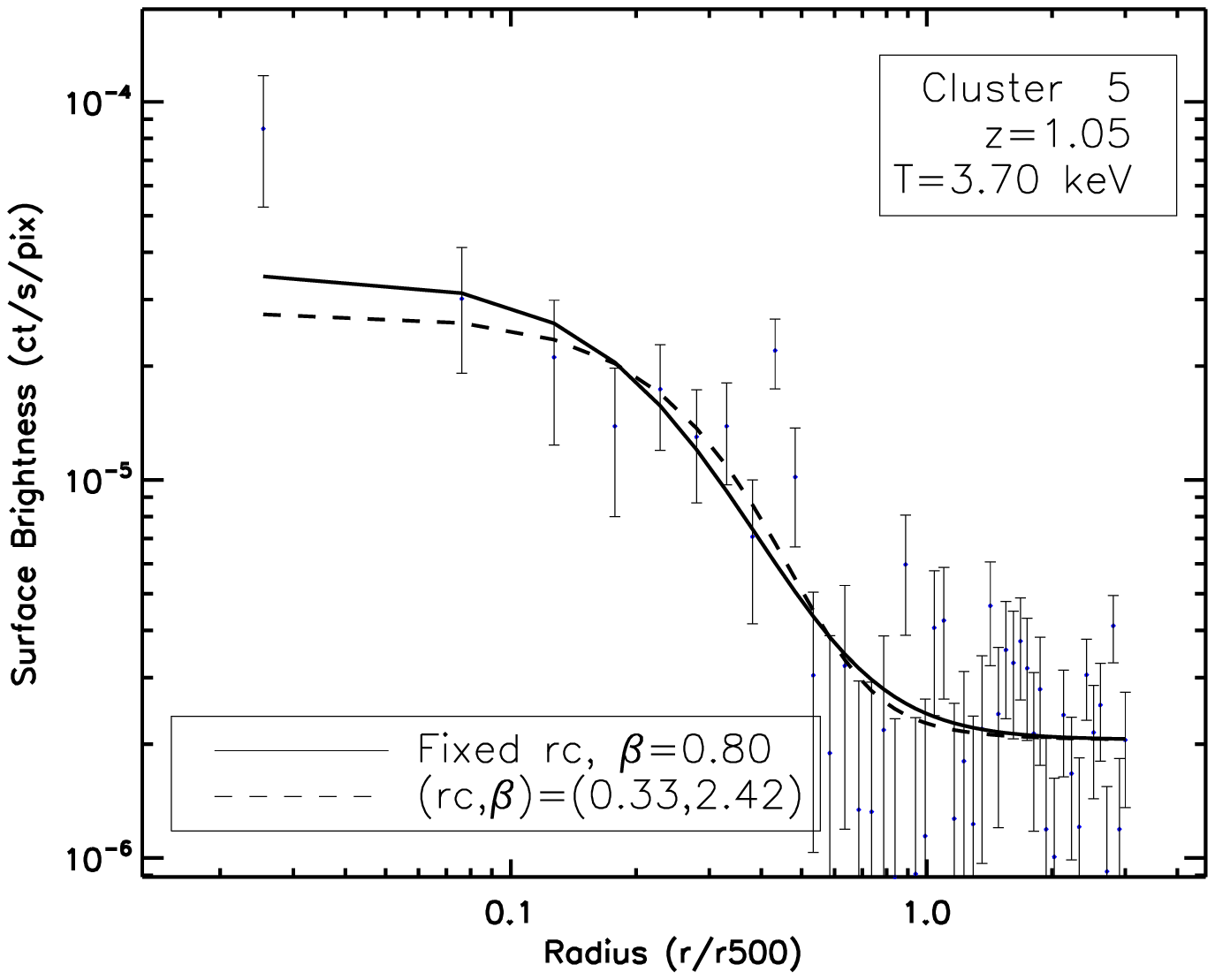,width=5.5cm,height=4.3cm}
\epsfig{file=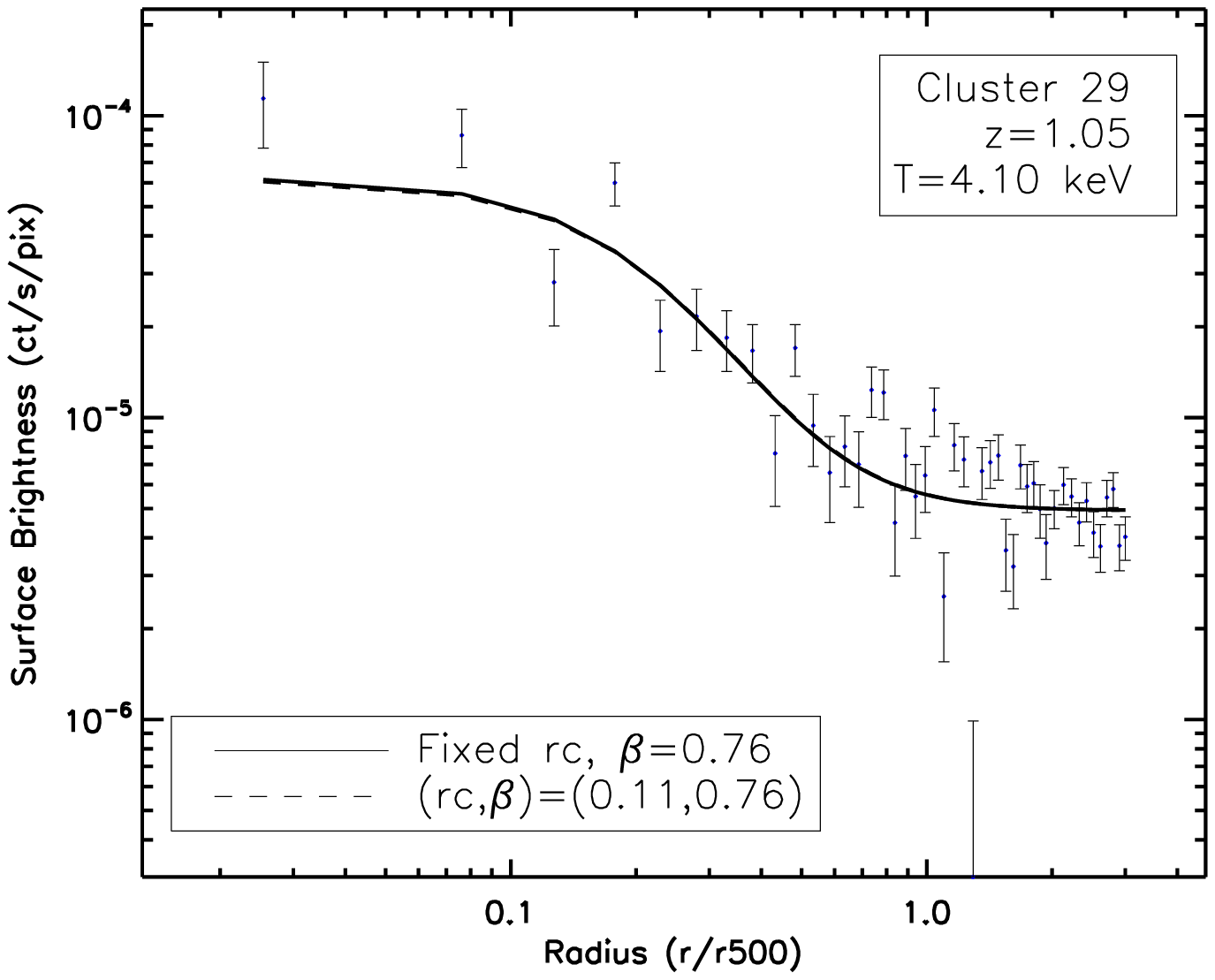,width=5.5cm,height=4.3cm}
\epsfig{file=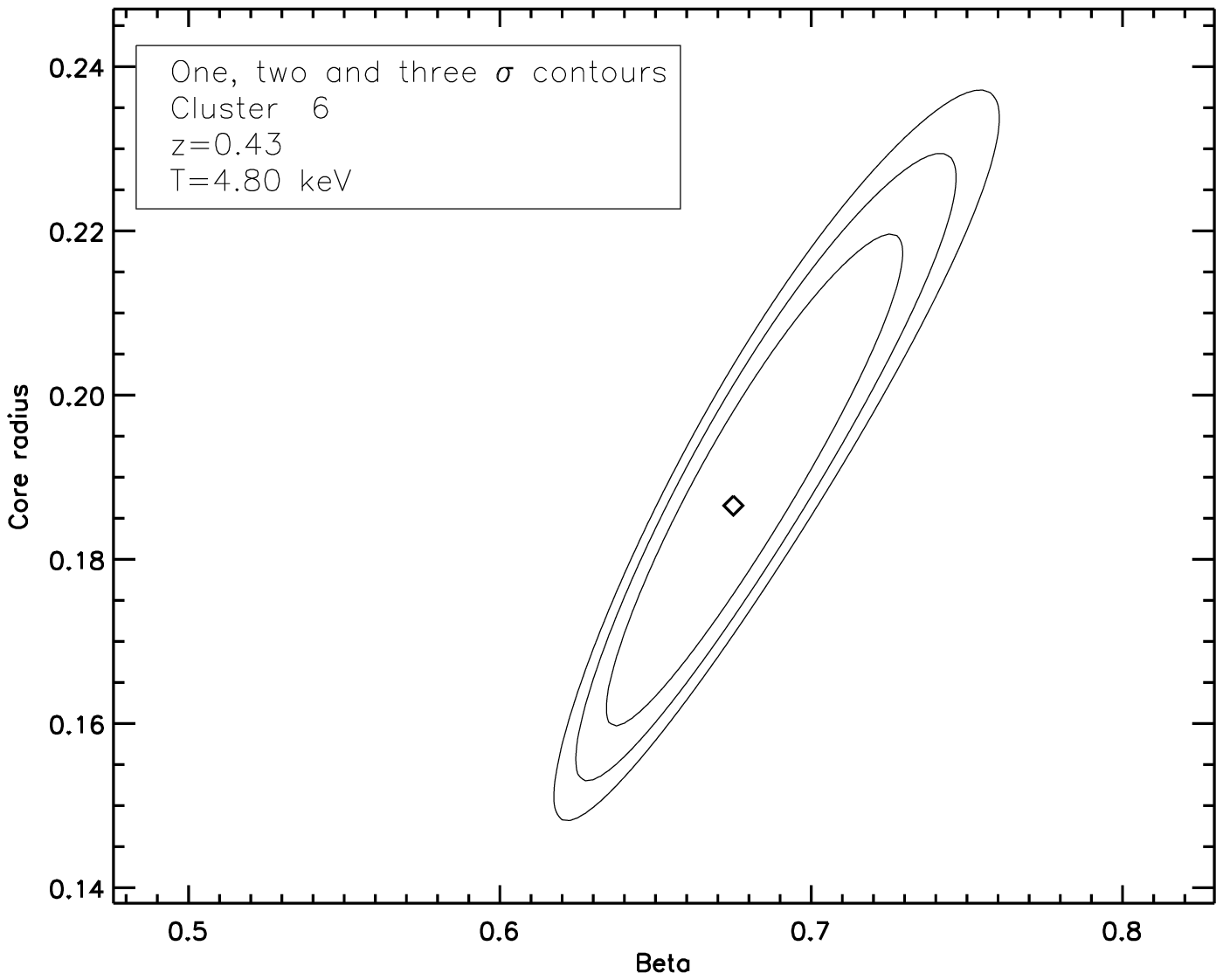,width=5.5cm,height=4.3cm}
\epsfig{file=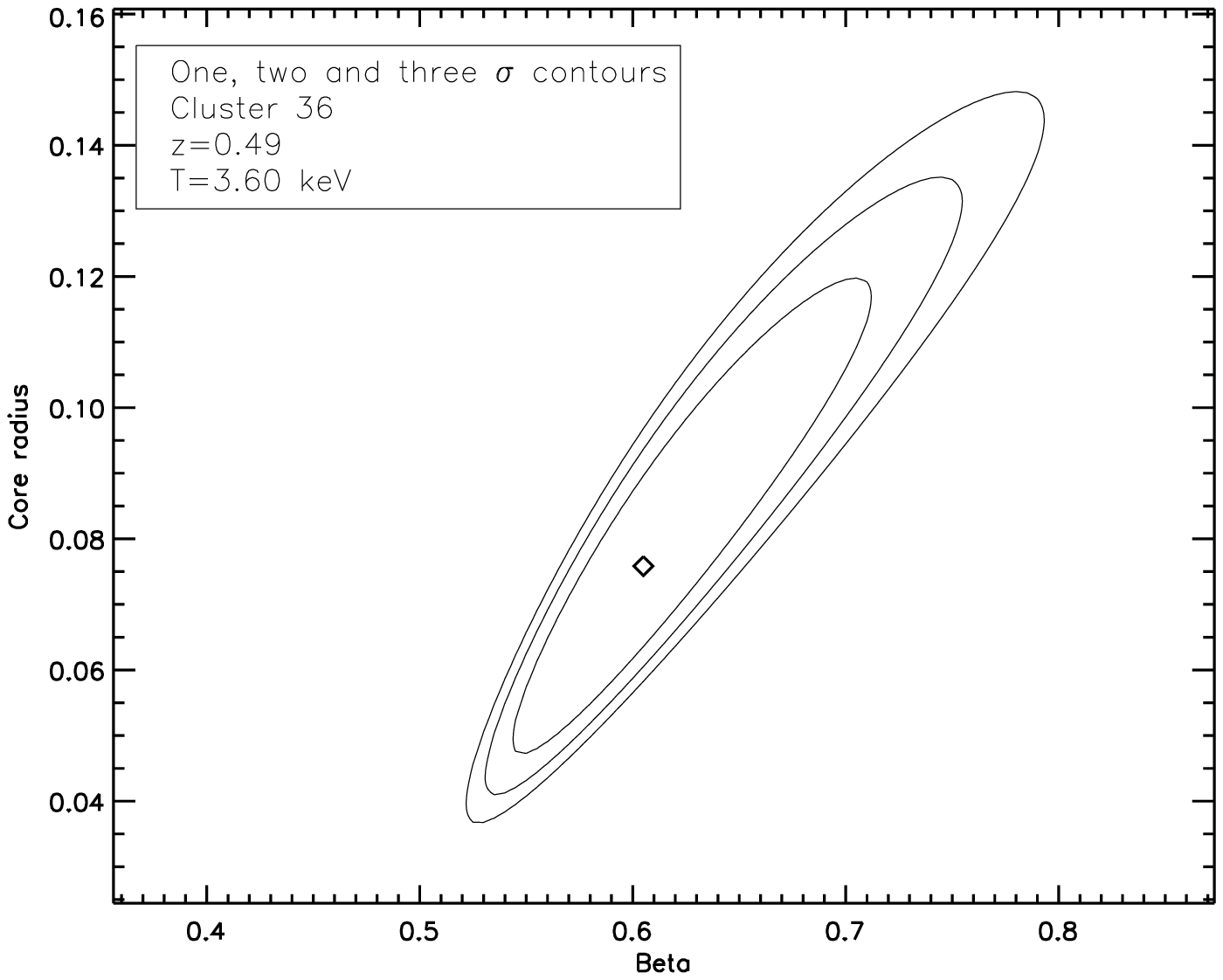,width=5.5cm,height=4.3cm}
\epsfig{file=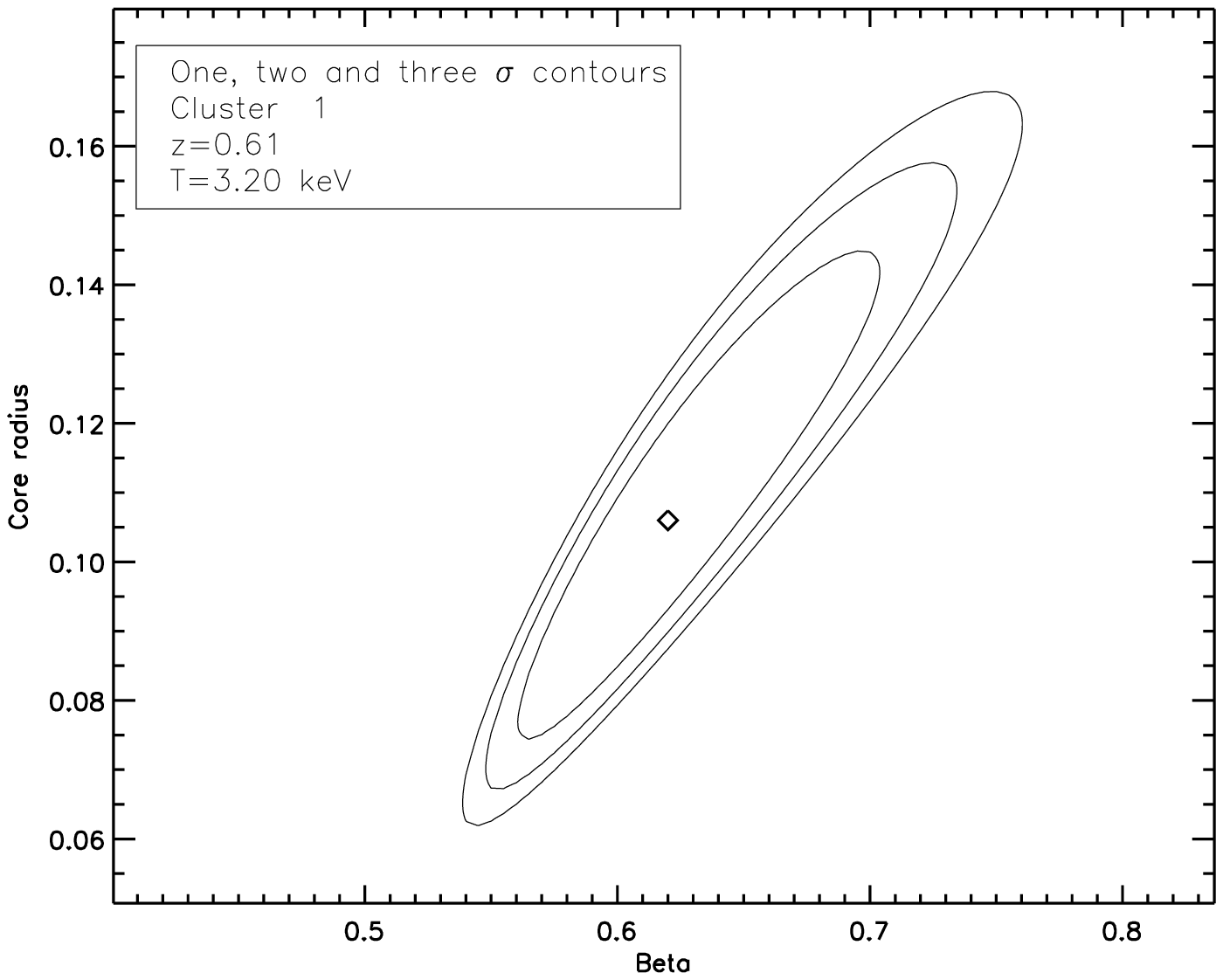,width=5.5cm,height=4.3cm}
\epsfig{file=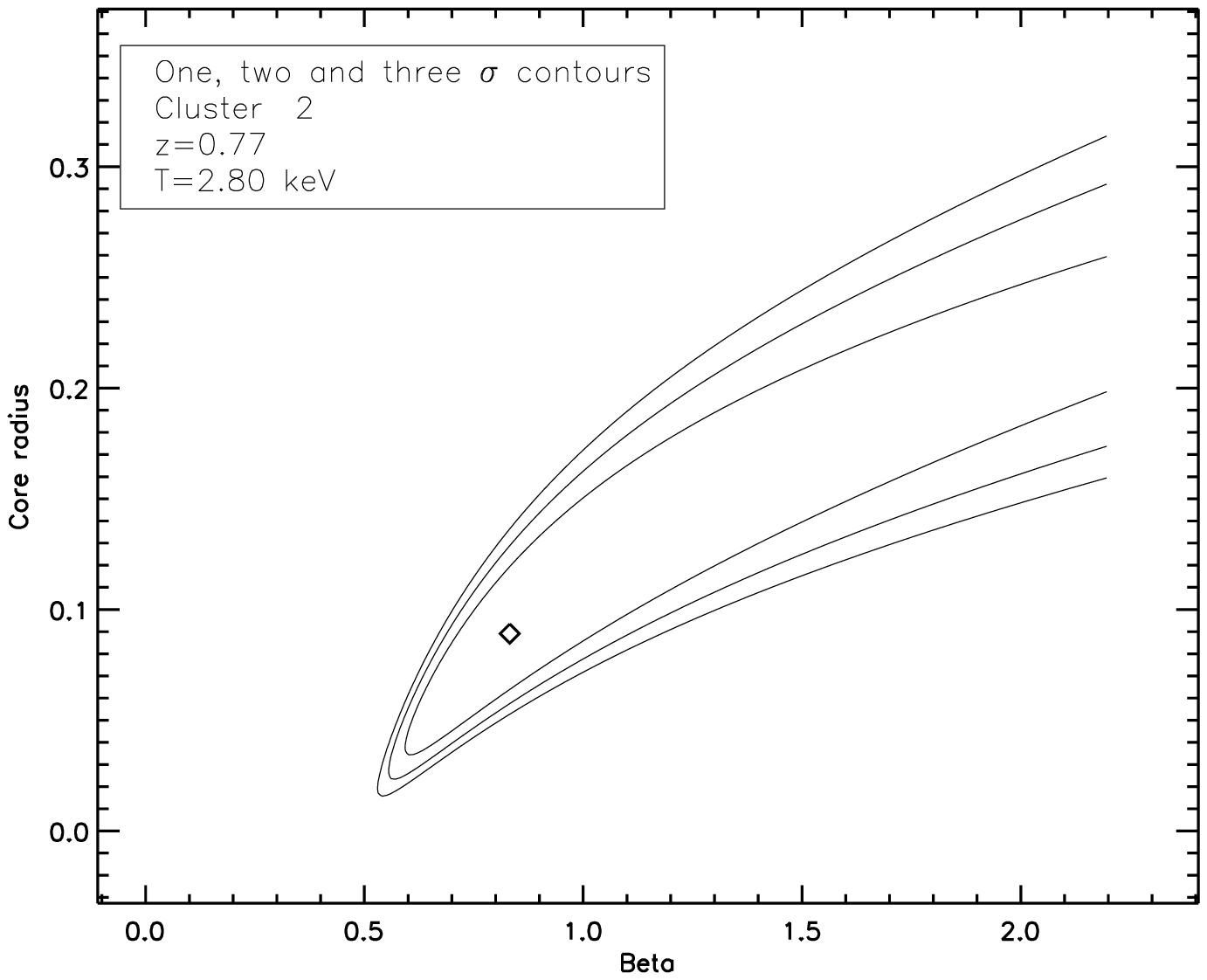,width=5.5cm,height=4.3cm}
\epsfig{file=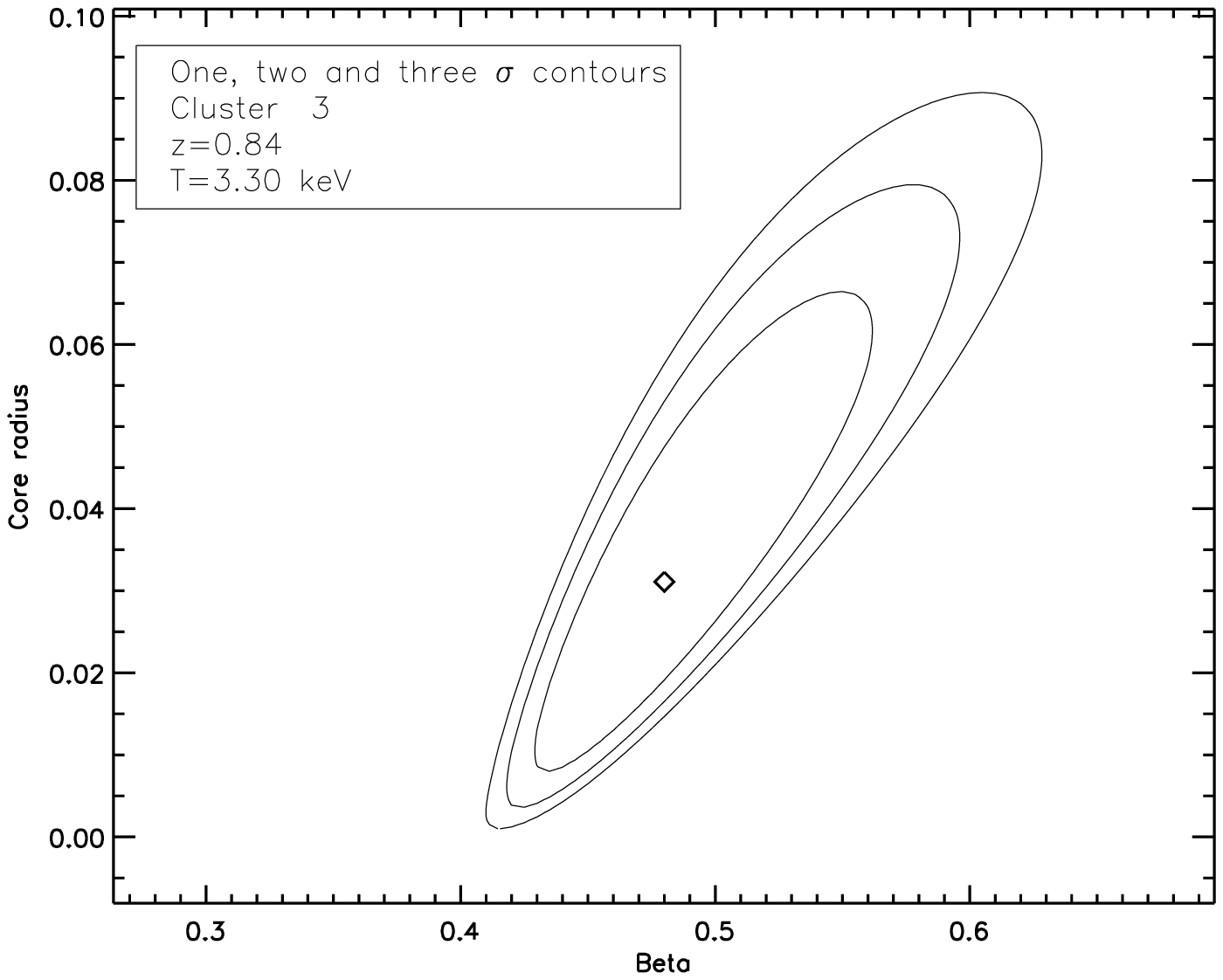,width=5.5cm,height=4.3cm}
\epsfig{file=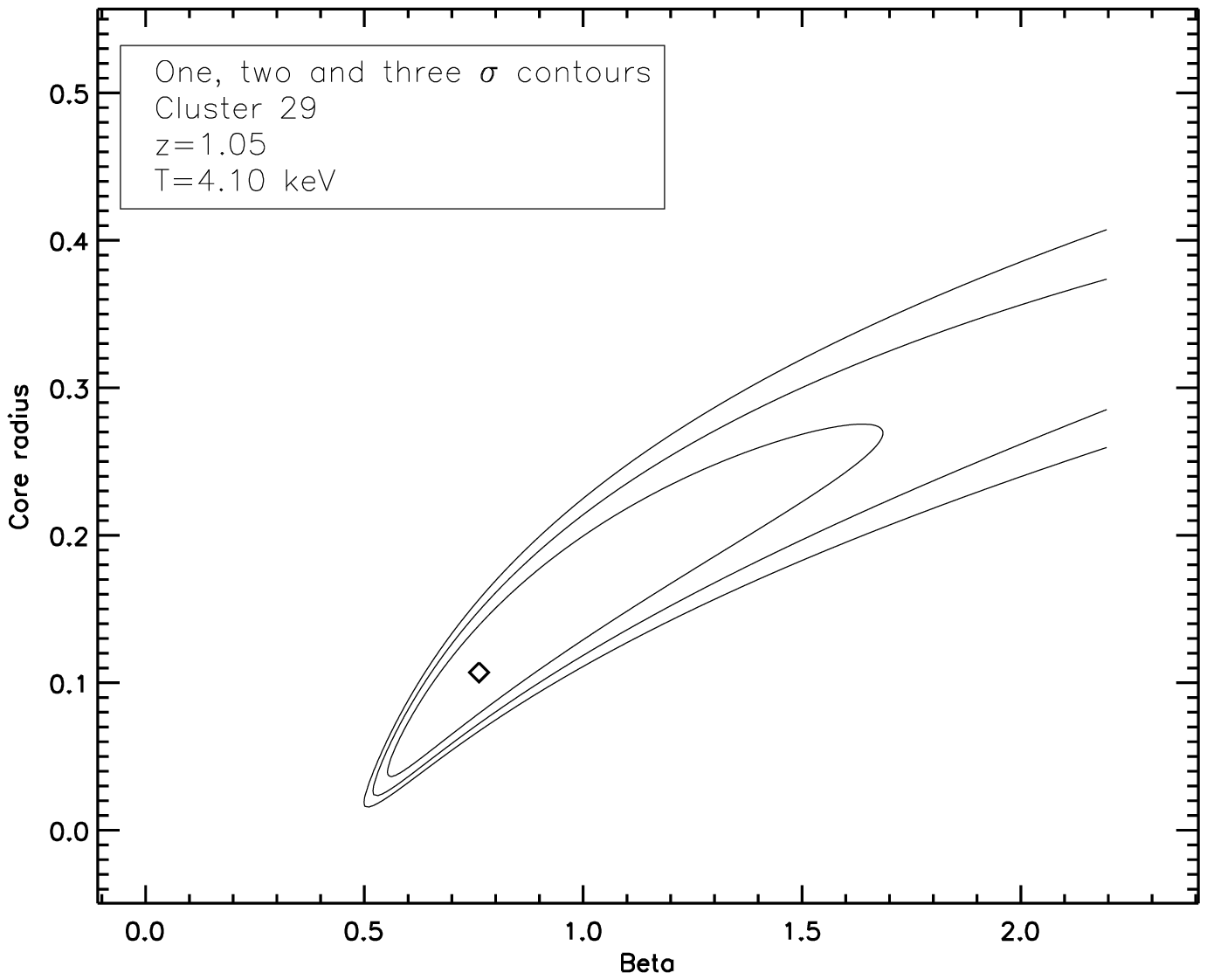,width=5.5cm,height=4.3cm}

\caption{X-ray surface brightness profiles of the individual C1 
clusters with redshift $0.43 \leq z \leq 1.05$, ordered according to 
redshift and the associated constrained $1 \sigma, 2 \sigma$ and $3 \sigma$ 
contours.  The dashed lines are the fitted \bmodel\ profiles with both \rc\ 
and $\beta$ freely fitted, while the solid lines are for the fitted profiles
with free $\beta$ and \rc\ fixed to \ffh. }
\label{ind_prof3}
\end{figure*}

The surface brightness profiles
with fits and associated $1\sigma$, $2\sigma$ and $3\sigma$ error
contours for individual clusters with redshift ranges 0.05-0.17,
0.26-0.33 and 0.43-1.05 are shown in Fig. \ref{ind_prof1},
\ref{ind_prof2} and \ref{ind_prof3} respectively, and numerical
results are given in Table~\ref{C1prop}.
The profile of cluster 50 is unusual; 
it has a remarkably large \rc\ value, of 1.12$\times$\rfh\ (see Fig.
\ref{ind_prof1}), and an elongated spatial extension indicating a
cluster in a state of merging, with a highly unrelaxed core. 

Amongst the 26 clusters with constrained fixed-\rc\ fits,
21 (81\%) possess CCs according to the criterion
outlined above (i.e. \fc\ $>1$), whilst the remaining five (19\%) are
non-cool core (NCC) (\fc\ $<1$) systems, as shown in Table \ref{C1prop}. 
However, for some of these, the classification must be regarded as uncertain,
since the error bar on \fc\ crosses unity. This is the case for
7 (from the 21) CC systems, and 2 (of the five) NCC
clusters. Hence, at the 1$\sigma$ level, 54\% (14 out of 26) of our
clusters show a central excess in surface brightness which may indicate
the presence of a cool core.

The median value of $\beta$  for free-\rc\ fits to the 26 clusters
is 0.61. A similar median value ($\beta$=0.63) is obtained from the
fixed--\rc\  fits. As for the \rc, its median value
is 0.08$\times$\rfh. As expected, this is rather smaller than
the canonical value of 0.105$\times$\rfh\ for the group scale component of
the emission, due to the influence on the fits of cuspy cores 
in many systems.


\subsection{X-ray surface brightness profiles of redshift-stacked clusters}

\begin{figure}
\center

\epsfig{file=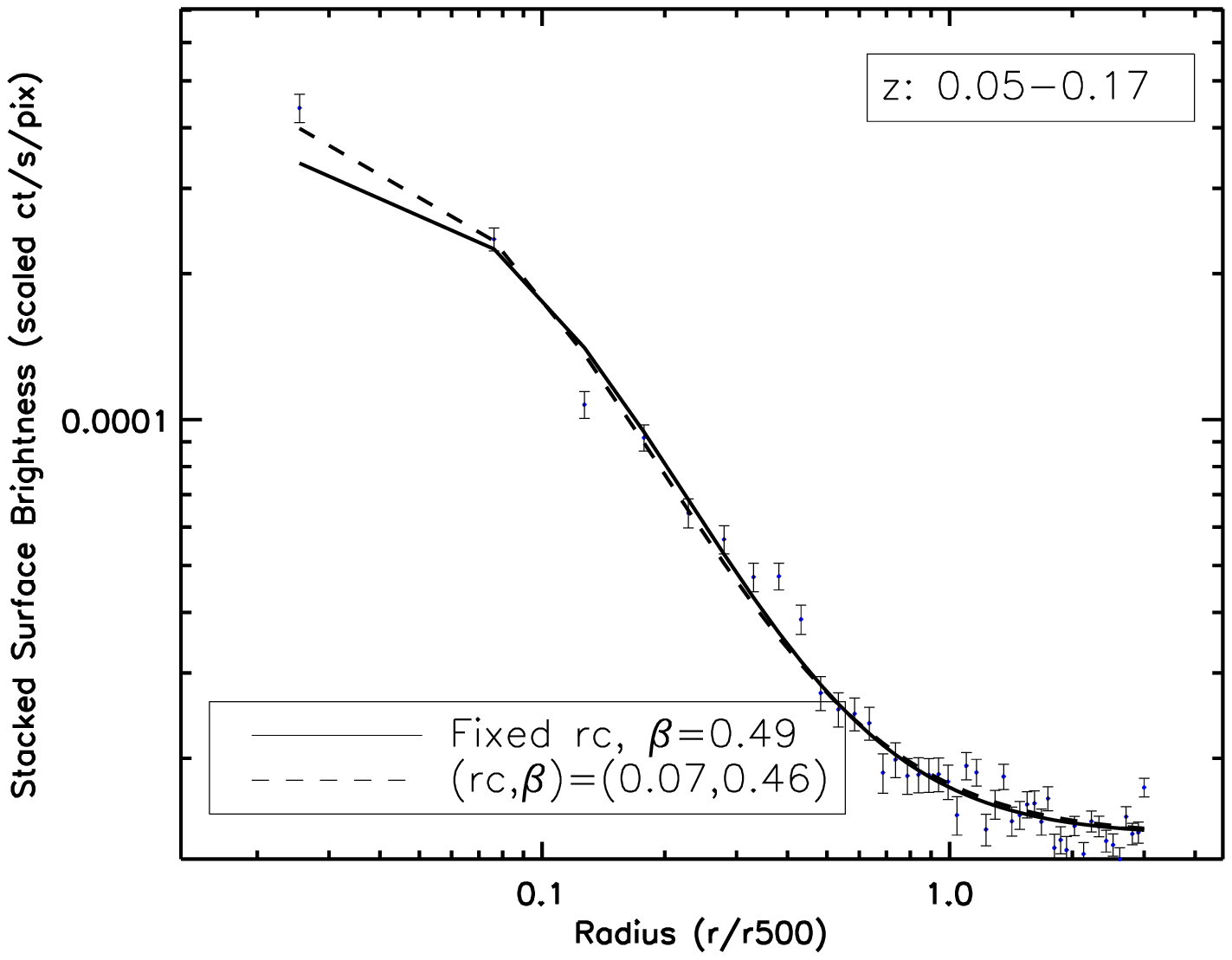,width=8.5cm,height=5.4cm}
\epsfig{file=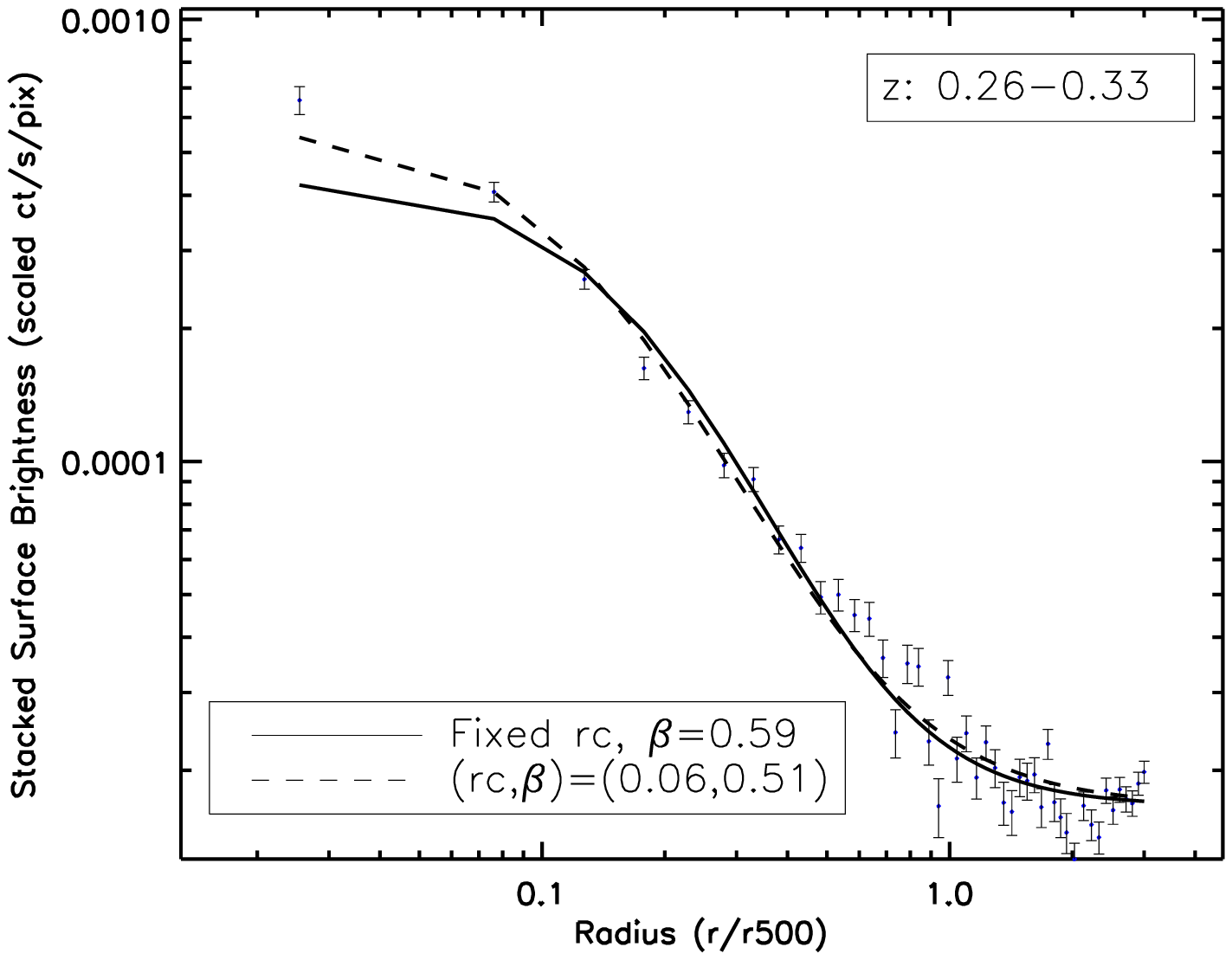,width=8.5cm,height=5.4cm}
\epsfig{file=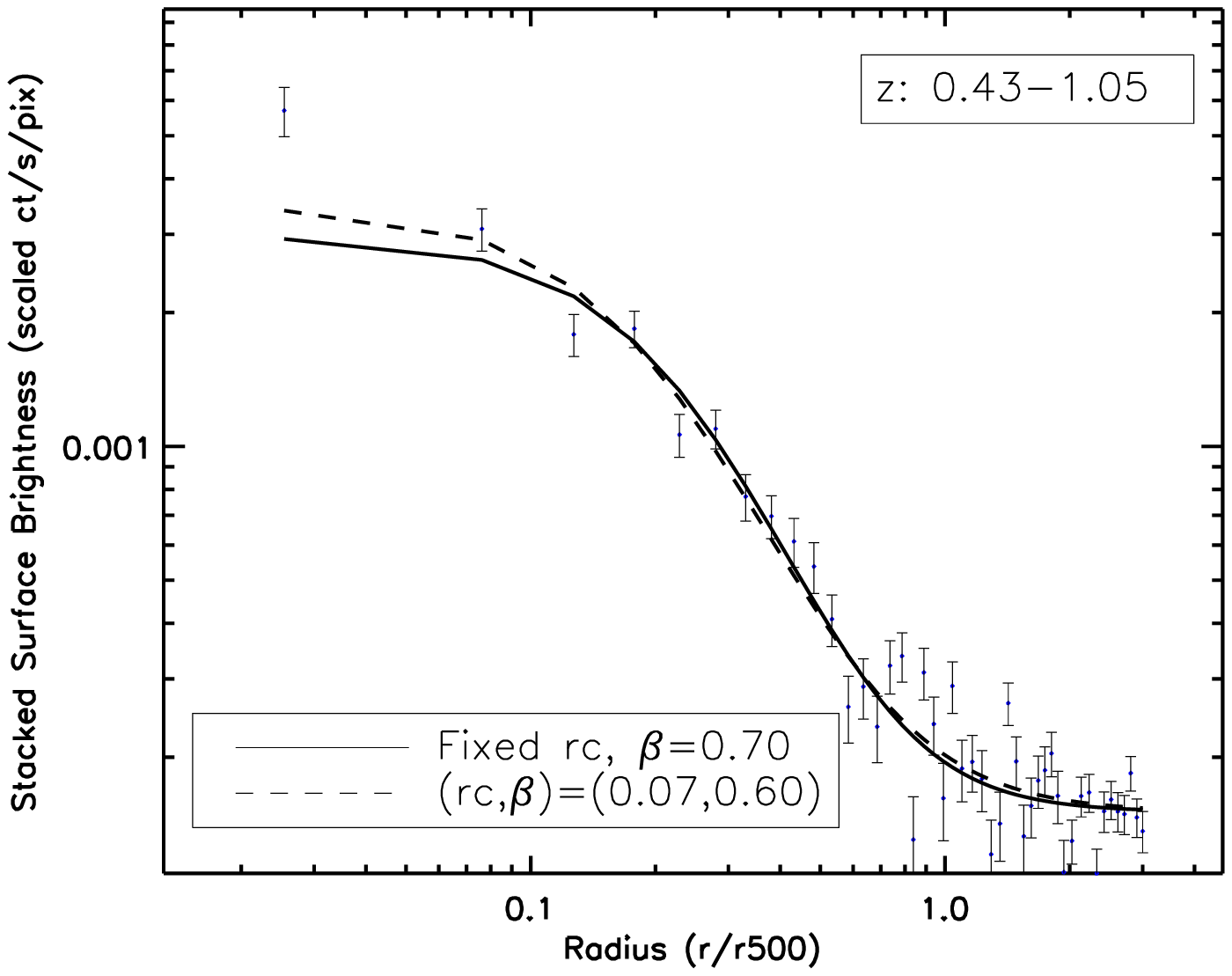,width=8.5cm,height=5.4cm}
\epsfig{file=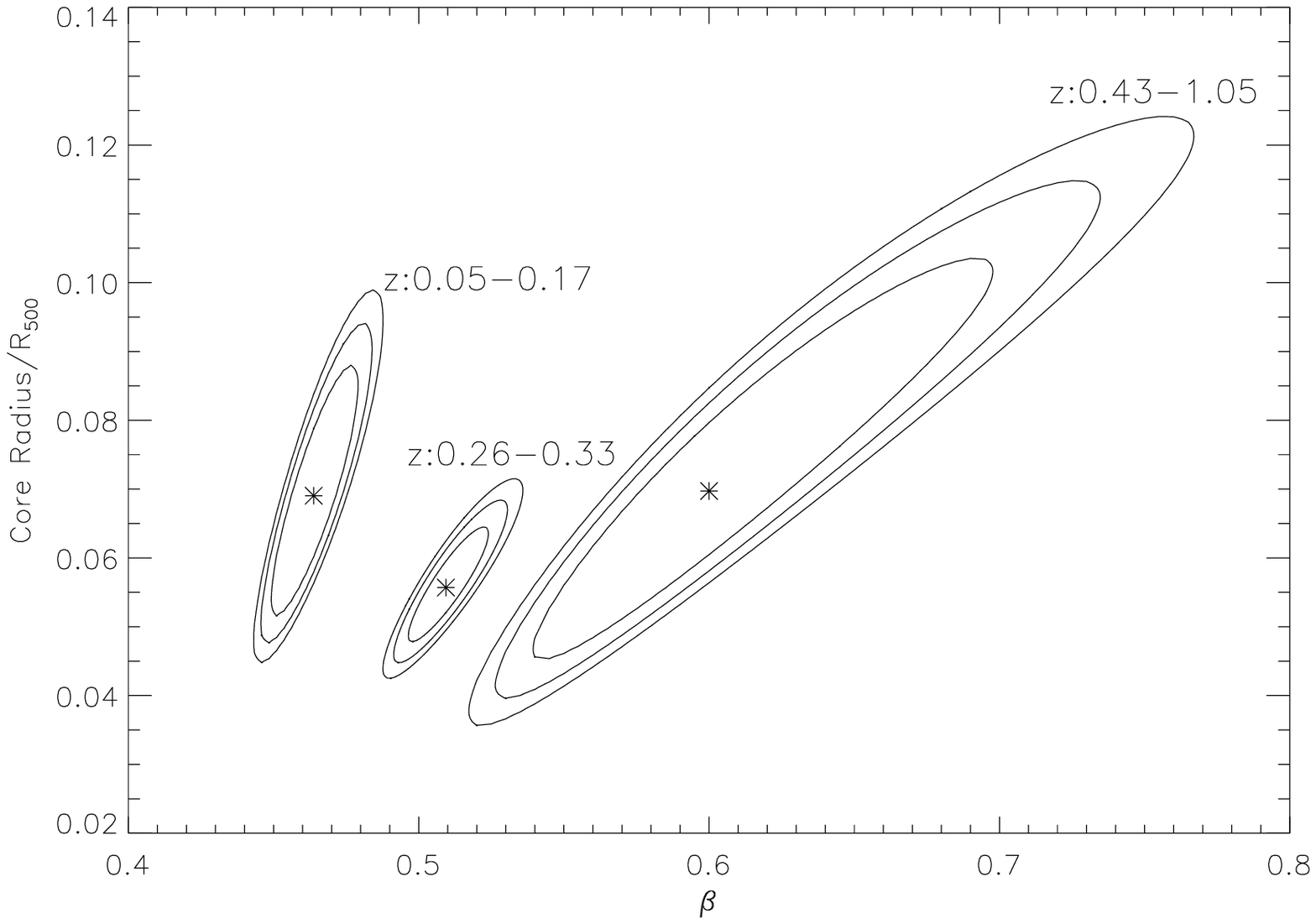,width=8.5cm,height=5.4cm}

\caption{X-ray surface brightness profiles of the redshift-stacked C1 clusters with the associated $1 \sigma, 2 \sigma$ and $3 \sigma$ contours of the free \rc\ fit.  The dashed lines are the fitted \bmodel\ profiles with both \rc\ and $\beta$ freely fitted, while the solid lines are for the fitted profiles with free $\beta$ and \rc\ fixed to \ffh. }
\label{zstack_prof}
\end{figure}

The C1 clusters span a redshift range of 0.05 to 1.05. To probe how
the X-ray surface brightness profiles evolve with redshift, we
divided the C1 sample into three redshift ranges: 0.05-0.17
(low--$z$), 0.26-0.33 (intermediate--$z$) and 0.43-1.05 (high--$z$).
The low--$z$ set consists of 6 clusters with an 
average redshift of 0.11 and a temperature range from 0.63 to 3.50~keV 
(average 1.33 keV). Only one of these (cluster 50) has $kT>2$~keV.
Twelve clusters fall in the
intermediate--$z$ stacked set, with average redshift and
temperature of 0.30 and 1.69~keV respectively. The high--$z$ set
contains 9 clusters spanning a redshift range 0.43 to 1.05 (average 0.72)
and having temperatures from 2.20 to 4.80~keV (average 3.51 keV). The profiles
and the error contours of the three stacked sets are presented in
Fig. \ref{zstack_prof} and the fitted parameters for the free and
fixed \bmodel\ fits with $1\sigma$ errors are shown in the first three
rows of Table \ref{results_stack}.

The central excess factor, \fc\ is seen to
increase with redshift; for the low--$z$ stack it is 1.30, increasing to
1.56 and 1.95 for the intermediate and high $z$ systems
respectively. Table \ref{results_stack} also shows that $\beta$ (for both
free and fixed \rc\ fits) increases with redshift, whilst the 
\rc\ values are essentially constant.

\begin{table*}
\begin{center}
\begin{tabular}{clcccccccccccccc}
\hline

Range	&	Stacked Clusters	&	Average z	&	Average $kT$	&	$\beta$	&	\rc$/R_{500}$	&	$\beta$	&	Central Excess	\\
	&	XLSSC	&		&	($keV$)	&	Fitted \rc	&		&	Fixed \rc	&	Factor ($f_c$)	\\
\hline															
\multicolumn{7}{|c|}{Redshift-stacked clusters (all clusters)} \\															
\hline															
z:0.05-0.17	&	11,52,21,41,50,35	&	0.11	&	1.33	&	$0.46_{-0.01}^{+0.01}$	&	$0.07_{-0.01}^{+0.01}$	&	$0.49_{-0.006}^{+0.003}$	&	$1.30  \pm 0.09$	\\
z:0.26-0.33	&	25,44,51,22,27,8,28,13,18,40,10,23	&	0.30	&	1.69	&	$0.51_{-0.01}^{+0.01}$	&	$0.06_{-0.01}^{+0.01}$	&	$0.590_{-0.011}^{+0.009}$	&	$1.56  \pm 0.11$	\\
z:0.43-1.05	&	6,36,49,1,2,47,3,5,29	&	0.72	&	3.51	&	$0.60_{-0.04}^{+0.06}$	&	$0.07_{-0.02}^{+0.02}$	&	$0.70_{-0.03}^{+0.02}$	&	$1.95  \pm 0.25$	\\
\hline															
\multicolumn{7}{|c|}{Temperature-stacked clusters (all clusters)} \\															
\hline															
kT:0.63-1.34	&	52,11,21,13,35,51,8,28,44,41	&	0.20	&	1.06	&	$0.49_{-0.02}^{+0.03}$	&	$0.07_{-0.02}^{+0.02}$	&	$0.52_{-0.01}^{+0.02}$	&	$1.48  \pm 0.14$	\\
kT:1.60-2.80	&	40,22,23,18,25,49,10,2,27	&	0.38	&	2.13	&	$0.65_{-0.05}^{+0.08}$	&	$0.07_{-0.02}^{+0.02}$	&	$0.79_{-0.04}^{+0.03}$	&	$2.17  \pm 0.28$	\\
kT:3.20-4.80	&	1,3,50,36,5,47,29,6	&	0.68	&	3.76	&	$0.60_{-0.05}^{+0.07}$	&	$0.07_{-0.02}^{+0.02}$	&	$0.69_{-0.03}^{+0.02}$	&	$1.81  \pm 0.27$	\\
\hline															
\multicolumn{7}{|c|}{Redshift-stacking for clusters with narrow temperature range: 1.20-1.34 keV} \\															
\hline															
z:0.14-0.26	&	41,35,44	&	0.19	&	1.28	&	$0.49_{-0.03}^{+0.03}$	&	$0.10_{-0.03}^{+0.03}$	&	$0.490_{-0.008}^{+0.012}$	&	$1.49  \pm 0.18$	\\
z:0.28-0.30	&	51,28,8	&	0.29	&	1.27	&	$0.56_{-0.07}^{+0.11}$	&	$0.14_{-0.05}^{+0.07}$	&	$0.52_{-0.03}^{+0.02}$	&	$1.98  \pm 0.45$	\\
\hline															
\multicolumn{7}{|c|}{Temperature-stacking  for clusters with narrow redshift range: 0.23-0.33} \\															
\hline															
kT:1.00-1.60	&	13,51,44,28,8,40	&	0.30	&	1.28	&	$0.54_{-0.04}^{+0.07}$	&	$0.11_{-0.04}^{+0.05}$	&	$0.53_{-0.02}^{+0.02}$	&	$1.83  \pm 0.32$	\\
kT:1.70-2.80	&	22,23,25,18,10,27	&	0.30	&	2.10	&	$0.55_{-0.01}^{+0.02}$	&	$0.07_{-0.01}^{+0.01}$	&	$0.620_{-0.006}^{+0.013}$	&	$1.39  \pm 0.10$	\\

\hline
\end{tabular}
\caption{Results of the \bmodel\ fit for the redshift-stacked and temperature-stacked clusters for all clusters and for redshift and temperature subsets. The $\beta$ and \rc\ values in the fifth and sixth columns are for the \bmodel\ with both \rc\ and $\beta$ freely-fitted. The $\beta$ values in the seventh column are for the \bmodel\ with fixed \rc. All errors are 1$\sigma$ errors.} 
\label{results_stack}
\end{center}
\end{table*}

\subsection{X-ray surface brightness profiles of temperature-stacked clusters}

\begin{figure}
\center
\epsfig{file=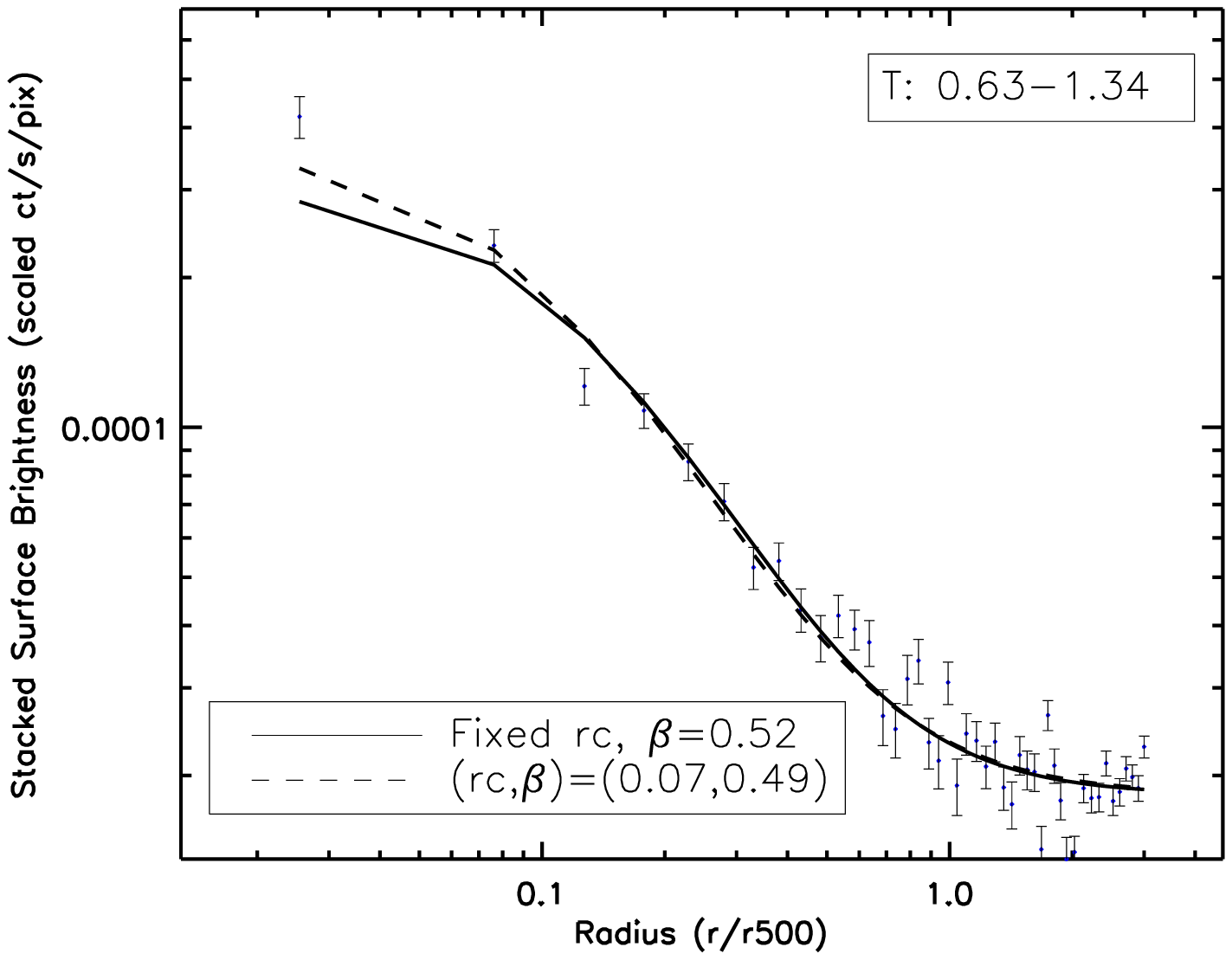,width=8.5cm,height=5.4cm}
\epsfig{file=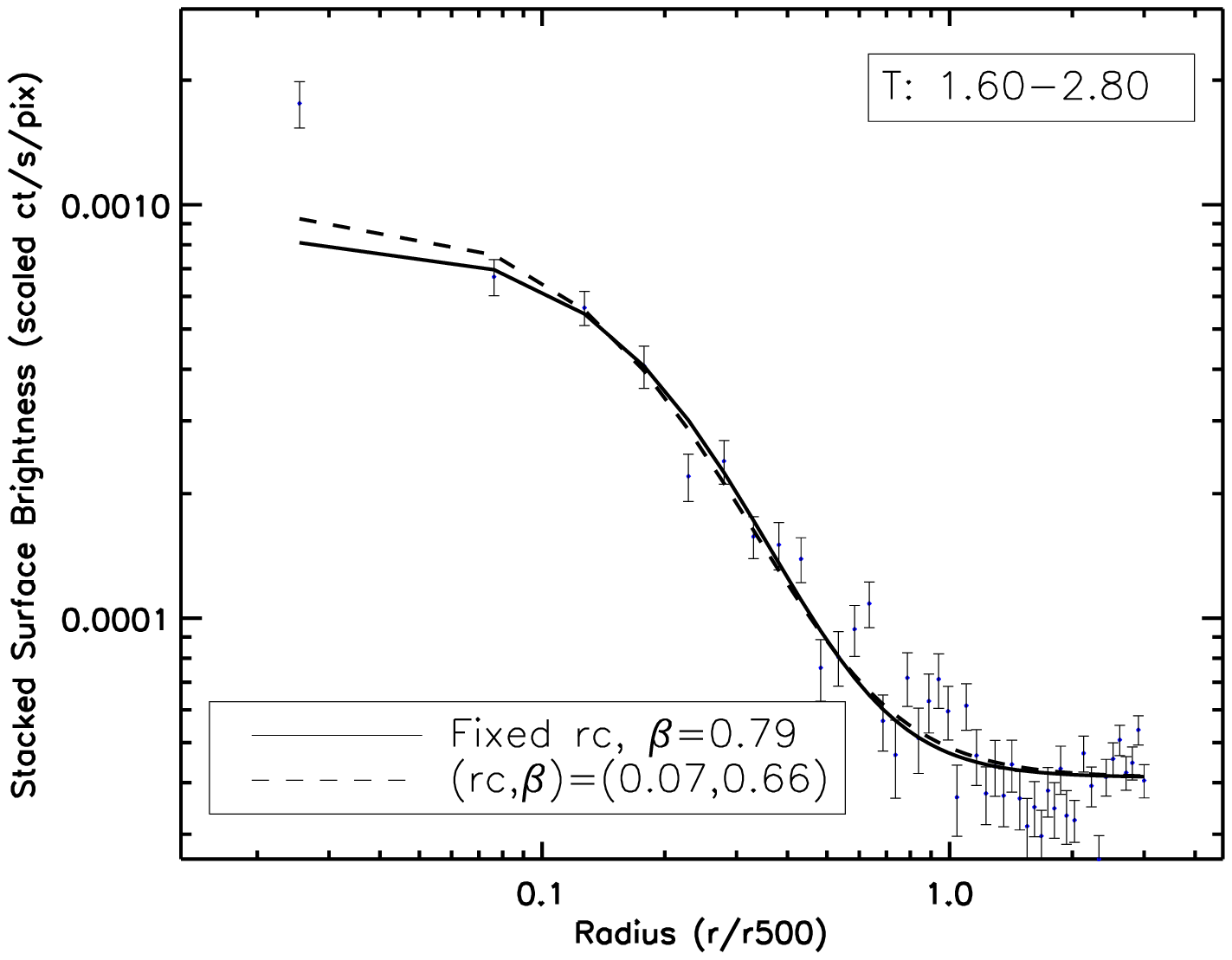,width=8.5cm,height=5.4cm}
\epsfig{file=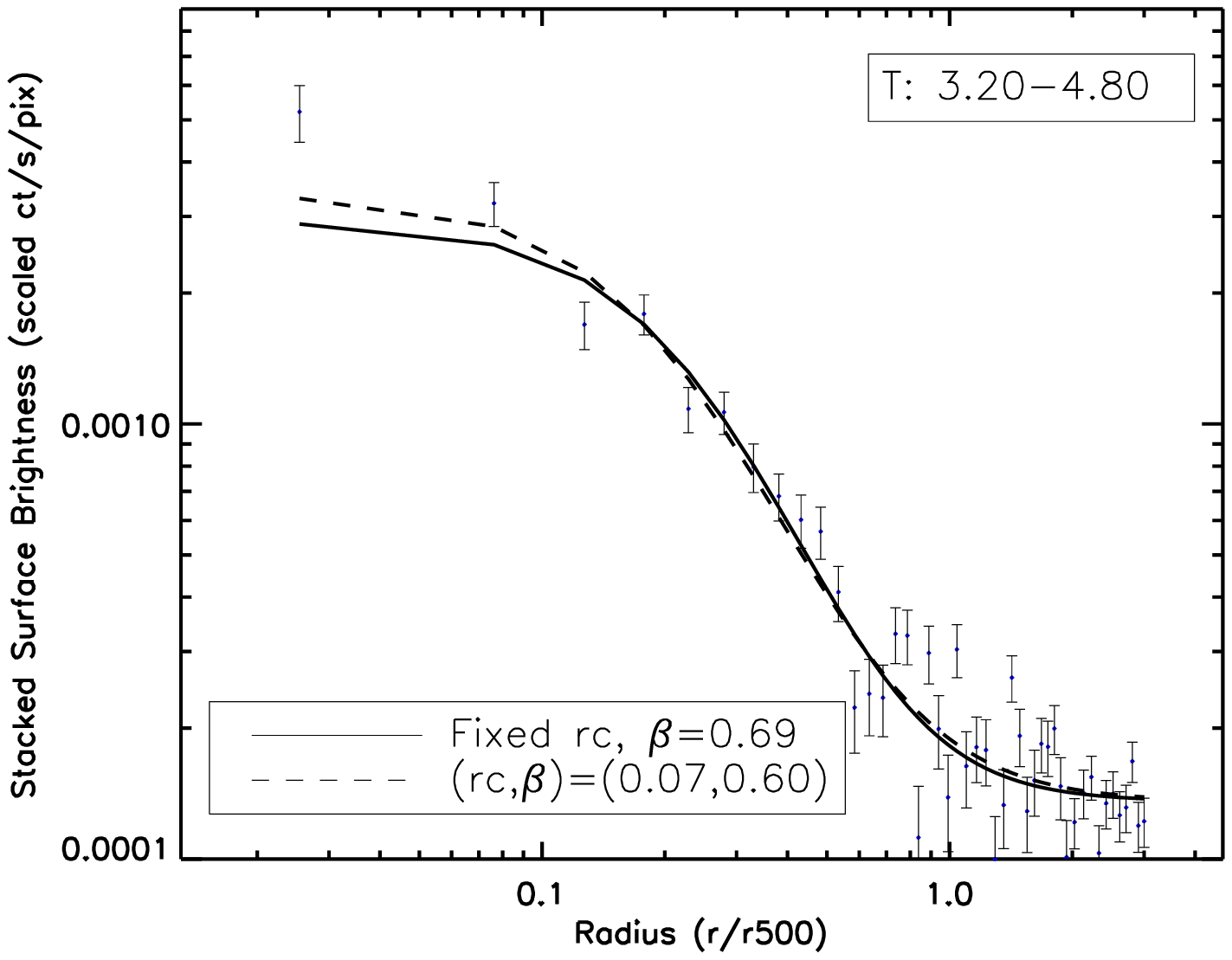,width=8.5cm,height=5.4cm}
\epsfig{file=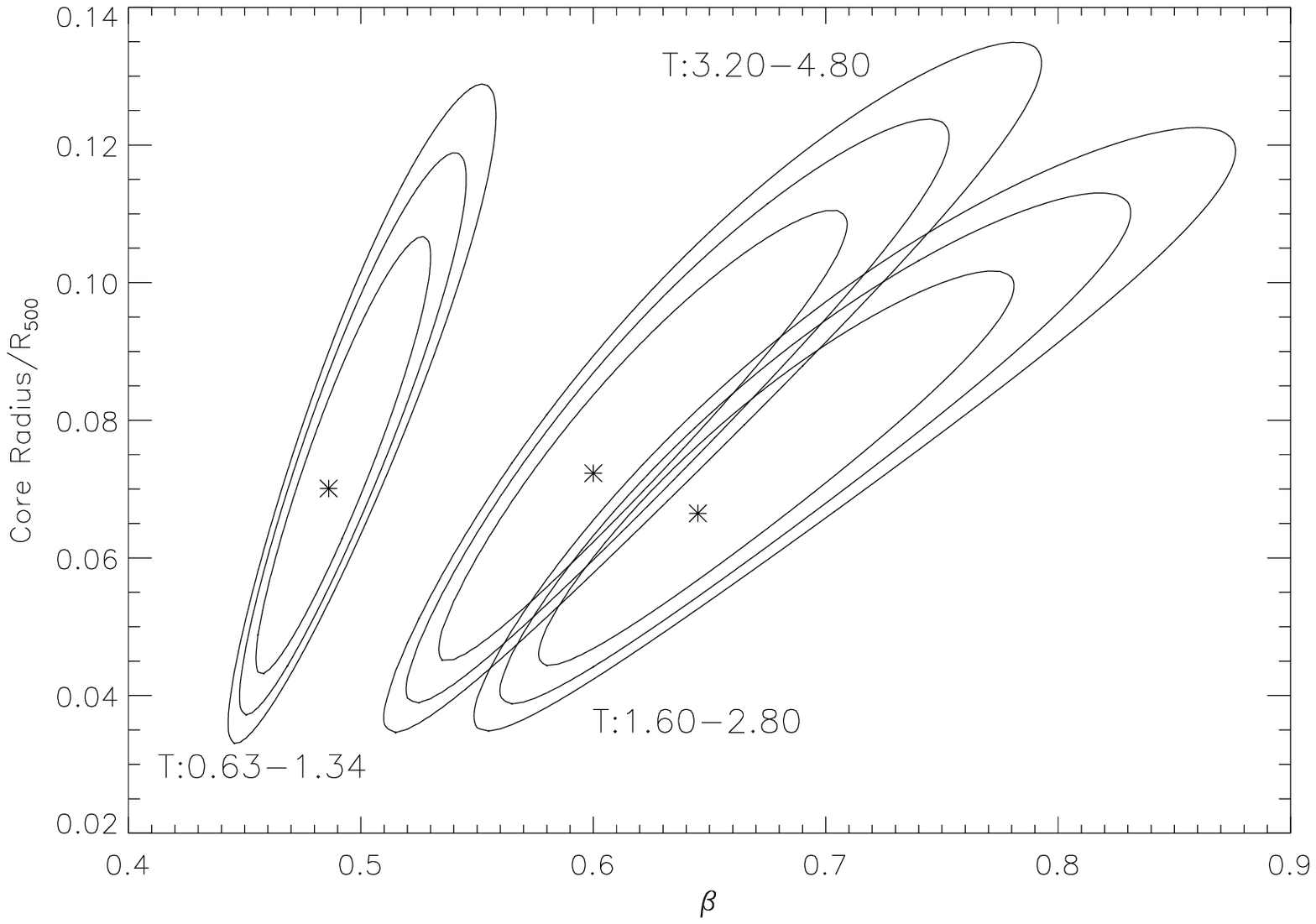,width=8.5cm,height=5.4cm}

\caption{X-ray surface brightness profiles of the temperature-stacked C1 clusters with the associated $1 \sigma, 2 \sigma$ and $3 \sigma$ contours of the free \rc\ fit.  The dashed lines are the fitted \bmodel\ profiles with both \rc\ and $\beta$ freely fitted, while the solid lines are for the fitted profiles with free $\beta$ and \rc\ fixed to \ffh. }
\label{tstack_prof}
\end{figure}

The full temperature range of our C1 sample (0.63 to 4.80 keV)
was divided into three subsets. The coolest set (0.63 keV
$\leq kT \leq$ 1.34 keV) contains ten clusters with average $kT$=1.06
keV and average $z$=0.20. There are 
nine clusters in the second set (1.60
keV $\leq kT \leq$ 2.80 keV) with averages $kT$=2.13 keV and
$z$=0.38, and the hottest set (3.20 keV $\leq kT \leq$ 4.80 keV) contains
eight clusters with average temperature and redshift of 3.76~keV and
0.68 respectively. The stacked profiles and the associated
$1\sigma$,$2\sigma$ and $3\sigma$ contours are shown in Fig.
\ref{tstack_prof} and the fitted parameter values in Table \ref{results_stack}.

All three temperature-stacked sets show evidence for CCs, with \fc\
$>1$. However, \fc\ does not show a monotonic trend with temperature as
was the case for the redshift-stacked clusters. The
intermediate-temperature set shows the strongest central excess,
with \fc=2.17$\pm0.28$.
Similarly, the $\beta$ values, for both fixed and free \rc\ fits,
do not show a monotonic trend across the full $T$ range of our sample,
although it is clear that the hotter systems ($kT>1.5$~keV) have 
$\beta$ values significantly higher than the groups in our coolest bin.
The core radius, \rc\, appears remarkably stable in these stacked clusters,
fitting at a value 0.07$\times$\rfh.
This is also essentially the case
for the redshift-stacked clusters, in which the high and low redshift
stackes fitted at \rc= 0.07$\times$\rfh\, whilst the 
intermediate-redshift stack gives \rc=0.06$\times$\rfh, which is the same
within errors.

\section{Discussion}

\subsection{X-ray surface brightness profiles of $z\sim0.3$ clusters}

\cite{Pacaud07} show (see their Fig. 3) that the redshift
distribution of the C1 clusters, which spans the redshift range 0.05
to 1.05, has a pronounced peak around $z\sim0.3$.  More than 40\% (12
systems) of our clusters are concentrated in the relatively small
redshift range $0.26\leq z \leq 0.33$. The average temperature of
these 12 clusters is 1.69 keV. 
Their average $M_{500}$ is 3.96 $\times 10 ^{13}
M_{\odot}$. This puts them in the realm of groups or poor clusters.
To our knowledge, this is the best sample of X-ray selected groups at
$z\sim0.3$ studied to date, and hence
our individual and stacked X-ray
profiles of these clusters provide the best available X-ray
profile of low-mass clusters at intermediate redshift, and should be
useful for future comparative studies. The individual X-ray
profiles with the the \rc--$\beta$ contours for these cluster are
shown in Fig. \ref{ind_prof2} and the stacked profile is the second
panel in Fig. \ref{zstack_prof}.

The free-\rc\ fit to the stacked profile for these clusters 
gives $\beta=0.51$ (see
Table \ref{results_stack}, second row) which is in agreement with
studies of low redshift groups (see for example, \citealt{Helsdon00} and
\citealt{Mulchaey03}). \cite{Helsdon00} attributed the fact that the 
slopes of group surface brightness profiles are flatter than the canonical
slope ($\beta$=0.67) for clusters, as a result of the effects of feedback
from galactic winds on
the intergalactic medium. The
stacked data show a central excess, with $f_c=1.56\pm0.11$,
indicating that these systems typically possess CCs. 
The individual profiles
of $z\sim0.3$ clusters in Fig. \ref{ind_prof2}, also support this
results; the central excess factor, $f_c$ is greater than
unity for 7 of the 12 clusters (25,51,22,8,28,40 and 10) and an
additional 4 clusters (44,13,18 and 23) have best fit $f_c > 1$, but with
error bars crossing unity. So, we conclude that 58-92\%
of our systems at $z\sim 0.3$
have CCs. The fitted \rc\ for the stacked profiles
is $0.06\times R_{500}$.

\subsection{Trends of \fc\ and $\beta$ with redshift and temperature}

The main results of our analysis are the presence of trends in the value of 
$\beta$, and in the incidence of cuspy cores, with redshift and temperature.
Our C1 clusters, as for any deep cluster survey, suffer from
{\it Malmquist} selection effects, which result in increasing mean cluster
luminosity with redshift, due to the fact that higher redshift clusters are
more difficult to detect than nearby ones -- see Fig. 3 in
\cite{Pacaud07}. Given the well-known correlation between X-ray
luminosity and temperature, there is a corresponding tendency for more
distant clusters in our sample to be hotter. This correletaion between
$z$ and $T$ within our sample, makes it difficult to establish whether
our observed trends in $\beta$ and \fc\ are evolutionary effects, or
whether they represent changes in cluster properties with system mass
(and hence temperature).

We attempt to address this issue in two ways:
firstly, we can examine whether the trends we see (in both individual and
stacked clusters) are stronger with respect to $T$ or $z$. Secondly, we
use the group of clusters at $z\sim 0.3$; subdividing these by temperature
allows us to check for trends with $T$ at essentially a single redshift.
Similarly, we also extract a subset of our clusters which cover a rather
narrow temperature range, but a larger spread in $z$.

\subsubsection{Trends of \fc}

\begin{figure}
\center

\epsfig{file=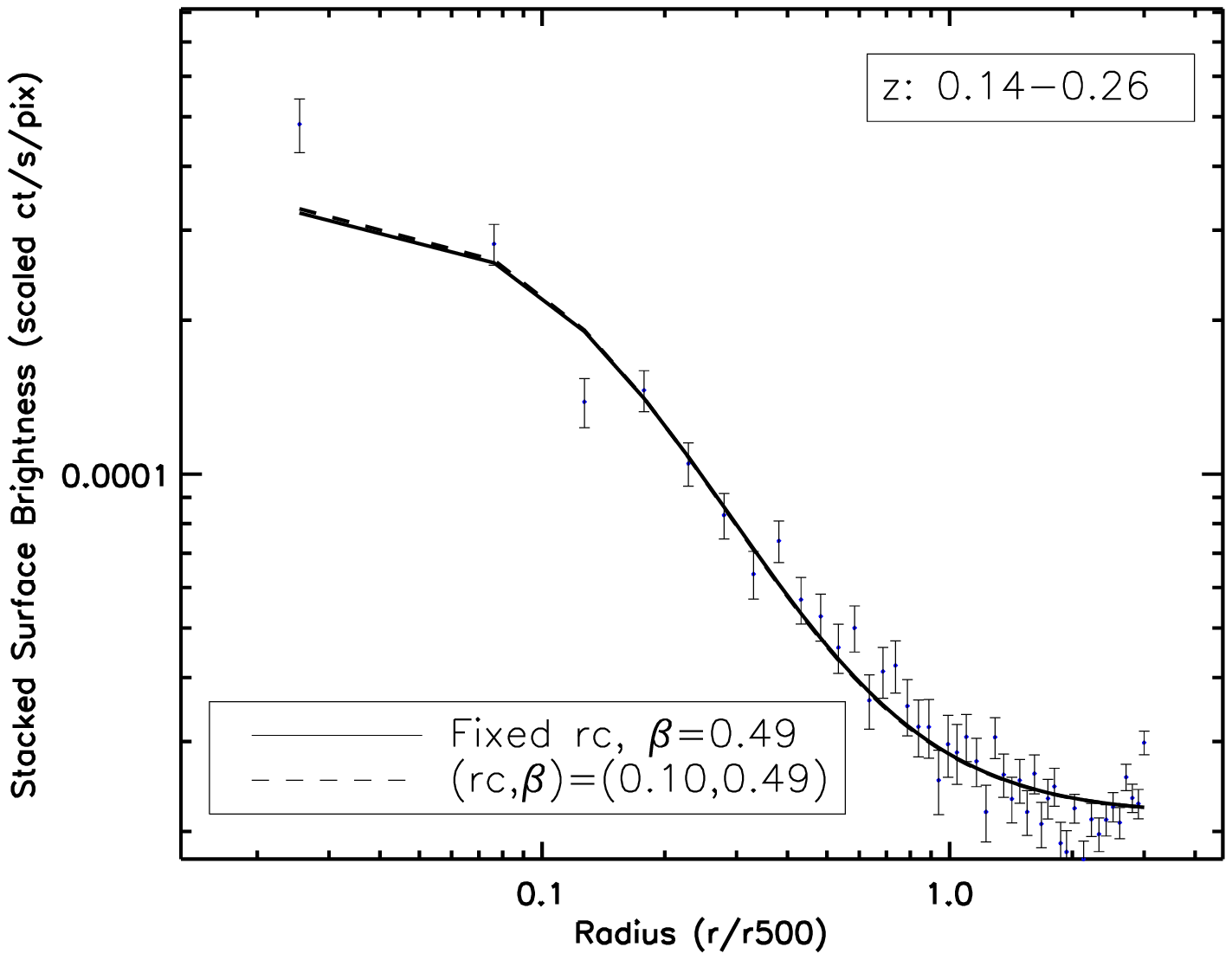,width=9cm,height=7cm}
\epsfig{file=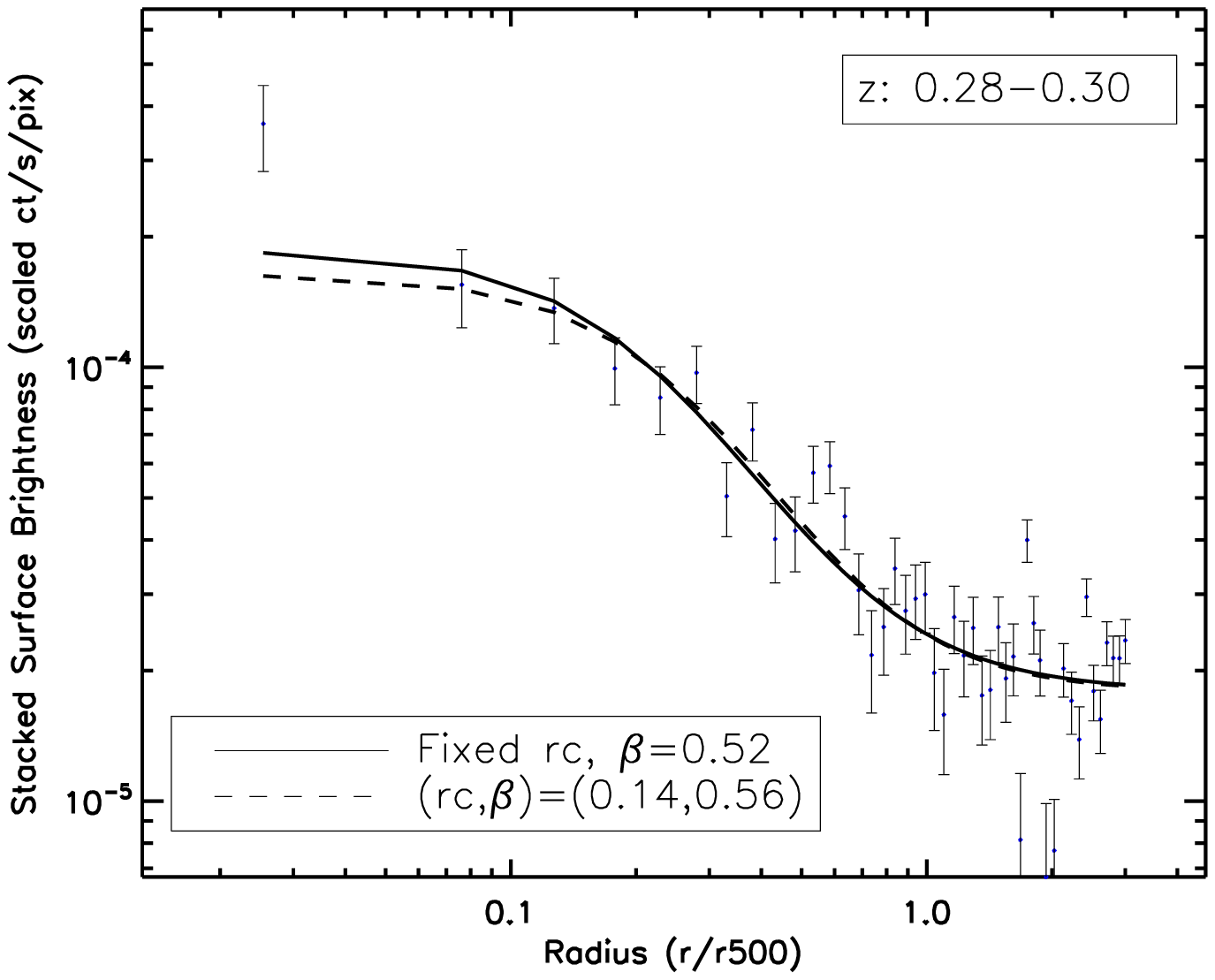,width=9cm,height=7cm}
\epsfig{file=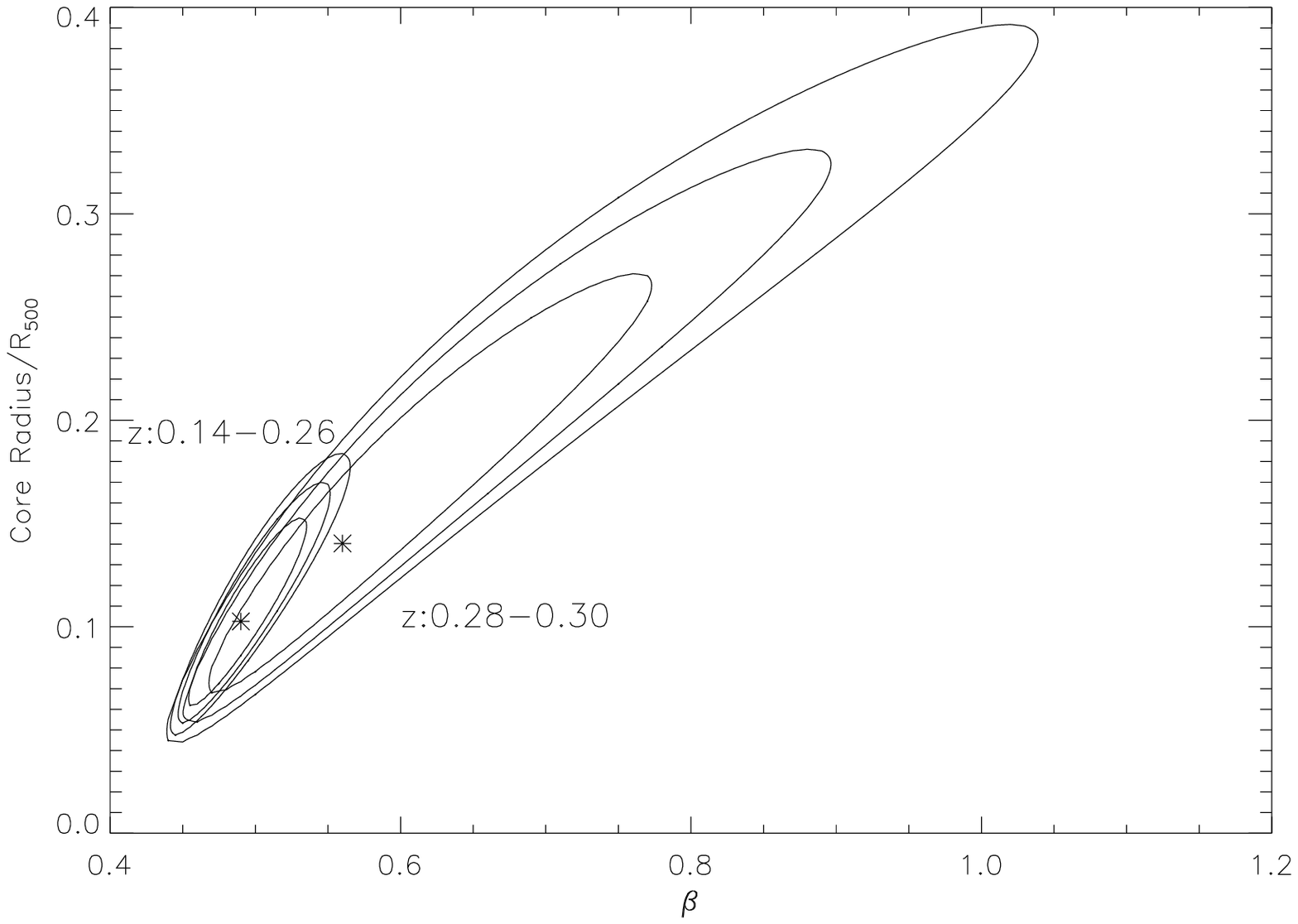,width=9cm,height=7cm}

\caption{X-ray surface brightness profiles of stacked C1 clusters with 
narrow temperature range (1.20-1.34 keV), grouped into two 
redshift bins: $0.14\leq z \leq 0.26$ (top panel) and 
$0.28\leq z \leq 0.30$ (middle panel). The bottom panel is the 
$1 \sigma, 2 \sigma$ and $3 \sigma$ contours.  The dashed lines are 
the fitted \bmodel\ profiles with both \rc\ and $\beta$ freely fitted, 
while the solid lines are for the fitted profiles with free $\beta$ 
and \rc\ fixed to \ffh. }
\label{zsub_prof}
\end{figure}

To investigate whether the trends we see in the
redshift-stacked sets are affected by the $T$-$z$ correlation in our sample,
we select six clusters with similar temperatures
($kT$=1.20 to 1.34 keV) but a relatively wide spread in redshift ($z$=0.14 to 0.30).
These are then divided into two subsets, each consisting of three systems: the 
first has $0.14 \leq z \leq 0.26$ (average $z$=0.19) and mean temperature
$\overline{kT}$=1.28 keV, the second has $0.28 \leq z \leq 0.30$ (average $z=0.29$)
and $\overline{kT}$=1.27 keV. 
The fit results for these subsets are shown in Fig. \ref{zsub_prof} and in
Table \ref{results_stack}.

\begin{figure}
\center

\epsfig{file=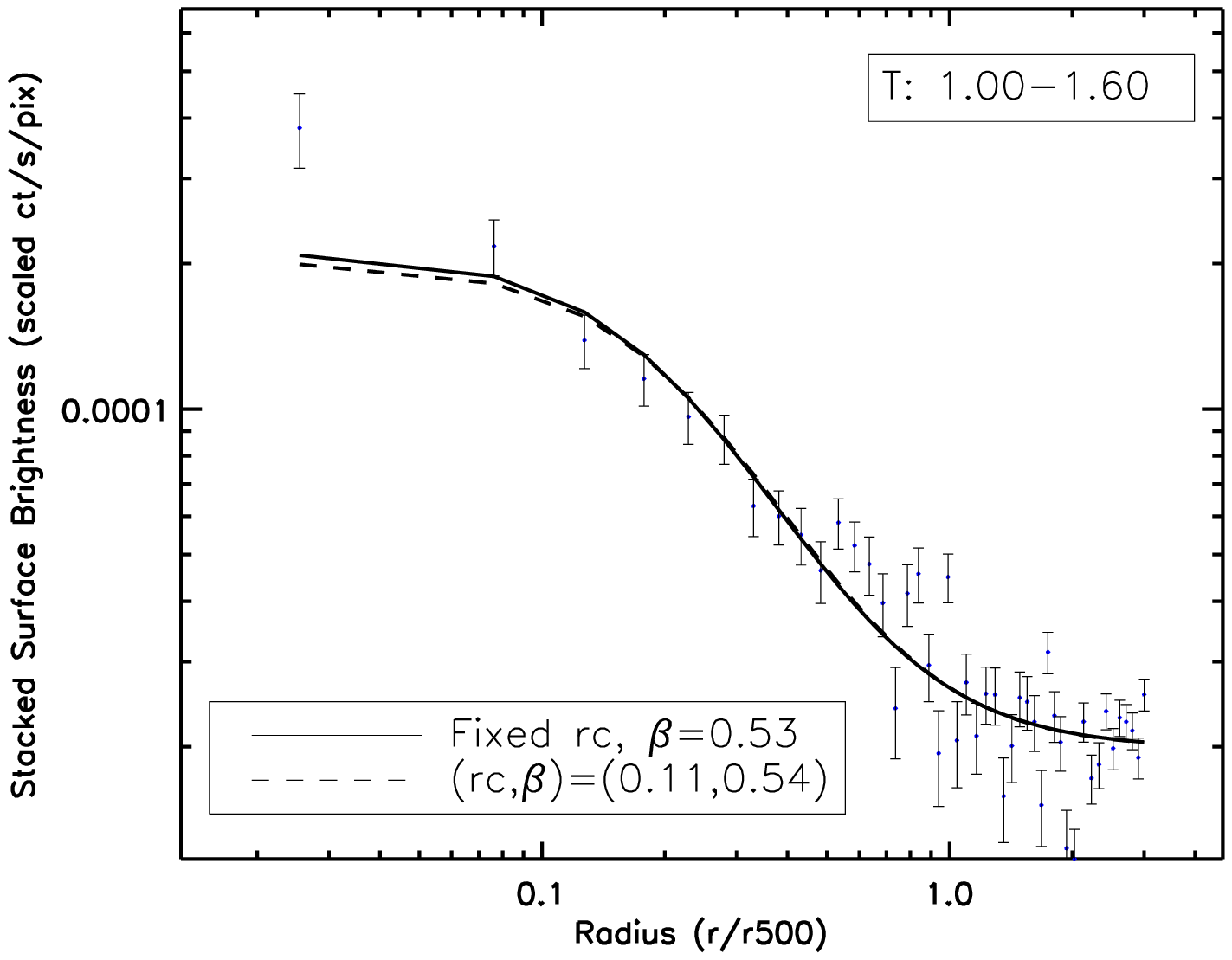,width=9cm,height=7cm}
\epsfig{file=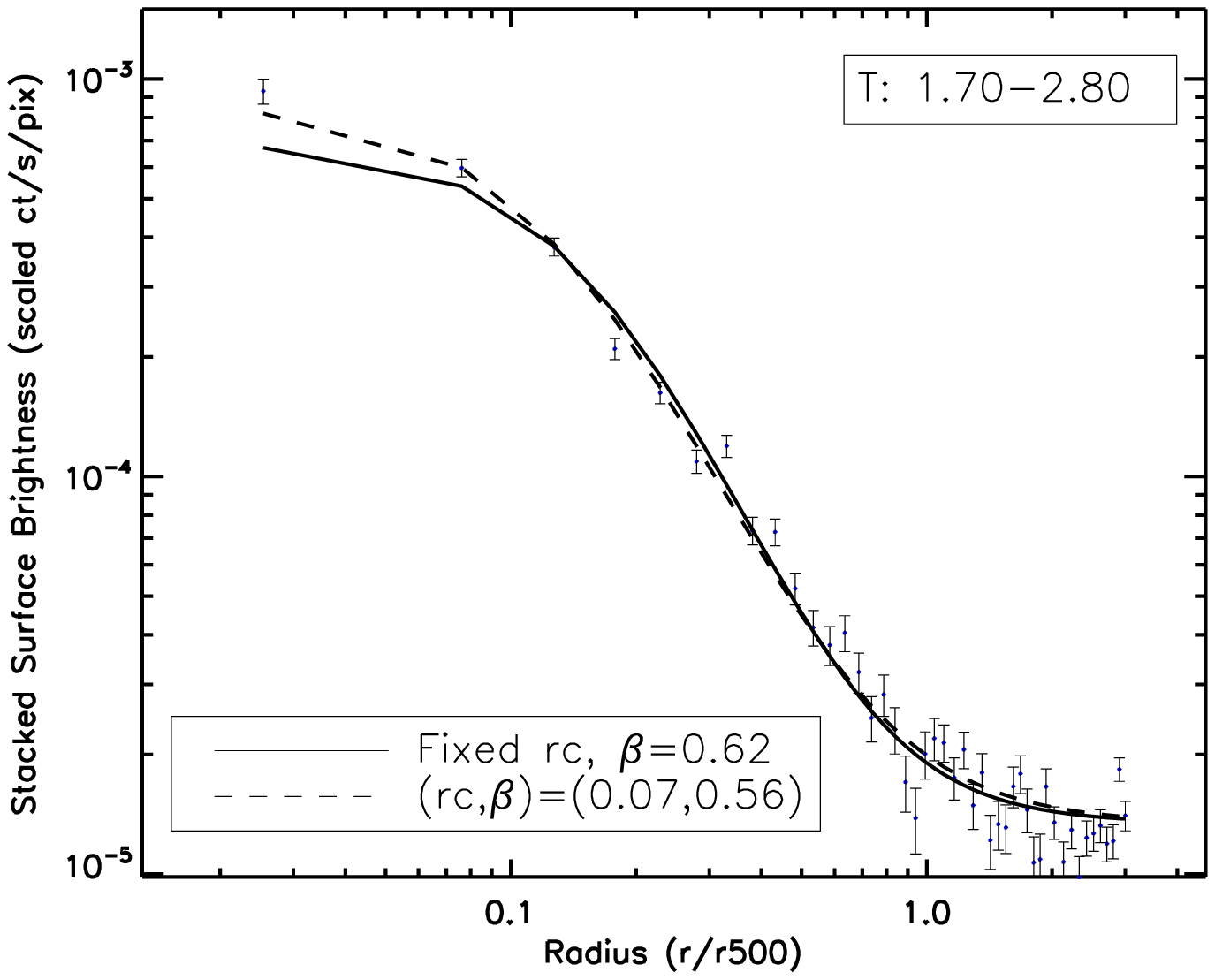,width=9cm,height=7cm}
\epsfig{file=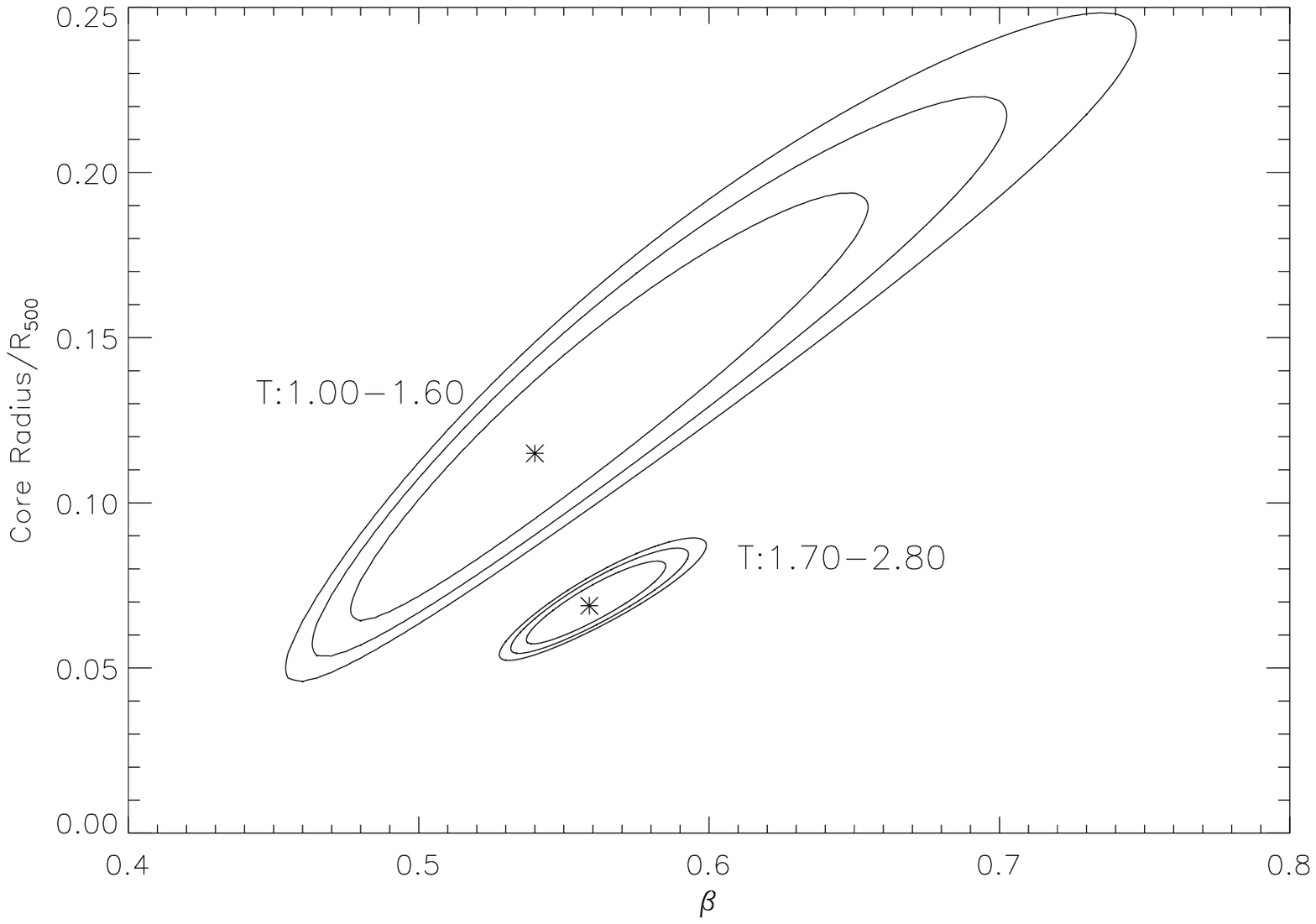,width=9cm,height=7cm}

\caption{X-ray surface brightness profiles of stacked C1 clusters with 
narrow redshift range (0.23-0.33), grouped into two temperature bins:
1.00-1.60 keV (top panel) and 1.70-2.80 keV (middle panel). The bottom
panel is the $1 \sigma, 2 \sigma$ and $3 \sigma$ contours.  The dashed
lines are the fitted \bmodel\ profiles with both \rc\ and $\beta$ freely
fitted, while the solid lines are for the fitted profiles with free $\beta$
and \rc\ fixed to \ffh. }
\label{tsub_prof}
\end{figure}

Similarly, we divided the twelve $z\sim0.3$ clusters, which span a
temperature range of 1.0-2.8 keV, into two temperature
bins: 1.0-1.6 keV and 1.7-2.8 keV with six clusters in each.
See Fig. \ref{tsub_prof} and the last two rows in
Table \ref{results_stack} for the results of stacking these subsets. 
The 1.0-1.6 keV clusters have an average
redshift $z=0.30$ and average temperature $\overline{kT}$=1.28 keV, while the 1.7-2.8 keV
clusters have the same average redshift $z=0.30$ and 
$\overline{kT}$=2.1 keV.

The results from these subsets, shown in Fig. \ref{zsub_prof} and 
\ref{tsub_prof}, reinforce the impression from Fig. \ref{zstack_prof}
and \ref{tstack_prof} that the increase in profile cuspiness
is a function of $z$ rather than $T$. In fact, the
temperature-stacked subset (Fig. \ref{tsub_prof}) actually shows a 
{\it decline} in \fc\ with temperature, whilst in the redshift-stacked
subset it increases from 1.49 for the $0.14 \leq z \leq 0.26$ clusters to 1.98 
for the $0.28\leq z \leq 0.30$ clusters (Fig. \ref{zsub_prof}).

\begin{figure}
\epsfig{file=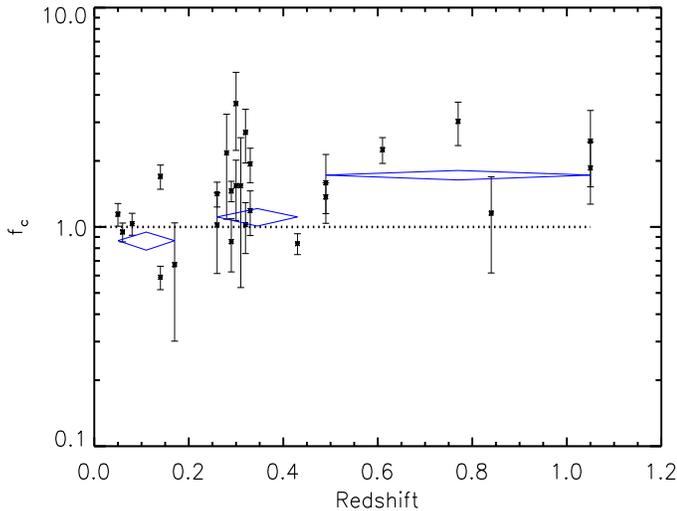,width=9cm,height=7cm}
\caption{Central excess factor, \fc\ (of the fixed \rc\ fits) plotted 
against redshift of the individual C1 clusters. \fc\ is defined as the
ratio of the observed surface brightness to the predicted (model) surface
brightness within $0.02 \times$\rfh\ (first radial bin). A value of \fc\
above 1 is an indication of a cool core cluster and vice versa. The
positions and sizes of the diamonds represent the weighted means and the
standard errors of the weighted means of the points as described in text
and appendix \ref{sewm}.}
\label{d_fc_z}
\end{figure}

We also tested the \fc--$z$ behaviour in the individual profiles. In Fig.
\ref{d_fc_z}, we plot \fc\ against redshift for the individual C1
clusters. The figure shows that the high--$z$ clusters tend to have
larger-than-unity values of \fc\ more often than clusters at lower redshift.
We tested for a correlation in this plot, using the Pearson correlation 
coefficient, which has a value 0.40 for 26 points, corresponding to
a Student t value of 2.12, which shows a poitive correlation at over 95\%
significance (2-tailed test). To visualise
the trend more clearly, we grouped adjacent data points into
three bins and computed their weighted mean and the standard error.
These binned results are shown as diamonds, though it should be noted
that resulting values are sensitive to the choice of bin boundaries.

\begin{figure}
\epsfig{file=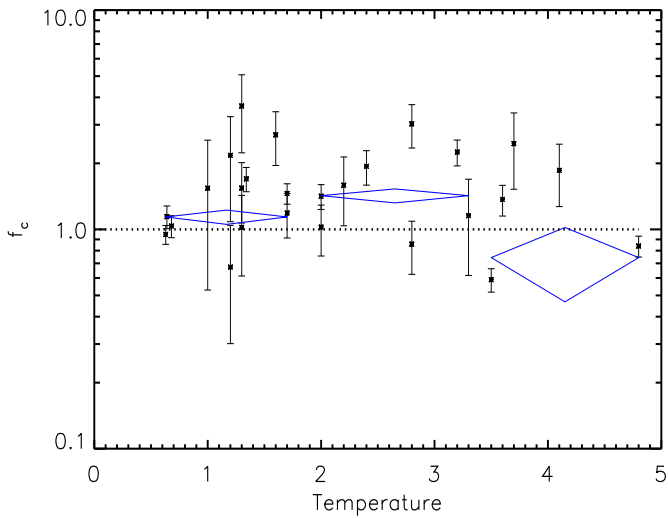,width=9cm,height=7cm}
\caption{Central excess factor, \fc\ (of the fixed \rc\ fits) plotted against temperature of the individual C1 clusters. The positions and sizes of the diamonds are calculated as described in \protect appendix \ref{sewm}.}
\label{d_fc_t}
\end{figure}

In contrast, the \fc\--temperature plot in Fig. \ref{d_fc_t},
shows no monotonic trend in \fc\ with $T$, with a Pearson
correlation coefficient of 0.01. The binned values (diamonds)
agree well the temperature-stacked results discussed earlier,
where we noticed that
intermediate-temperature clusters had the highest central excess.

The present work is the first study of the {\it evolution} of
CCs within galaxy groups, although previous work
(e.g., \citealt{Helsdon00}) has shown that CCs are common in X-ray
bright groups at low redshift.  Richer clusters have received much
more study.  CC clusters are found to be common at low and
moderate redshifts, see \cite{Bauer05}, but \cite{Vikhlinin07} found
only a very small fraction of clusters at $z>$0.5 to have cuspy X-ray
brightness profiles, which were taken as an indication of cool
cores. \cite{Santos08} found moderate CC clusters out to
$z=1.4$, but noted an absence of {\it strong} CCs at
redshifts higher than 0.7.

Our results therefore suggest that groups behave differently to clusters,
in that cuspy cores are actually {\it more} prominent at higher $z$
in these poorer systems. How can we understand this difference?
One possibility is that the central excess seen in groups
at moderate-high $z$ is not due to CCs at all, but to
the presence of central AGN. We can immediately rule out the possibility that
the effect is due to just a few groups with bright AGN contaminating our
stacked profiles by noting (cf. Fig.~\ref{d_fc_z}) that a central
excess is seen in the {\it majority} of systems at $z>$0.3. Hence, any effect
from central AGN would have to be widespread and moderate. 

\begin{figure}
\epsfig{file=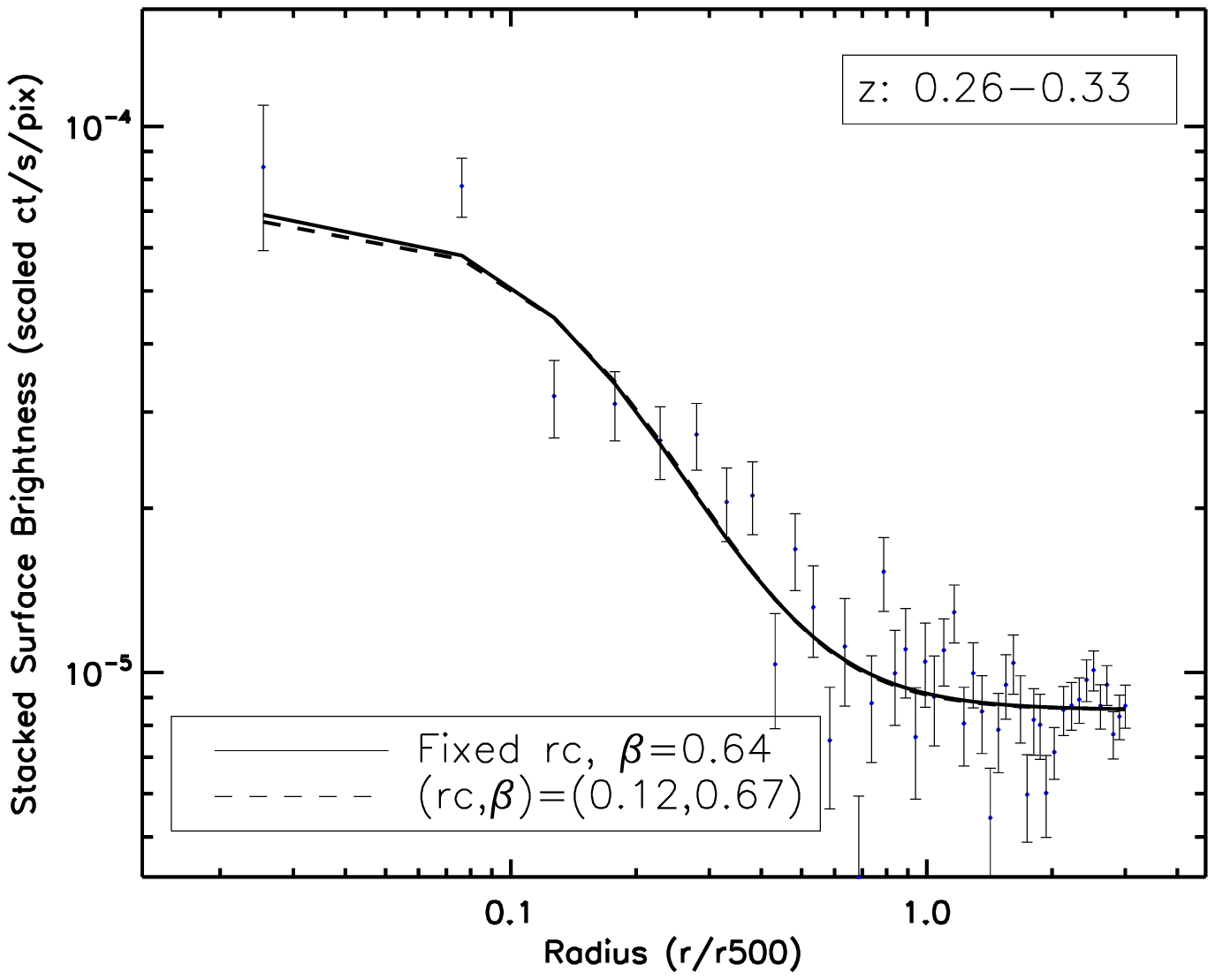,width=9cm,height=7cm}
\epsfig{file=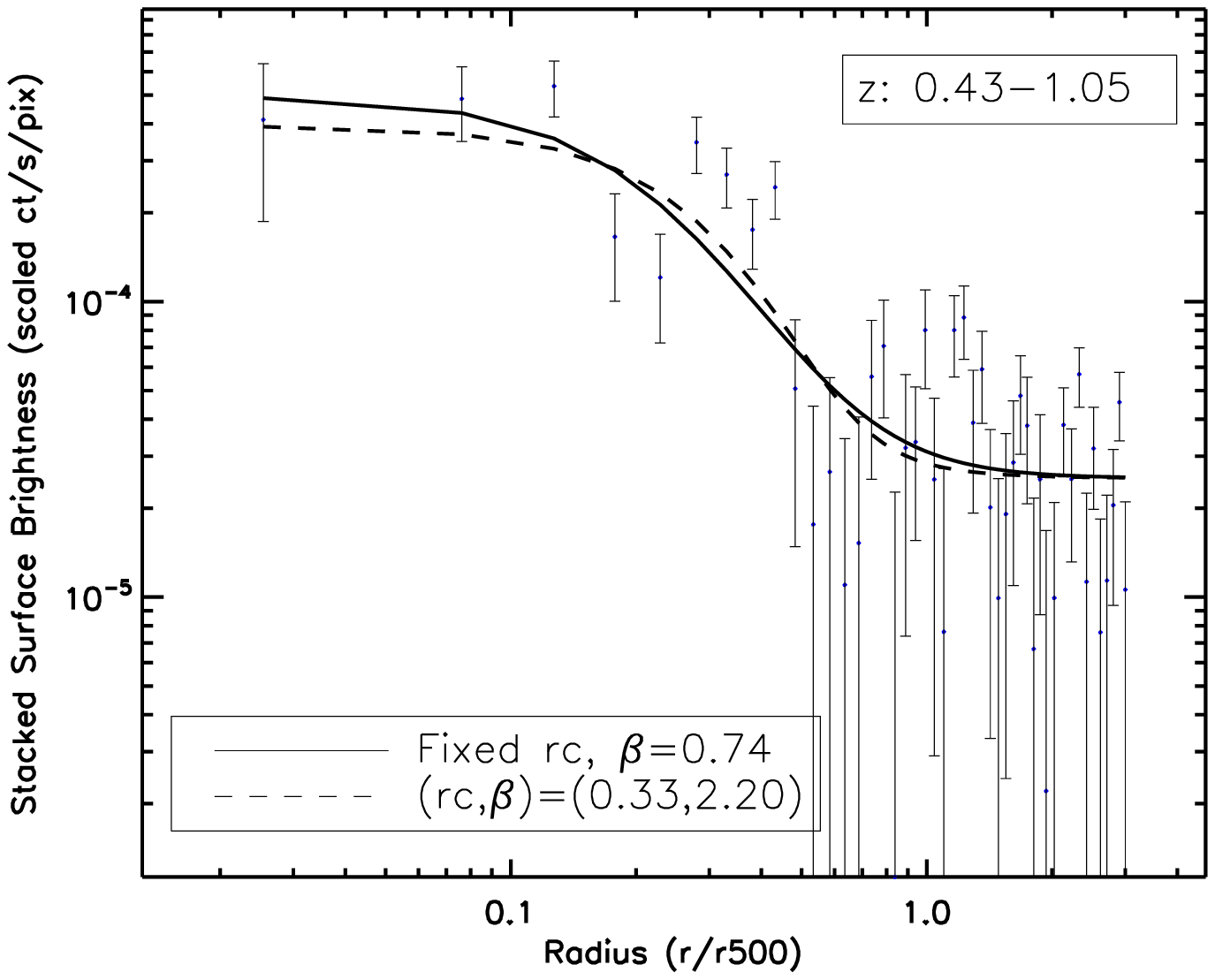,width=9cm,height=7cm}
\caption{Stacked profiles for the intermediate- and high-redshift 
cluster subsamples extracted from hard band (2.0-4.5 keV). These can
be compared directly with the corresponding soft band stacks
shown in Fig. \ref{zstack_prof}, and shown that the central
excess above the fixed-\rc\ \bmodel\ is not present in the hard band.}
\label{hardprofiles}
\end{figure}

There would be significant spectral differences between central excesses
generated by CCs and AGN, since the thermal emission from cool cores
is much softer than the X-ray spectra of AGN.
The limited statistics for individual clusters in our sample do
not permit us to investigate whether the core emission is soft or hard.
However, this can be investigated using the stacked data.
We therefore repeated the 
stacking analysis for intermediate- and
high-$z$ clusters using X-ray images derived from the hard energy band,
2.0-4.5 keV. The results are shown in Fig. \ref{hardprofiles}. Comparing
with the corresponding soft (0.5-2.0 keV) band profiles in
Fig. \ref{zstack_prof}, we notice the disappearance of the central excess
above the \bmodel\ in the hard-energy profiles in both the intermediate
and high redshift stacks. This provides strong evidence that this central
excess does not arise from AGN in cluster cores.

Assuming that the cuspy profiles really do indicate the presence of
CCs, the observation of such cores in groups at high redshift can help
to constrain the reasons for their absence in high $z$ clusters.
The decline in CC clusters with redshift could result from 
disruption of CCs due to the higher
merging rates at high redshifts (e.g., \citealt{Cohn05} and \citealt{Jeltema05})
or from the effects of preheating (\citealt{McCarthy04}), which
can raise the gas entropy and prevent cooling. In the latter case,
the impact of a given entropy boost will be larger in cooler
systems (see e.g. \citealt{Borgani05}), especially
at high redshift, so one would expect if anything to see a {\it drop}
in the incidence of CCs within groups at high $z$, at least as large
as pronounced as that in richer clusters.

The hypothesis that CCs are destroyed by cluster mergers appears
more promising, since this
effect might be stronger in more massive systems. For example,
\cite{Burns08} find that CCs are more common
in low mass clusters, and attribute the lack of CCs in more massive
systems to their destruction by early mergers in systems destined
to grow into large clusters. On the other hand, these authors
caution that their model does not reproduce the observed reduction with
redshift in CCs within massive clusters. In fact, no numerical simulations
have yet succeeded in adequately reproducing the properties of
cluster cores.

\subsubsection{Trends of $\beta$}

\begin{figure}
\epsfig{file=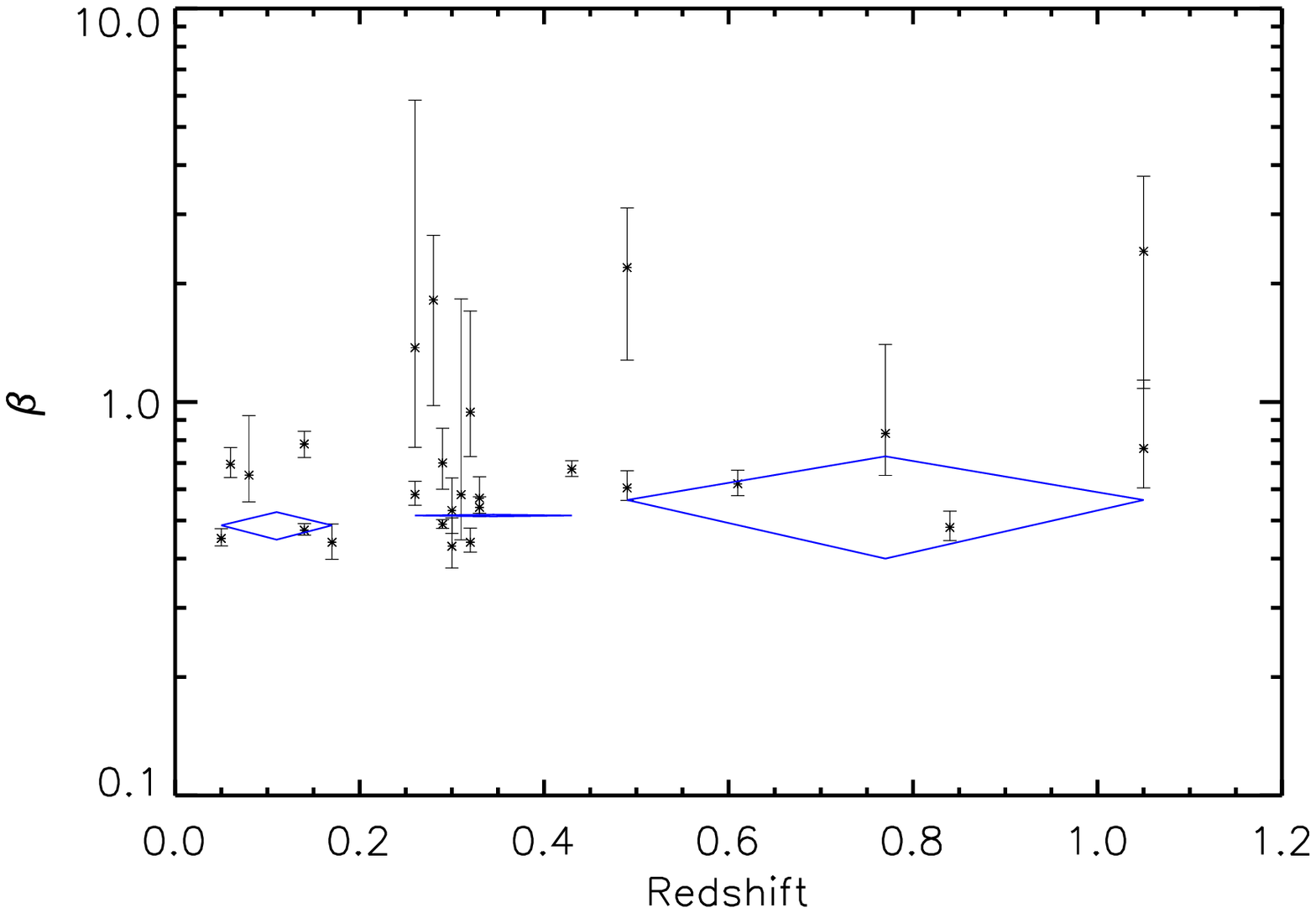,width=9cm,height=7cm}
\caption{$\beta$ values (of the free \rc\ fits) versus redshift of the individual C1 clusters.  The positions and sizes of the diamonds are calculated as described in \protect appendix \ref{sewm}.}
\label{d_beta_z}
\end{figure}

\begin{figure}
\epsfig{file=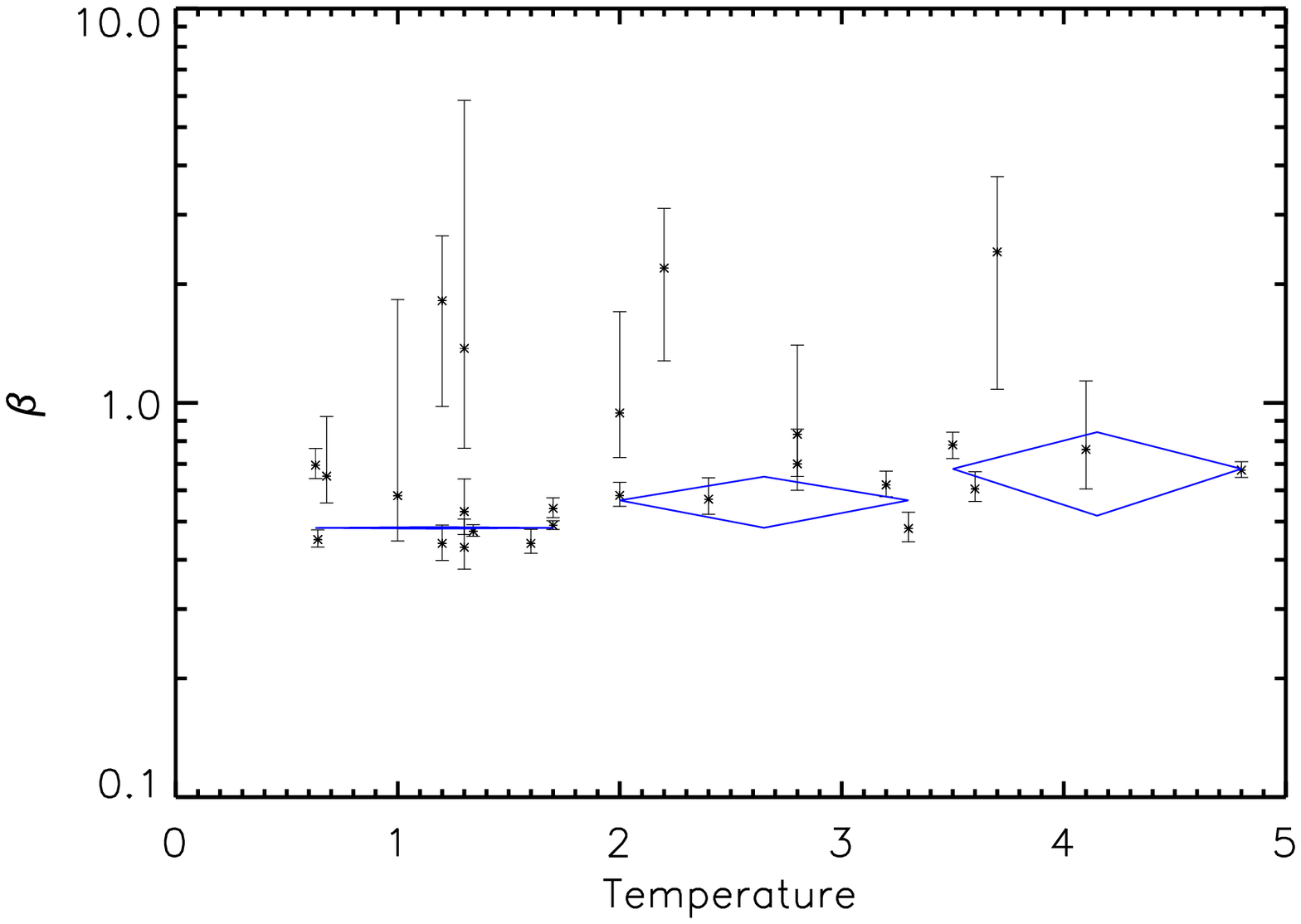,width=9cm,height=7cm}
\caption{$\beta$ values (of the free \rc\ fits) versus temperature of the individual C1 clusters. The positions and sizes of the diamonds are calculated as described in \protect appendix \ref{sewm}.}
\label{d_beta_t}
\end{figure}

Whilst it seems quite clear, as discussed above, that the trend
in central cuspiness is primarily related to redshift, rather
than temperature, the situation with regard to $\beta$ is not so 
straightforward. In the stacked datasets, comparison of 
Fig.~\ref{zstack_prof} and \ref{tstack_prof} suggest that the relationship
with redshift is stronger: $\beta$ rises monotonically through the
three redshift intervals, whilst in temperature the only clear result
is that the cool systems have lower $\beta$. In contrast, the
fits to individual clusters (Fig.  \ref{d_beta_z} and \ref{d_beta_t})
show a more pronounced trend with temperature than with redshift.
The stacked subsets of the narrow  temperature and redshift ranges
(Fig. \ref{tsub_prof} and \ref{zsub_prof}) produce ambiguous results:
$\beta$ increases with temperature if \rc\ is fixed, but with $z$ if 
\rc\ is left free to fit.

So, on the basis of our data alone, we are unable to say whether
the general trend in $\beta$ is
driven by the temperature or the redshift. However, evidence from studies
of low redshift groups and clusters is very relevant here.
Such studies provide clear evidence of a positive correlation between
$\beta$ and temperature in local systems systems --
e.g., \cite{Osmond04b} and Fig. 7 in the study of \cite{Croston08}, who
analysed clusters with redshift $<0.2$. Combining these previous results with
ours, favours a trend in $\beta$ with temperature (and hence cluster mass)
rather than an evolutionary effect. Our results
then demonstrate that this trend is still present in groups and clusters
at $z \sim 0.3$.

\section{Summary and conclusions}

In this paper, we used \textit{XMM-Newton} observations of 27 X-ray selected
galaxy clusters spanning the redshift range ($0.05 \leq z \leq 1.05$) to study
the spatial properties of their ICM. Most of these clusters fall in the realm
of low-mass clusters or groups, with ICM temperatures from 0.63 to 4.80~keV.
The XMM data provide typically a few hundreds X-ray source counts. We extracted and vignetting-corrected the
profiles to 3$\times$\rfh\ where they flattened and reached the photon
background values, which were estimated locally for each cluster.

In addition to the individual profiles, we also stacked the profiles into
three redshift and temperature bins. To explore the effects of \textit{Malmquist
bias}, we further stacked clusters with similar redshifts/temperatures into
two subsets each with different averaged temperature/redshift. Both
individual and stacked profiles were fitted with blurred (to account for the
PSF errors of the \textit{XMM-Newton} cameras) \bmodel s with both
free and fixed core radii.
The fixed-\rc\ \bmodel\ fits were used to test
whether a profile showed evidence of a cuspy core,
making this study the first to probe the evolution of CCs out to
$z\gsim0.3$ within poor clusters.

Our main conclusions are:
\begin{itemize}
\item We find that 54\% of our sample show evidence for cool cores, in the
 form of a central excess (at $>1\sigma$ significance) above a standard
 \bmodel.  
\item For the free-\rc\ fits to individual clusters, the median value of
 $\beta$ is 0.61, and the median \rc\ is 0.08$\times$\rfh.
\item For the fixed-\rc\ fits to individual clusters, the median $\beta$
 is 0.63.  
\item Twelve systems in our sample (with $\overline{kT}=1.69$ keV) have $z\sim0.3$, allowing us to characterise the X-ray surface
 brightness profiles of intermediate redshift X-ray selected groups.  The
 free-fit parameters to the stacked data from these 12 systems gives
 $\beta=0.51\pm0.01$ and \rc=$0.06\pm0.01\times$\rfh. This stacked
 profile indicates the presence of CCs ($f_c=1.56\pm0.11$), with
 7 of the 12 systems showing a significant central excess in
 their individual profiles.
\item Stacked and individual profiles for our sample of poor galaxy
 clusters show that the CCs do not disappear at high redshift, but rather
 become more prominent, though one would like
 to confirm this result with higher spatial resolution observations.
\item The slope parameter, $\beta$, shows a positive trend with
 both redshift and temperature in our data. Combining this results with 
 previous findings, we incline towards a trend with temperature 
 (and hence mass) rather than redshift. The present study then demonstrates 
 for the first time, that the $\beta$-$T$ trend seen at low $z$ is also 
 present in groups and clusters at $z\gsim0.3$.
\end{itemize}

\section{Acknowledgements}
We thank Monique Arnaud for providing us with the \textsc{Fortran} code which
was used to calculate the point spread function redistribution matrix, and 
Ewan O'Sullivan for useful discussions about AGN contamination in cluster
cores. XMM-Newton is an ESA observatory.


\appendix

\section{Standard error on weighted mean in presence of real scatter}
\label{sewm}
When averaging data of variable statistical quality, a more robust mean
is obtained by weighting the averaged values by their inverse
variances. Standard formulae for the standard error on such a 
weighted mean assume that statistical errors represent the {\it only}
source of variance. For our application, this is not true, since
there are real cluster-to-cluster variations, in additional
to statistcal scatter. Here we derive an expression for the standard
error of a weighted mean in these circumstances.

For a data set $x_i=x_1,x_2,..,x_n$ with variable statistical errors
$\sigma_i=\sigma_1,\sigma_2,..,\sigma_n$,
the weighted mean $\bar{x}$ is
\begin{equation} \bar{x}=\frac{\displaystyle\sum_{i=1}^{n}w_{i}x_{i}}{\displaystyle\sum_{i=1}^{n}w_{i}}, \end{equation}
where $w_{i}=1/\sigma_{i}^{2}$ are the weights. This weighted mean will
properly take into account the varying statistical quality of the data. 

The variance in this
the weighted mean, $var(\bar{x}),$ is
\begin{equation} var(\bar{x})=\frac{\displaystyle\sum_{i=1}^{n}w_{i}^{2}
var(x_{i})}{\left(\displaystyle\sum_{i=1}^{n}w_i\right)^2}. \end{equation} 

In the presence of real, non-statistical scatter in the $x$ values,
the expected variance for the \textit{ith} data point is 
\begin{equation} var(x_i)=\left\langle
(x_i-\mu)^2 \right\rangle= \sigma_i^2+\sigma_t^2, \end{equation} 
where $\sigma_t$ is the \textit{true} (non-statistical) variance of the population 
and 
\begin{equation} \left\langle (x_i-\mu)^2
\right\rangle=\frac{n}{n-1} \left\langle (x_i-\bar{x})^2 \right\rangle. \end{equation}
So an estimate $\hat{\sigma}^2_t$ of $\sigma^2_t$ is obtained from
\begin{equation} \hat{\sigma}^2_t=\frac{1}{n-1}\displaystyle\sum_{i=1}^{n}\left[
(x_i-\bar{x})^2-\sigma_i^2 \right], \end{equation} 
and the variance of the weighted
mean becomes \begin{equation} var(\bar{x})=\frac{\displaystyle\sum_{i=1}^{n}w_i^2\left[
\sigma_i^2+\frac{1}{n-1} \displaystyle\sum_{i=1}^{n}\left[
(x_i-\bar{x})^2-\sigma_i^2 \right] \right]
}{\left(\displaystyle\sum_{i=1}^{n}w_i\right)^2}. \end{equation} 
Substituting for $w_i=1/\sigma_i^2$ we get
\begin{equation} var(\bar{x})=\frac{\displaystyle\sum_{i=1}^{n}\left[
\frac{1}{\sigma_i^2}+\frac{\hat{\sigma}_t^2}{\sigma_i^4} \right]
}{\left(\displaystyle\sum_{i=1}^{n}\frac{1}{\sigma_i^2} \right) ^2}, \end{equation} 
and finally the standard error of the weighted mean (SEWM), $S\!E_{\bar{x}}$,
is
\begin{equation} S\!E_{\bar{x}}=\sqrt{\frac{var(\bar{x})}{n}}=\sqrt{\frac{\displaystyle\sum_{i=1}^{n}\left[\frac{1}{\sigma_i^2}+\frac{\hat{\sigma}_t^2}{\sigma_i^4}
\right] }{n\left(\displaystyle\sum_{i=1}^{n}\frac{1}{\sigma_i^2}
\right)^2}}. \end{equation}


\bsp
\label{lastpage}
\end{document}